\documentclass[final]{foresj}
\numberwithin{equation}{section}
\numberwithin{figure}{section}
\numberwithin{table}{section}
\usepackage{tabularx}
\usepackage{subcaption}
\usepackage{float}
\usepackage{multicol}
\usepackage[table,xcdraw]{xcolor}
\usepackage{cancel}
\usepackage{listings}
\usepackage{amsmath}
\usepackage[mathscr]{euscript}
\usepackage[utf8]{inputenc}
\usepackage{multirow}
\usepackage{tikz}
\usepackage{chngcntr}

\usepackage{fancyvrb,newverbs}
\definecolor{cverbbg}{gray}{0.93}
\newverbcommand{\cverb}
  {\setbox\verbbox\hbox\bgroup}
  {\egroup\colorbox{cverbbg}{\box\verbbox}}

\nolinenumbers  

\DeclareSymbolFont{rsfs}{U}{rsfs}{m}{n}
\DeclareSymbolFontAlphabet{\mathscrsfs}{rsfs}

\Year{}

\usepackage[square,numbers,sort&compress,comma]{natbib}
\usepackage{breqn}
\newcommand\changecol[1]{
\newline
\newline
\newline
\newline
\newline
\newline
\newline
\newline
\newline
\newline
\newline
\newline
\newline
\newline
\newline
\newline
\newline
\newline
\newline
\newline
\newline
\newline
\newline
\vspace{1.5mm}
\newline}
\begin{document}

\title{Modelling of human exhaled sprays and aerosols to enable real-time estimation of spatially-resolved infection risk in indoor environments}

\author[\emph{Imperial College London}]{Daniel \surname{Fredrich}$^*$, Aliyah M. \surname{Akbar}, Muhammad Faieq \surname{bin Mohd Fadzil}, Afxentis \surname{Giorgallis}, \\Alexander \surname{Kruse}, Noah \surname{Liniger}, Lazaros \surname{Papachristodoulou} and Andrea \surname{Giusti}$^{*}$ \\}

\address{Imperial College London, Department of Mechanical Engineering, London SW7 2AZ, UK}

\date{2 September 2021}

\begin{abstract}
A numerical framework for the `real-time' estimation of the infection risk from airborne diseases (e.g., SARS-CoV-2) in indoor spaces such as hospitals, restaurants, cinemas or teaching rooms is proposed. The developed model is based on the use of computational fluid dynamics as a pre-processor to obtain the time-averaged ventilation pattern inside a room, and a post-processing tool for the computation of the dispersion of sprays and aerosols emitted by its occupants in `real time'. The model can predict the dispersion and concentration of droplets carrying viable viral copies in the air, the contamination of surfaces, and the related spatially-resolved infection risk. It may therefore provide useful information for the management of indoor environments in terms of, e.g., maximum occupancy, air changes per hour and cleaning of surfaces. This work describes the fundamentals of the model and its main characteristics. The model was developed using open-source software and is conceived to be simple, user-friendly and highly automated to enable any potential user to perform estimations of the local infection risk.

\vskip1pt
\end{abstract}

\maketitle

\noindent \rule{0.49\textwidth}{0.5pt}
\vspace{-2cm}

\section*{Contents}
\vspace{3mm}
1~~Introduction \dotfill 1 \\
\\
2~~Droplet dispersion model \dotfill 4 \\
\\
3~~Effect of thermal plumes \dotfill 9 \\
\\
4~~Modelling the dispersion of breath \dotfill 17 \\
\\
5~~Ventilation field in indoor environments \dotfill 24 \\
\\
6~~Droplet dispersion under thermal plumes \dotfill 31 \\
\\
7~~Infection risk estimation \dotfill 36 \\
\\
8~~Summary and future work \dotfill 38 \\

\vspace{0.6mm}

\noindent \rule{0.49\textwidth}{0.5pt}

\noindent $^*$\emph{Corresponding authors}:

Dr Andrea Giusti, a.giusti@imperial.ac.uk \\
\indent (ORCID: 0000-0001-5406-4569) \\
\indent Dr Daniel Fredrich, df1615@imperial.ac.uk \\
\indent (ORCID: 0000-0003-2207-4679)

\section{Introduction}
\vspace{3mm}
\noindent By \emph{A. Giusti and D. Fredrich}
\vspace{3mm}

\noindent The Covid-19 (SARS-CoV-2) pandemic and the related measures adopted by governments around the world to contain the spread of the virus have drastically changed our daily lives. Full and partial lockdowns have become quite frequent to keep infection rates under control and prevent hospitals and intensive care units from being overwhelmed. At the same time, social interactions have drastically decreased and a new way of working and communicating has been established. Smart working (work from home) and meetings through online platforms are now part of our everyday routine. Considering how fast these changes have been introduced, one could argue we are part of a revolution that will probably change our lives forever.

Although some of the changes determined by the current pandemic have had a positive impact on our lives, for example the reduction of transportation and related pollutant emissions as well as tailored and flexible working hours, the lack of direct human interaction has also had its downsides. Lockdowns and restrictions to mobility have affected many business activities causing huge losses (e.g., restaurants, pubs, cinemas, theatres, to name a few) and have driven many families into financial crises. Universities, schools and other educational institutions are one of the most affected areas. With the transition to remote lectures, teaching has become less effective. Successful learning is a combination of several factors and requires interaction of students with their peers, as well as interaction with the teachers. An environment that lets students focus on the new material proposed by the modules and the possibility of interacting with course mates immediately after the lecture to verify the understanding or simply to compare notes are fundamental elements of the learning process. Remote lectures have made the learning experience `flat' and `dry'. All of the lectures happen in the same room (the students' bedroom in most cases) on a two-dimensional screen. The students are not fully immersed in the lecture, compared to when they were in a lecture room and it is more difficult to focus as well as to associate each lecture to a specific environment. In other words, the students are not immersed into the lecture anymore and keeping up the attention is more difficult. Students also do not have the possibility of catching up with their class mates, slowing down the learning process and decreasing motivation.

Teaching in university requires more than ever to `go back to normality'. Ideally, we would like to see again lecture rooms and tutorial rooms fully populated by students. However, at the same time we have to guarantee the students' and teachers' safety in terms of infection risk. Although there is still an ongoing debate on what the possible routes for viral transmission are, it is quite widely accepted that the most important contamination routes are (i) direct contact with contaminated surfaces and (ii) inhalation through the air of saliva drops or aerosols emitted by an infected person. The second route has opened a long debate on the role of big and small drops. The former (diameter above 5~microns, as a rule of thumb) are likely to settle down on the floor within a few meters under the action of gravity. In this case, the so called `2-meter' rule could be effective to protect us from infection in a environment with relatively low levels of turbulence and ventilation. However, the smaller droplets usually stay suspended for a longer time, forming an aerosol. Depending on the room's ventilation pattern, they may be transported throughout the entire room making any social-distance measure ineffective. In the case of aerosols, protection is mainly based on keeping the viral load (e.g.,~amount of inhaled virus over a given time) below a given threshold. From a building management point of view, to decrease the infection risk in a closed (indoor) environment, one could act on the number of people in the room, the time they stay in the room, or on the ventilation (e.g.,~changes of volume per hour). To take effective actions, it is of primary importance to have tools that are capable of estimating the infection risk for a given ventilation pattern and based on the position of occupants in the room. The objective of this work is, therefore, to develop a computational framework to estimate the infection risk in indoor environments that takes into account both the ventilation pattern and the location of the occupants.

\begin{figure}[t]
    \centering
    \includegraphics[width=\linewidth]{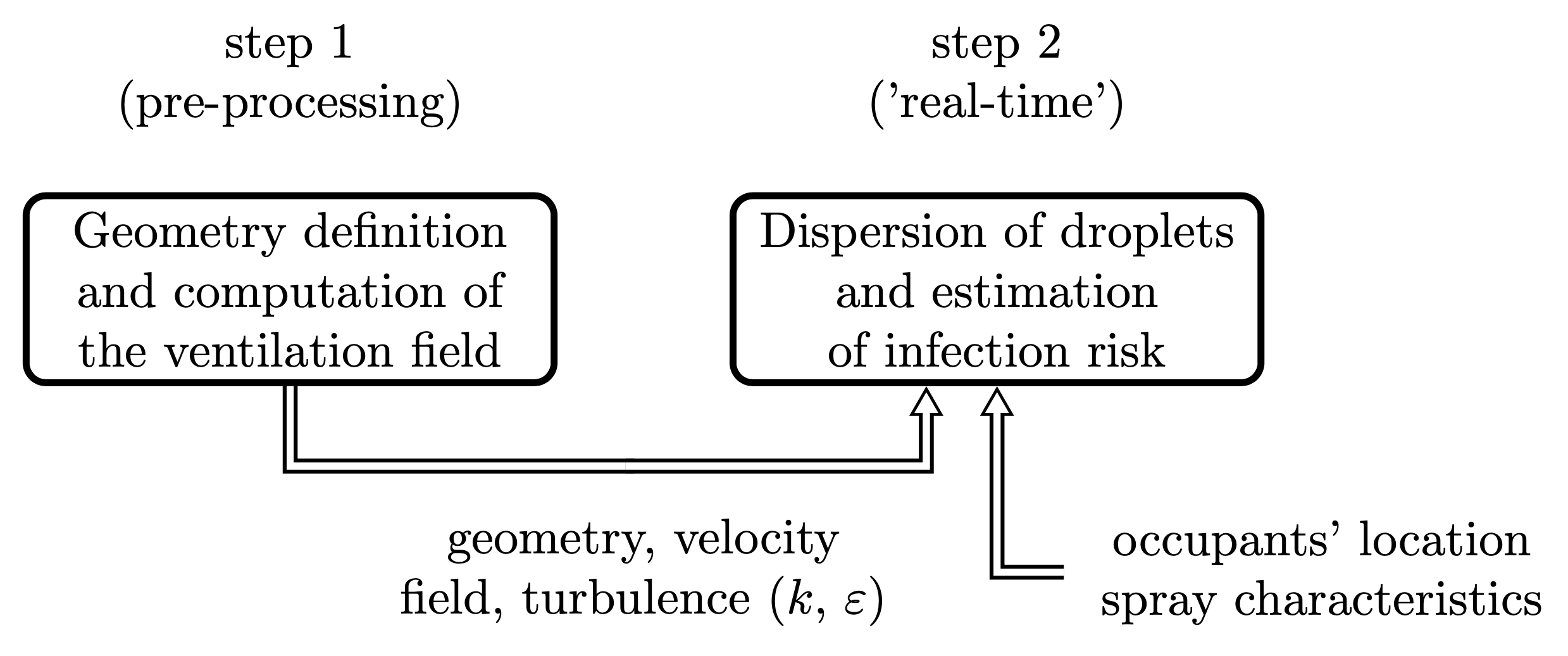}
    \vspace{-10pt}
    \caption{Schematic of the two main elements (pre-processor and post-processor) of the proposed numerical method.}
    \label{fig:modelSchematic}
\end{figure}

This report discusses the fundamental components of the model and their assessment and validation. The long-term aim is to make real-time estimations of the infection risk using computers available in the lectern of teaching rooms. Therefore, the model is conceived to be simple, highly automated and computationally affordable. This means that some elements of the model, although based on careful analysis with advanced tools (such as unsteady computational fluid dynamics), have been developed with some degree of `engineering' to represent the physics with sufficient accuracy but at the same time with affordable computational cost. This is for example the case of correction of the background ventilation flow field to consider pulsed jets issued by the mouth of the occupants. The basic idea of the model is to split the computation of the dispersion of saliva droplets into two steps, as schematically shown in Figure~\ref{fig:modelSchematic}. The first step is run in pre-processing and is the computation of the ventilation pattern of the room using computational fluid dynamics. The model proposed here is fully automated. Both the geometry and the simulation can be run with user-friendly input parameters. Results of the simulations are used to build a database of ventilation patterns (velocity and turbulence fields) to be used for the `real-time' prediction of the dispersion of droplets. The strategy used to model both lecture and tutorial rooms will be discussed in this report together with the effect of turbulence modelling on the predicted flow field. In principle, simulations both with occupants fixed in space and without occupants can be run. All possible combinations of occupants/ventilation flow rates can be run to build a database, although the number of simulations could become very large. Our recommendation is to first run simulations with no occupants and then inject droplets from the location of the occupants in the second stage of the model. To make this approximation more reliable, we have been studying how to change the background ventilation pattern to consider the presence of people in terms of both an exhaled plume from the mouth and a thermal plume created by buoyancy due to the temperature difference between the human body and the surrounding air. Once the ventilation flow field is computed and stored in a database, the second step of the model consists in the computation of the dispersion of droplets issued by each of the occupants. The computation, based on Lagrangian tracking of a statistically representative sample of droplets, starts with the injection of a spray in correspondence of the mouth of each occupant. Droplets are then tracked in time and space to build statistics of their dispersion in the room as well as contamination of the surfaces. The dispersion model uses the turbulence characteristics from the computational fluid dynamics simulations to model the turbulent dispersion of the droplets. Different turbulent dispersion models have been tested and they will be discussed. In addition, models to estimate the carbon dioxide (CO$\mathrm{_2}$) dispersion in the room, an important factor for air quality independently of viral presence, have been implemented to allow to monitor the air quality. Models based on both continuous jets and pulsed puffs will be discussed.

The report is organised as follows. Section~\ref{sec:dispersion} describes the fundamentals of the modelling of dispersion of saliva droplets. The statistical distribution of saliva droplets is discussed together with different approaches to model the effect of turbulence on the motion of individual droplets. Section~\ref{sec:tPlume} provides more insights into the thermal plume generated by heated bodies with the main aim of quantifying the effect of the buoyancy air flow from human bodies on the ventilation field. The questions we want to address are: does the thermal plume generated by human bodies significantly affect the velocity field? And if so, in what conditions? To give an answer to these questions, computational fluid dynamics simulations are performed. The numerical solver used to reproduce buoyancy effects is first validated against experimental data from the literature. Then, the solver is applied to the study of heated human bodies. Section~\ref{sec:Breath} introduces models to reproduce the evolution and dispersion of CO$\mathrm{_2}$ exhaled by occupants. The models considered in this report are inherited from the modelling of dispersion of atmospheric pollutants released by stacks. In particular, the evolution of both continuous and pulsed jets are explored. In addition, the modelling of the velocity flow field in the vicinity of the mouth is addressed. Such model is used to locally modify the computed ventilation flow field (in pre-processing) to take into account the effect of breathing. Section~\ref{sec:cfd} discusses the approach used to model the geometry of both tutorial and lecture rooms. Computational fluid dynamics simulations are then performed to study mesh independence and sensitivity to the turbulence modelling. Recommendations on the grid refinement and turbulence model for a reliable prediction of the ventilation field are identified. Section~\ref{sec:combo_plumes} combines the velocity field computed for thermal plumes with the spray dispersion computations to study the effect of a thermal plume on the dispersion of aerosols. Finally, Section~\ref{sec:infectionRisk} describes the model used to compute the infection risk and shows an example of a computation of infection risk in a model tutorial room. A summary of the current achievements and recommendations for future work are provided in Section~\ref{sec:outlook}.

The numerical tools have been developed using open-source software to facilitate their use without the necessity for licenses. For computational fluid dynamics simulations, the OpenFOAM suite has been used. The code for the computation of droplet and CO$\mathrm{_2}$ dispersion has been developed in Python. Also, the scripts for pre-processing (e.g.,~generation of the lecture room geometry and mesh) have been written in Python. It is important to note that the developed tools can also be applied to any indoor environment and activity where occupants stay in a given location for a relatively long period of time. Examples of such activities include cinemas, theatres, and restaurants. The proposed model therefore has the potential to assist such businesses in developing tailored strategies for the management of indoor spaces, e.g., in terms of occupancy, with a direct impact on their economy and the economy of the country as a whole.
More information on the underlying code, including the coupling of the different tools as well as the graphical user interface, is available from the corresponding authors upon request.

\clearpage

\section{Droplet dispersion model}\label{sec:dispersion}

\vspace{3mm}

\noindent By \emph{A. Kruse, N. Liniger, A. Giusti and D. Fredrich}

\vspace{3mm}

\noindent This section describes the fundamentals of the model used to evaluate the dispersion of droplets in the indoor environment. The tracking of droplets follows the typical strategy used in Eulerian-Lagrangian methods. Droplets are injected at locations corresponding to the mouth of occupants and then tracked as Lagrangian material points. One-way coupling (i.e., the gas phase affects the droplet motion but not vice-versa) is assumed given the very low volume fraction of saliva droplets emitted through breathing. This assumption can be considered reasonable since only sporadic events like sneezing or coughing could lead to relatively high saliva volume fractions in the vicinity of the mouth. Note that the one-way coupling approximation, together with the steady-state assumption, also allows us to compute the ventilation flow field in pre-processing and keep it constant for the computation of the droplet dispersion. Following common practice in Eulerian-Lagrangian methods, droplets are represented by numerical parcels, that is `clusters' of droplets with the same characteristics (e.g., diameter, temperature). A statistically representative sample of parcels is injected and tracked until they completely disappear, i.e., exit the domain or deposit on a surface. In the following, the main models used in the computation of the dispersion of saliva droplets are discussed. 

\subsection*{Droplet size distribution and evaporation}

The droplet size is an important factor that determines the behaviour of the dispersion of the exhaled droplets. Hence, a model of size distribution was implemented, so that the size of each injected droplet is sampled from a statistical distribution representative of speaking or coughing activities.

During exhalation, different sizes of droplets exit the mouth ranging from 1~$\mu$m to 1~mm. The size distribution is given by a probability density function, which varies according to the mode of exhalation; either speaking or coughing. The functions were adopted from Ref.~\cite{Oliviera} and are based on the tri-modal log-normal distribution provided in Ref.~\cite{Johnson}. The model is named the bronchiolar-laryngeal-oral tri-modal model, which combines the droplet release from three different areas within the respiratory system: the lower respiratory tract, the larynx and the upper respiratory tract/oral cavity. Equation \ref{eq:SizeModel} provides an expression for the logarithmic number concentration gradient for different droplet sizes:
\begin{multline}
\label{eq:SizeModel}
    \frac{\mathrm{d} \mathrm{Cn}_{k}}{\mathrm{d} \log_{10}(d_k)} =
    \ln{(10)} \sum ^{3}_{i = 1} \Bigg[\left(\frac{\mathrm{Cn}_{i}}{\sqrt{2\pi}\ln{\mathrm{GSD}_i}}\right) \times
    \\
    \exp{\left(-\frac{(\ln{(d_k)} - \ln{(\mathrm{CMD}_i)})^2}{2(\ln{\mathrm{(GSD_i)}})^2}\right)}\Bigg].
\end{multline}
The number of droplets of size $d_k$ is given by the sum over each mode $i$. Cn$_i$ is the droplet number concentration, $d$ is the diameter, GSD$_i$ is the geometric standard deviation and CMD$_i$ is the count median diameter. The values of these coefficients are given in Table~\ref{tab:BLOParams}~\cite{Johnson}.

\begin{table}[t]
\caption{BLO parameters from Refs.~\cite{Johnson,Oliviera}.}
\label{tab:BLOParams}
\begin{tabular}{lccc}
\hline
\multicolumn{1}{c}{\textbf{i}} & \textbf{1}        & \textbf{2}        & \textbf{3}        \\
\multicolumn{1}{c}{}                   & {(B mode)} & {(L mode)} & {(O mode)} \\ \hline
\multicolumn{4}{l}{\textbf{Speaking}}                                             \\ \hline
Cn$_i$ ($\mathrm{N}_p$ cm$^{-3}$)                   & 0.069    & 0.085    & 0.001    \\ 
CMD$_i$ ($\mu$m)                        & 1.6      & 2.5      & 145      \\ 
GSD$_i$                                 & 1.3      & 1.66     & 1.795    \\ 
Cm$_i$ ($\mu$g cm$^{-3}$)               & 0.21     & 2.2      & 7500     \\ \hline
\multicolumn{4}{l}{\textbf{Coughing}}                                             \\ \hline
Cn$_i$ ($\mathrm{N}_p$ cm$^{-3}$)                   & 0.087    & 0.12     & 0.016    \\ 
CMD$_i$ ($\mu$m)                        & 1.6      & 1.7      & 123      \\ 
GSD$_i$                                 & 1.25     & 1.68     & 1.837    \\ 
Cm$_i$ ($\mu$g cm$^{-3}$)               & 0.22     & 1.09     & 69000    \\ \hline
\end{tabular}
\end{table}

\begin{figure}[t]
    \centering
    \includegraphics[width=\linewidth]{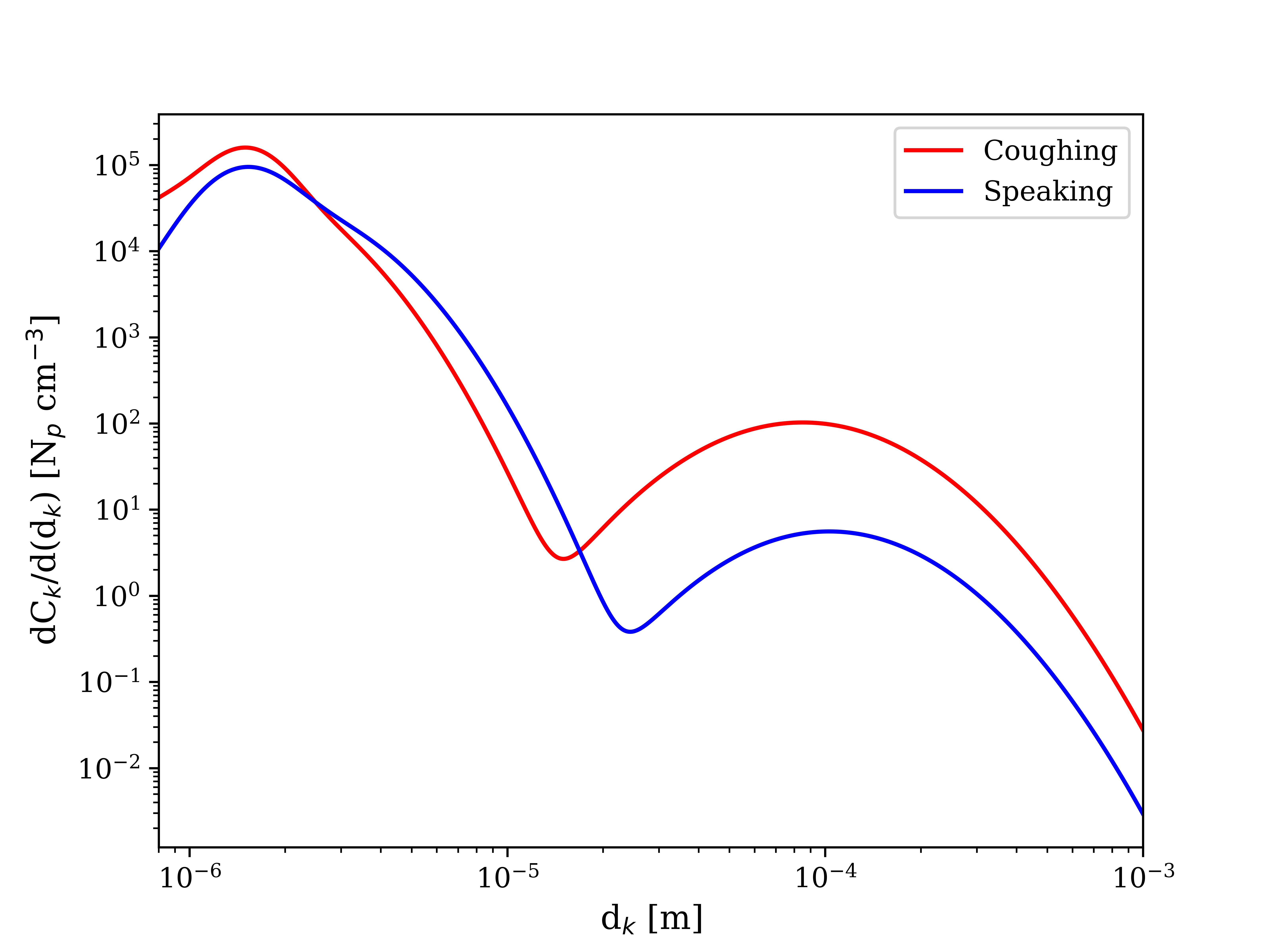}
    \vspace*{-6mm}
    \caption{Droplet number concentration for different diameters.}
    \label{fig:concentrationgrad}
\end{figure}

\begin{figure}[t]
    \centering
    \includegraphics[width=\linewidth]{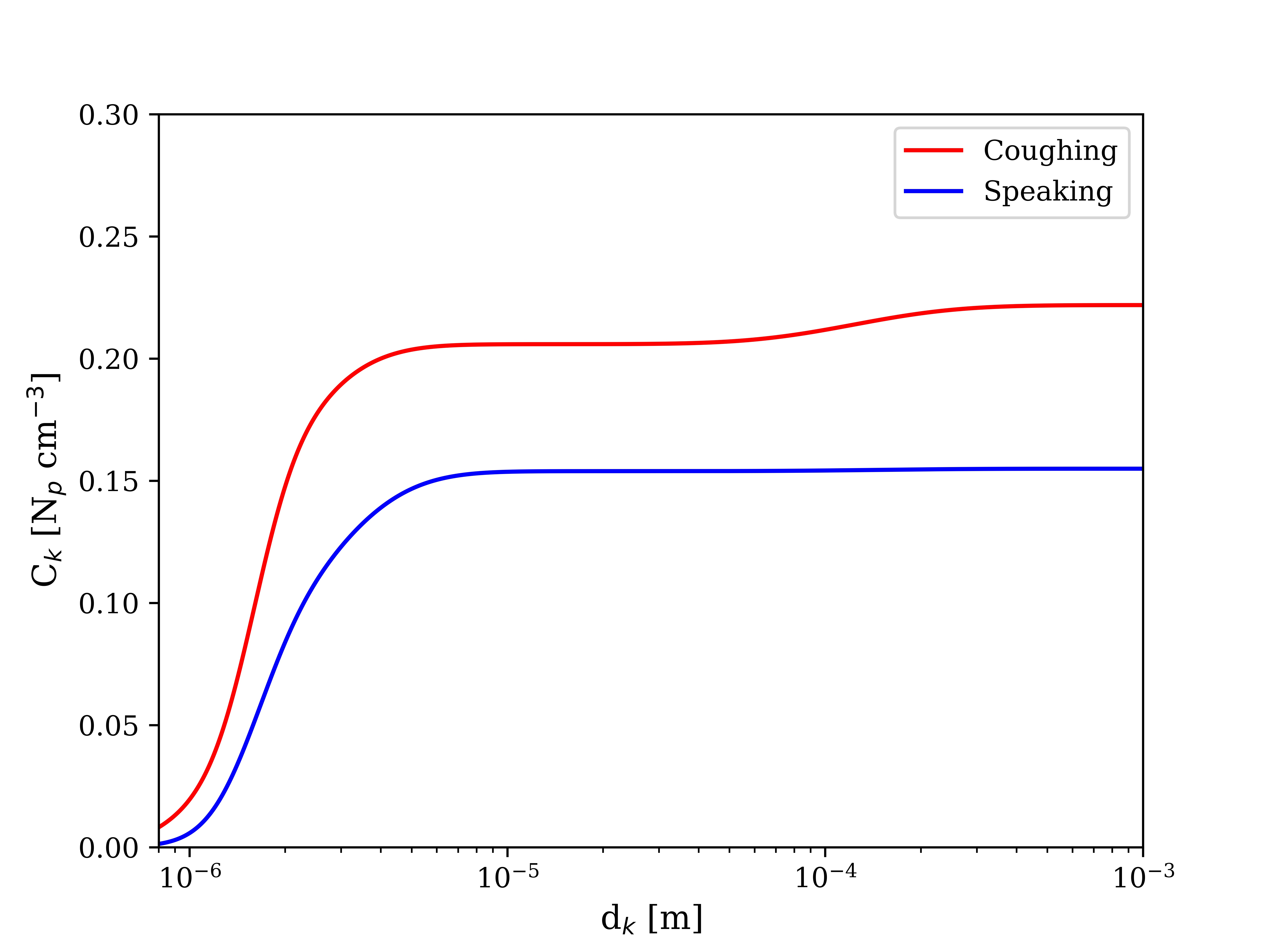}
    \vspace*{-6mm}
    \caption{Cumulative distribution function.}
    \label{fig:cumulative}
\end{figure}

Figure~\ref{fig:concentrationgrad} shows the droplet number concentration plotted against the droplet diameter, as described by Equation~\ref{eq:SizeModel}. The distributions for speaking and coughing were calculated independently. The expression for concentration cannot be integrated analytically, so a numerical trapezium integration rule was implemented to obtain the final cumulative distribution for coughing and speaking, shown in Figure~\ref{fig:cumulative}. For each injected parcel, the initial droplet diameter was determined through a random sampling from the cumulative number distribution. The sampling was performed by assigning a random value (generated from a uniform distribution) between the minimum and the maximum of the cumulative distribution, and then extracting the corresponding diameter from the curve in Figure~\ref{fig:cumulative}. Consequently, whilst speaking, droplets approximately the size of 2~$\mu$m are the most likely to be injected.

\subsubsection*{Mass concentration}
During the process of exhalation, the droplets released are modelled as parcels, i.e., a collection of droplets which are exhaled together as packages. The rate at which droplets are released is proportional to the volumetric flow rate leaving the mouth multiplied by the concentration of droplets per volume. This is summarised in Equation~\ref{eq:Np.vs.VP}:
\begin{equation}
    \label{eq:Np.vs.VP}
    \dot{\mathrm{N}}_p = \mathrm{C}_k^f \dot{V},
\end{equation}
where $\dot{\mathrm{N}}_p$ is the rate of droplets released, $\dot{V}$ is the volumetric flow rate and $\mathrm{C}_k^f$ is the asymptotic value of the cumulative distribution function for large values of $\mathrm{d}_k$. Multiplying the droplet injection rate by the period of a breath $T_b$, the number of droplets injected during a breath $\mathrm{N}_p$ may be obtained using Equation~\ref{eq:Np}:
\begin{equation}
    \label{eq:Np}
    \mathrm{N}_p = \dot{\mathrm{N}}_p \ T_b .
\end{equation}
Therefore, the number of droplets per injected parcel, $\mathrm{N}_p^*$, is given by Equation~\ref{eq:N_p*} as:
\begin{equation}
    \label{eq:N_p*}
    \mathrm{N}_p^* = \frac{\mathrm{N}_p}{\mathrm{N}_{inj}},
\end{equation}
where $\mathrm{N}_{inj}$ is the total number of parcels injected during the period of exhalation. From Equation~\ref{eq:N_p*}, the mass of the parcel, $m^*$, may be found as given by Equation~\ref{eq:m*}:
\begin{equation}
    \label{eq:m*}
    m^* = \mathrm{N}_p^*\frac{\pi}{6} d_k^3 \rho_p ,
\end{equation}
where $d_k$ is the diameter of the droplets within the parcel, sampled from the cumulative distribution function in Figure~\ref{fig:cumulative}, and $\rho_p$ is the density of the droplet.

\subsubsection*{Evaporation}

The exhaled droplets may evaporate whilst suspended in the air and therefore their diameter and hence mass may decrease over time. As the droplets reach a diameter of approximately 0.1~$\mu$m, their behaviour of motion changes and becomes comparable to a Brownian motion, as outlined further below in this section. It is therefore of importance to consider the effects of evaporation on the dispersion of droplets. In practical engineering applications a variety of rather simple models may be used to account for the effects of evaporation. Finding a compromise between accuracy and computational cost is also essential in the present work, given the aim of developing tools to be used in `real-time'.

\subsubsection*{The $\mathscrsfs{D}^2$-law}

A widely-used model for droplet evaporation is the so-called $\mathscrsfs{D}^2$-law. It proposes that the decrease in droplet diameter squared is linearly related to the time a droplet is suspended in the continuous phase. Therefore, the rate at which the surface area decreases is constant. This assumption correlates well with the observation that an object with a larger surface area dries faster~\cite{Jakubczyk12}.

\begin{figure}[t]
    \centering
    \includegraphics[width=\linewidth]{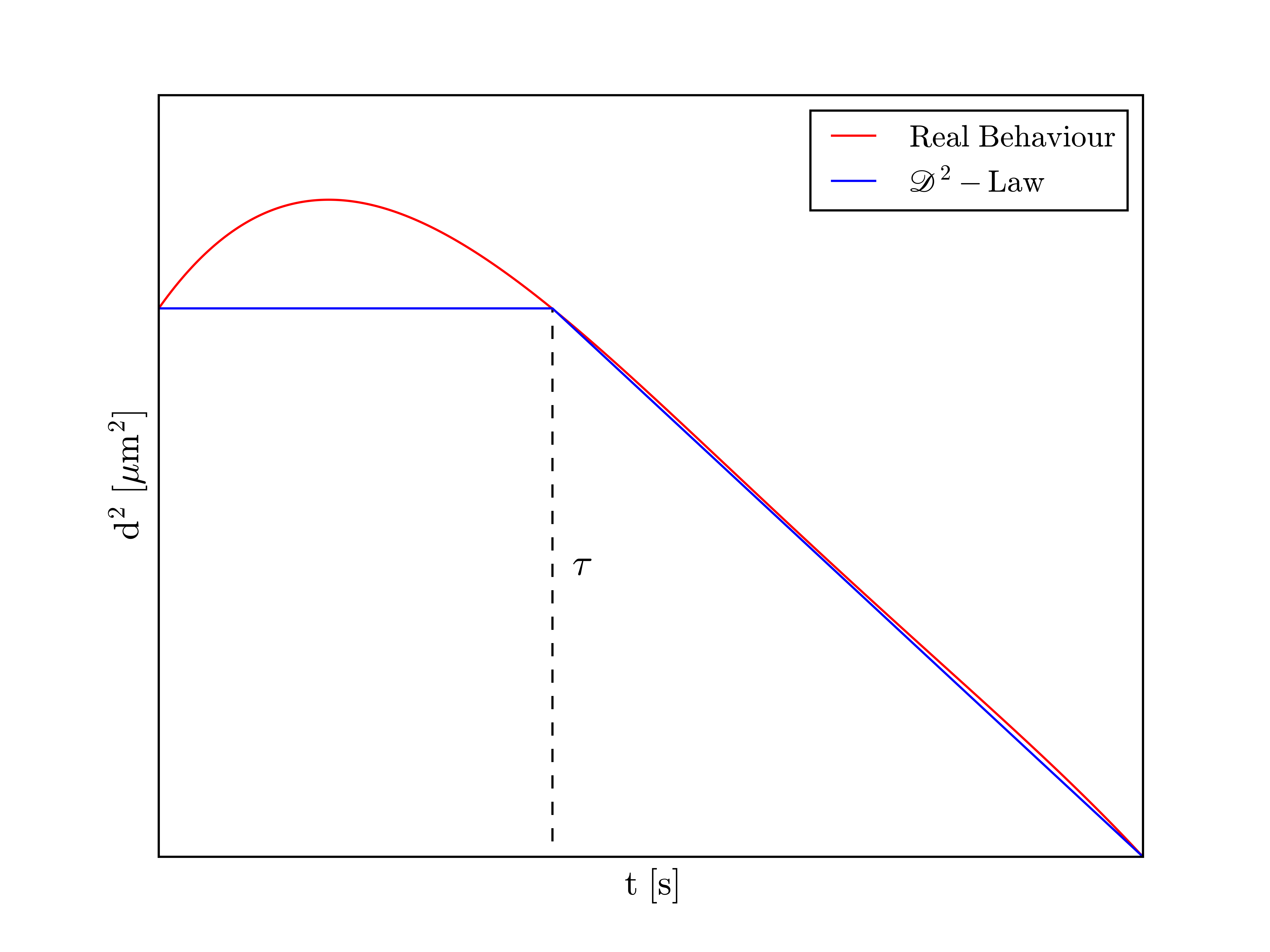}
    \vspace{-6mm}
    \caption{Qualitative plot of diameter evolution over time.}
    \label{fig:DSquared}
\end{figure}

In reality, the diameter squared usually follows a path shaped similar to a parabola, as can be seen in Figure~\ref{fig:DSquared}, where the initial part of the transient, usually referred to as `heat-up period', is caused by an increase of the droplet temperature. The duration of the `heat-up period' compared to the total evaporation time could change depending on the initial temperature of the droplet and its composition. Note that, in the case of saliva droplets, the initial temperature could also be higher compared to the room environment. In that case, during the initial transient, the droplet temperature could decrease and condensation of water vapor on the droplet surface may be observed. The behaviour of the droplet diameter could thus be different from the schematic shown in Figure~\ref{fig:DSquared}.

For the sake of simplicity, we model the initial transient of duration $\tau$ by assuming a constant droplet diameter. After this initial transient, the classical $\mathscrsfs{D}^2$-law, characterised by a linear decrease of the diameter squared, is used. This is described by Equation~\ref{eq:D2Law}:
\begin{equation}
\centering
\label{eq:D2Law}
d(t) =
\begin{cases}
d_0, & 0 \leq t \leq \tau,\\
\left(d_0^2 - A \left(t-\tau\right)\right)^{1/2}, & \ \ \ \ \ \, t > \tau,
\end{cases}
\end{equation}
where $d_0$ is the initial diameter at time $t = 0$ and $A$ is a constant which is determined from the droplet and gas-phase properties.

Saliva is a fluid manly consisting of water but also amounts of sodium, potassium, calcium, magnesium, bicarbonate and phosphates \cite{Saliva14}, which are considered impurities that affect the material constant $A$. As shown by Ref.~\cite{Jakubczyk12}, impurities can lead to a dramatic departure from the $\mathscrsfs{D}^2$-law. This was shown to be most significant when the concentration of impurities grows due to evaporation. The increased concentration causes the relation to deviate and creates a non-linear region. The evaporation of saliva droplets should be further investigated in the future at conditions relevant to ventilated indoor environments. If deviation from the $\mathscrsfs{D}^2$-law is deemed important, a more accurate evaporation model should be developed and implemented. Further studies are also required to get an accurate estimate of $A$ and $\tau$.

\subsection*{Dispersion models} \label{subsection: Dispersion Models}

All dispersion models implemented in the tool are based on the same underlying force, $\vec{F}$, balance of Equation~\ref{eq:Newton2ndLaw} (Newton's second law) applied in the Lagrangian frame of reference. Solving the force balance in the Lagrangian frame of reference allows us to easily model the dispersion of a polydisperse spray (in the Eulerian frame of reference an equation for each class of diameters must be solved, which could become computationally expensive -- see for example methods based on the population balance equation):
\begin{equation}
\label{eq:Newton2ndLaw}
    \frac{\mathrm{d}\vec{u}_p}{\mathrm{d}t} = \frac{\vec{F}}{m_p},
\end{equation}
\noindent where $m_p$ is the mass of the droplet. To find the position of the droplet at a given time, simple kinematics can be applied as in Equation~\ref{eq:PositionParticle}:
\begin{equation}
    \label{eq:PositionParticle}
    \frac{\mathrm{d}\vec{x}_p}{\mathrm{d}t} = \vec{u}_p.
\end{equation}
Equations~\ref{eq:Newton2ndLaw} and~\ref{eq:PositionParticle} are numerically integrated over time. At each time step, the new droplet velocity, $\vec{u}_p$, and position, $\vec{x}_p$, are obtained and used as initial conditions for the subsequent step. The presence of turbulence increases the dispersion of the droplets. The effect of turbulence is included in the simulation through dispersion models, which give a time-dependent contribution to the drag force. The evaluation of the force depends on the specific dispersion model selected for the simulation. The following sections describe the implemented dispersion models, i.e., the Langevin random walk and the Gosman model.

\subsubsection*{Langevin random walk}

For small droplets with diameters below 0.1~$\mu$m, the forces in Equation~\ref{eq:Newton2ndLaw} are due to microscopic collisions with fluid particles, which cause changes in the droplet velocity. At this order of magnitude of droplet size, the inertia is negligible and hence the instantaneous gas velocity is equal to the force acting on the droplet. The instantaneous gas velocity can be sampled from a continuous random walk (CRW) model adapted from Ref.~\cite{Mofakham} with a non-normalised Langevin type equation and a no drift correction term given by:
\begin{equation}
\label{eq:Langevin}
    \delta\vec{u}_p = (\vec{U}_g - \vec{u}_p)\frac{2 \, \varepsilon}{K} \, \delta t + \sqrt{2 \,  \varepsilon \, \delta t} \, \vec{\xi},
\end{equation}
where $\delta \vec{u}_p$ is the change in the droplet velocity over time step $\delta t$, $\vec{U}_g$ is the mean gas velocity vector, $\varepsilon$ is the turbulence dissipation rate, $K$ is the turbulence kinetic energy and $\vec{\xi}$ is a vector whose components are randomly sampled real numbers from a normal distribution with a mean of 0 and a standard deviation of 1.

Equation~\ref{eq:Newton2ndLaw} is used at each $\delta t$ to evaluate the change of $\vec{u}_p$ due to random collisions. A gravitational term may also be added, which however is not significant for droplets of such size. In the numerical integration, the updated droplet velocity is calculated explicitly from Equation~\ref{eq:LangevinPart} for every time step:
\begin{equation}
\label{eq:LangevinPart}
    \vec{u}_{p}\left( t + \delta t \right) = \vec{u}_{p} \left( t \right) + \delta \vec{u}_p .
\end{equation}

\subsubsection*{Gosman model}

For droplets above 0.1~$\mu$m in size, inertia is not negligible and therefore other forces need to be considered. The force in Equation~\ref{eq:Newton2ndLaw} is assumed to consist of the forces due to drag, gravity and buoyancy, as expressed in Equation~\ref{eq:Forces}, which was adopted from Ref.~\cite{Oliviera}. All other forces (e.g.,~Basset force, Saffman force, virtual mass) are neglected:
\begin{equation}
\label{eq:Forces}
    \vec{F} = \vec{F}_D + \vec{F}_g + \vec{F}_b .
\end{equation}
The drag force for a spherical droplet is: 
\begin{equation}
    \label{eq:dragforce}
    \mid \vec{F}_D \mid = \frac{1}{8} C_D \mid \vec{U}_g - \vec{u}_p \mid ^2 \pi d^2 \rho_g,
\end{equation}
\noindent where $\rho_g$ is the gas density. The droplet size is in the regime of Stokes drag, i.e., $\mathrm{St} \ll 1$, which implies a drag coefficient of~\cite{StokesDrag}:
\begin{equation}
    C_D = \frac{24}{\mathrm{Re}_p},
\end{equation}
\noindent where:
\begin{equation}
    \mathrm{Re}_p = \frac{\rho_{g} \, \mid \vec{U}_g - \vec{u}_p \mid \, d}{\mu_{g}},
\end{equation}
is the droplet Re number including the dynamic viscosity of the gas phase, $\mu_{g}$. The forces due to gravity, $\vec{g}$, and buoyancy may be expressed by Equations~\ref{eq:Gravity} and~\ref{eq:Buoyancy}, respectively:
\begin{equation}
    \label{eq:Gravity}
    \vec{F}_g = \frac{1}{6} \pi d^3 \rho_p \vec{g} ,
\end{equation}
\begin{equation}
    \label{eq:Buoyancy}
    \vec{F}_b = -\frac{1}{6} \pi d^3 \rho_g \vec{g} .
\end{equation}
\noindent By substituting all force terms into Equation~\ref{eq:Newton2ndLaw}, we obtain a general transport equation for an inertial droplet:
\begin{equation}
    \label{eq:ParticleEQ}
    \frac{\mathrm{d}\vec{u}_p}{\mathrm{d}t} = \frac{3 C_D}{d} \frac{\rho_g}{\rho_p} \mid \vec{U}_g - \vec{u}_p \mid (\vec{U}_g - \vec{u}_p) + \left( 1- \frac{\rho_g}{\rho_p} \right)\vec{g} .
\end{equation}
\noindent In order to solve this equation numerically, the derivatives are discretised and solved implicitly. Equation~\ref{eq:PositionParticle} requires the velocity to be integrated over time. To include the effect of turbulence, a Monte-Carlo model adopted from Ref.~\cite{Gosman} is used. The model analyses and compares turbulent eddy time scales to estimate the time aerosols and droplets remain within an eddy. Following that, the force on the droplet is integrated on the estimated time scale by keeping turbulent fluctuations constant. Turbulent fluctuations used to evaluate the drag force are updated as the droplet leaves the eddy.

According to the Reynolds decomposition of turbulent flows, the instantaneous gas velocity, $\vec{u}_g$, can be written as the sum of the time-averaged gas velocity, $\vec{U}_g$, and a fluctuating velocity component, $\vec{u}'_g$. This is expressed as:
\begin{equation}
\label{eq:ReynoldsDecomposition}
    \vec{u}_g = \vec{U}_g + \vec{u}'_g .
\end{equation}
\noindent The time-averaged velocity is directly computed by the computational fluid dynamics simulation of the ventilated room (see Section~\ref{sec:cfd}). To estimate the fluctuating component, $u'_g$, a normal distribution is used, with a mean $\mu = 0$ and a standard deviation equal to~\cite{Gosman}:
\begin{equation}
\label{eq:sigma}
    \sigma = \sqrt{\frac{2 \, K}{3}} .
\end{equation}
\noindent The time interval for which the droplet interacts with an eddy, $t_{\mathrm{int}}$, is determined from the smallest possible interaction time scale. According to Ref.~\cite{Gosman}, the interaction time is the minimum of the eddy transit time, $t_{\mathrm{R}}$, and the eddy lifetime, $t_{\mathrm{e}}$, expressed in Equation~\ref{eq:t_int}. This is the time scale over which the velocity fluctuation is kept constant during the numerical integration. When $t_{\mathrm{int}}$ has elapsed, a new value of $u'_g$ is calculated:
\begin{equation}
\label{eq:t_int}
    t_{\mathrm{int}} = \mathrm{min}(t_{\mathrm{R}} , t_{\mathrm{e}}) .
\end{equation}
\noindent The eddy lifetime, $t_{\mathrm{e}}$, is computed by:
\begin{equation}
\label{eq:EddyLife}
    t_{\mathrm{e}} = \frac{l_{\mathrm{e}}}{\mid \vec{u}'_g \mid},
\end{equation}
\noindent where $l_{\mathrm{e}}$ is the eddy length scale estimated as:
\begin{equation}
\label{eq:EddyLengthScale}
    l_{\mathrm{e}} = \frac{\mathrm{C}_\mu^{1/2} K^{3/2}}{\varepsilon},
\end{equation}
\noindent where $\mathrm{C}_\mu$ is a constant. The transit time scale is calculated from:
\begin{equation}
\label{eq:Transittime}
    t_{\mathrm{R}} = - \tau_r \ln \left(1 - \frac{l_{\mathrm{e}}}{\tau_r \mid \vec{u}_g - \vec{u}_p \mid }\right),
\end{equation}
\noindent where $\tau_r$ is the so-called droplet relaxation time, which is given by~\cite{Gosman}:
\begin{equation}
\label{eq:DropRelax}
    \tau_r = \frac{4}{3} \rho_p \frac{d}{\rho_g C_D \mid \vec{u}_g - \vec{u}_p \mid }
\end{equation}

\subsubsection*{Validation of the Gosman model and effects of increased turbulence}

\begin{figure}[t]
    \centering
    \includegraphics[trim={0 1cm 0 1cm}, width=\linewidth]{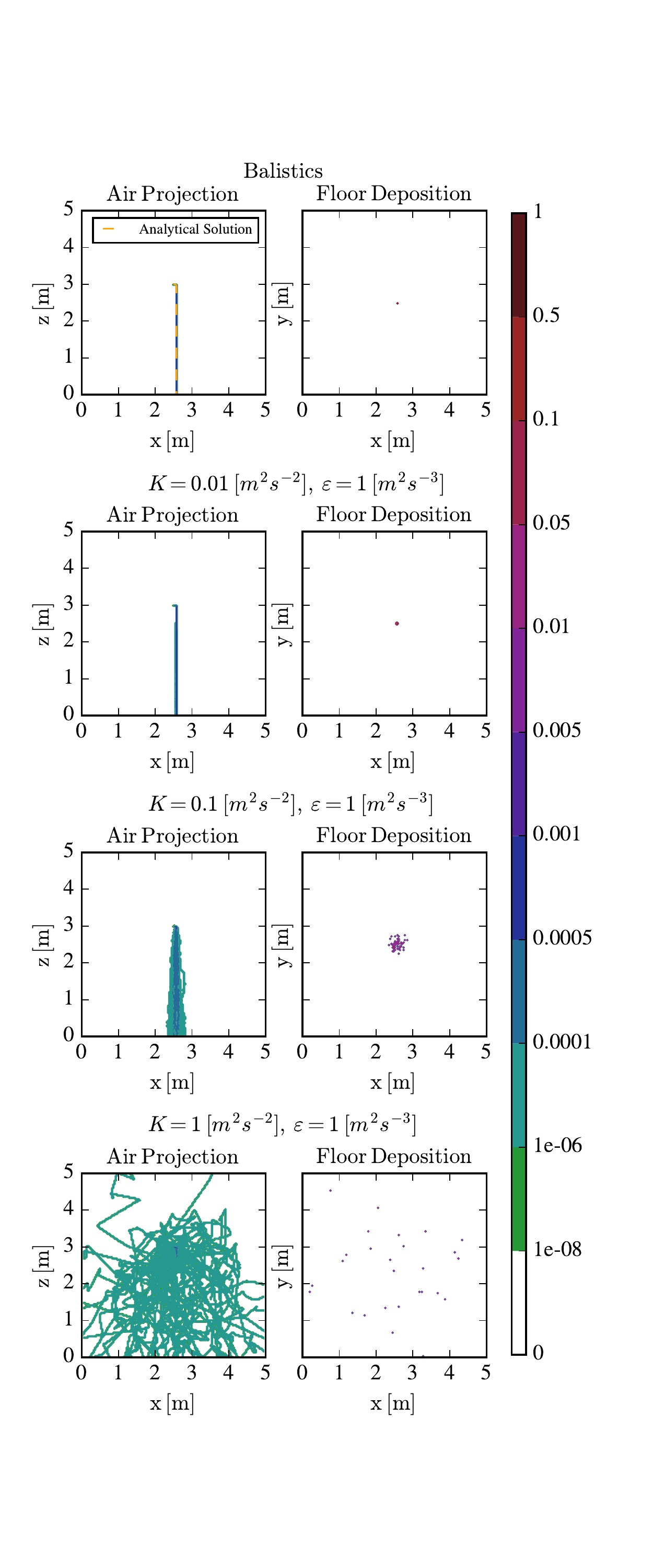}
    \caption{Droplet location probability from the Gosman model for 50 droplets with diameters of 100~$\mu$m injected at different turbulence settings.}
    \label{fig:GosmanValid}
\end{figure}

To validate the Gosman model, the numerical results were compared to an analytical trajectory. With no turbulence and background velocity, i.e.~a uniform velocity field of $\vec{U}_g =\vec{0}$, and neglecting buoyancy, the transport Equation~\ref{eq:ParticleEQ} becomes:
\begin{equation}
    \label{eq:AnalytParticle}
    \frac{\mathrm{d}\vec{u}_p}{\mathrm{d}t} = \frac{-3 \pi d \mu}{m_p} \vec{u}_p + \vec{g} .
\end{equation}
\noindent Equation~\ref{eq:AnalytParticle} can be solved analytically. When solving the differential equation with the initial conditions:
\begin{equation}
\centering
\label{eq:InitialCond}
\begin{cases}
\vec{u}_p(t=0) = [u_0, 0, 0]^T & \textrm{[m/s]}, \\
\vec{x}_p(t=0) = [x_0, y_0, z_0]^T & \textrm{[m]},
\end{cases}
\end{equation}
\noindent the following solution is obtained:
\begin{equation}
\label{eq:AnSol}
    \begin{bmatrix}
    x(t) \\
    y(t)\\
    z(t)
    \end{bmatrix}_p = 
    \begin{bmatrix}
    \frac{u_0 m_p}{-3 \pi d \mu} \left(\exp\left(\frac{-3 \pi d \mu}{m_p}t\right)-1\right) + x_0 \\
    y_0\\
    g \left(\frac{m_p}{-3 \pi d \mu}\right)^2 \left(1-\exp\left(\frac{-3 \pi d \mu}{m_p}t\right)\right) + g \frac{m_p}{-3 \pi d \mu} t + z_0
    \end{bmatrix} .
\end{equation}
\noindent Figure~\ref{fig:GosmanValid} shows the numerical dispersion results obtained in a closed room under different levels of turbulence compared to the analytical solution obtained for the trajectory of a large parcel.
The respective colour bars indicate the probability that a droplet injected at $x$ = 2.5~m and $z$ = 3~m will pass through a certain location.

As can be seen in Figure~\ref{fig:GosmanValid}, the numerical solution is in good agreement with the analytical solution. When the turbulence setting is increased, droplet dispersion becomes stronger and the trajectories are no longer comparable to the analytical solution. Current social distancing rules (e.g.,~the so called `2-meter rule') are typically based on the ballistic trajectory of large saliva droplets, which are valid at very low levels of turbulence. However, as indicated in the numerical results, these rules become ineffective for surroundings with higher turbulence. Hence, to implement more accurate social distancing measures, simulations based on the human occupation and room geometry need to be performed. Such evaluations are affected by the local turbulence and ventilation, which have to be carefully evaluated, especially in the case of high-turbulent environments.

\begin{figure}[t]
    \centering
    \includegraphics[trim={0 1cm 0 1cm},width=\linewidth]{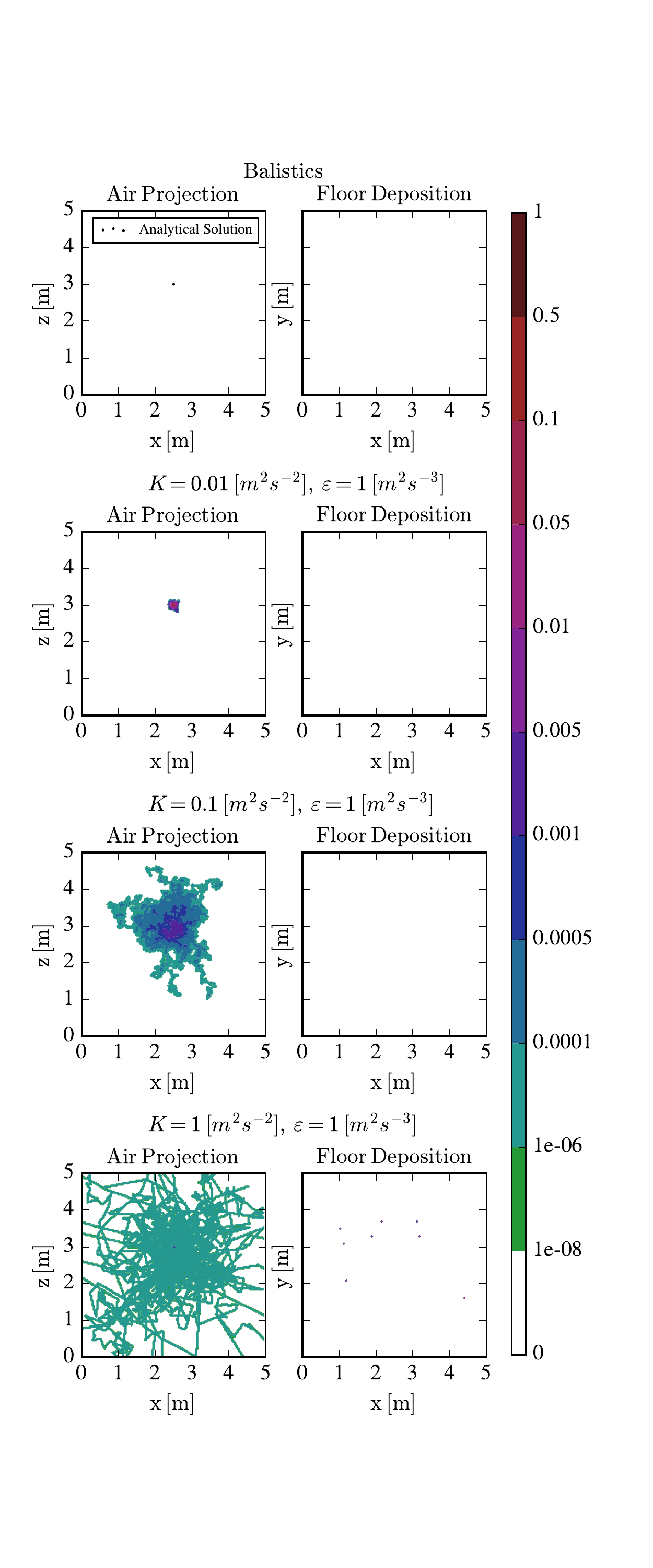}
    \caption{Droplet location probability from the Langevin random walk model for 50 droplets with diameters of 0.1~$\mu$m injected at different turbulence settings.}
    \label{fig:LangValid}
\end{figure}

For the sake of completeness, Figure~\ref{fig:LangValid} shows the trajectory of a parcel with small diameter computed using the Langevin random walk. The top left plot shows that the numerical solution obtained is again in agreement with the analytical solution. Since the parcel stays suspended in the air for a long time, the solutions indicate that the effect of gravity is indeed negligible and therefore validate the assumption that the inertia of parcels smaller than 0.1 $\mu$m can be neglected. Additionally, it is evident that in environments with higher turbulence, the spray dispersion largely deviates from the analytical solution, and that droplets disperse into many directions, which could not have been predicted with the analytical model.

\clearpage
    
\section{Effect of thermal plumes}\label{sec:tPlume}

\vspace{3mm}

\noindent By \emph{A. Giorgallis, L. Papachristodoulou, A. Giusti and \\
    D. Fredrich}

\vspace{3mm}

\noindent The main objective of this investigation was to determine whether the thermal plume produced due to the temperature difference between the human body and the surrounding air is significant in the dispersion of saliva droplets, possibly affecting the spreading of viral particles in indoor environments. This topic is classified as of low certainty according to the Aerosol Society \cite{Aersosolsociety} and requires further study. Therefore, the present work aims to provide more insights on this phenomenon.

The dispersion of saliva droplets is mainly determined by the time-averaged ventilation pattern and the local level of turbulence (see Section~\ref{sec:dispersion}). Therefore, to determine the impact of a human buoyant plume on the dispersion behaviour, it is of paramount importance to have an accurate prediction of the flow field generated by the thermal plume. In this study, computational fluid dynamics (CFD) simulations have been used to predict the flow field. Simulations were performed using the steady-state `buoyantSimpleFoam' solver in OpenFOAM. The results were later also used to predict the dispersion of saliva droplets to provide a full assessment of the thermal plume's effect on dispersion (see Section~\ref{sec:combo_plumes}). This section focuses on the CFD results and the effect of the thermal plume on the velocity field inside the room.

Our work was based on previous studies dealing with thermal plumes \cite{Computational_investigation_of_plumes,Numerical_investigation_plumes,Covid_plume_paper}. Compared to these previous studies, our main research focus here is to develop a better understanding of the behaviour of human thermal plumes and assess their significance under different room conditions, such as different ventilation rates. 

\subsection*{Thermal plume investigation}

Thermal plumes arise due to the temperature difference between the human body and the surrounding air. Heat transfer between the body and the air occurs, increasing the temperature of the air. The density of the air surrounding the body is thus reduced relatively to the surrounding air, giving rise to buoyant forces. The buoyancy effect is due to the combined presence of a fluid density gradient and a body force that is proportional to density \cite{Incropera}. This results in the formation of a thermal plume.

\subsubsection*{Non-dimensional numbers affecting thermal buoyant plumes}

In most of the simulations carried out and outlined in this report, different dimensions and temperature differences were used, making direct comparisons between the results difficult. Instead of using dimensional parameters to describe each simulation, two non-dimensional numbers are employed to make comparisons between the results: the Prandtl and Rayleigh numbers.

The Prandtl number, Pr, represents the ratio of momentum diffusivity to thermal diffusivity. It is given by:
\begin{equation}\label{eq:prandtl}
    \mathrm{Pr}=\frac{\upnu}{\upalpha},
\end{equation}
where $\upnu$ is the kinematic viscosity and $\upalpha$ is the thermal diffusivity of the fluid. A large Prandtl number leads to greater momentum relative to the thermal transfer. Therefore, as a result, the thickness of the thermal boundary layer becomes smaller compared to the momentum boundary layer.

The Rayleigh number, Ra, describes the ratio of buoyancy and thermal diffusivity. It is given by:
\begin{equation}\label{rayleigh}
    \mathrm{Ra}=\frac{\upbeta g S^{3}(T_w-T_{\inf})}{\upalpha\upnu},
\end{equation}
where $\upbeta$ is the volumetric thermal expansion coefficient, $g$ is the gravitational acceleration, $S$ is the characteristic length, $T_w$ is the temperature of the body and $T_{\inf}$ is the temperature of the fluid. An increase in the Ra number causes an increase in the magnitude of buoyant forces, relative to the magnitude of viscous forces, leading to higher velocities inside the momentum boundary layer. Note that the Ra number also involves the thermal diffusivity of the flow, which is the measure of the rate at which heat disperses throughout an object or body~\cite{thermal_diff}.
    
\subsubsection*{Validation}

\begin{figure}[t]
    \centering
    \includegraphics[height=1.8in, width=2in]{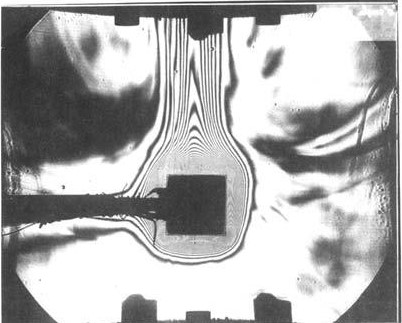}
    \caption{Convection around the isothermal cube in an experimental set-up \cite{Cube_glycerine}.}
    \label{fig:cube_glycerine_exp_set_up}
\end{figure}

In order to validate the accuracy of the selected solver in OpenFOAM, a simulation of the experiment performed by Cha and Cha~\cite{Cube_glycerine} was performed. In the experiment, a heated isothermal cube was submerged in a square tank containing glycerine and an image of the thermal plume was recorded using a holographic interferometer. This experiment was selected because of the simple set-up and geometry of the object immersed in the fluid. The set-up of the experiment is shown in Figure~\ref{fig:cube_glycerine_exp_set_up}.

\begin{figure}[t]
    \centering
    \includegraphics[height=1.8in, width=2.8in]{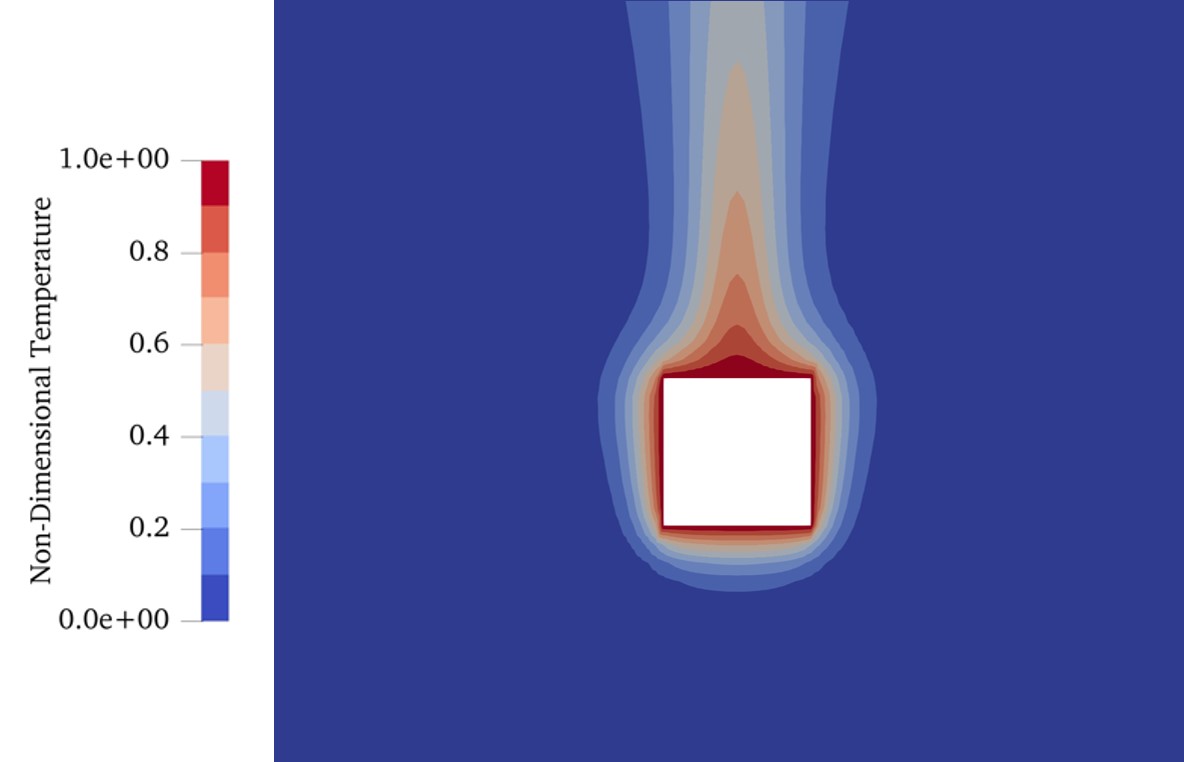}
    \caption{Non-dimensional temperature, $\theta$, in a vertical cross-section predicted by the CFD simulation.}
    \label{fig:exp_recr_sim}
\end{figure}

\begin{figure}[t]
    \centering
    \includegraphics[width=0.8\linewidth]{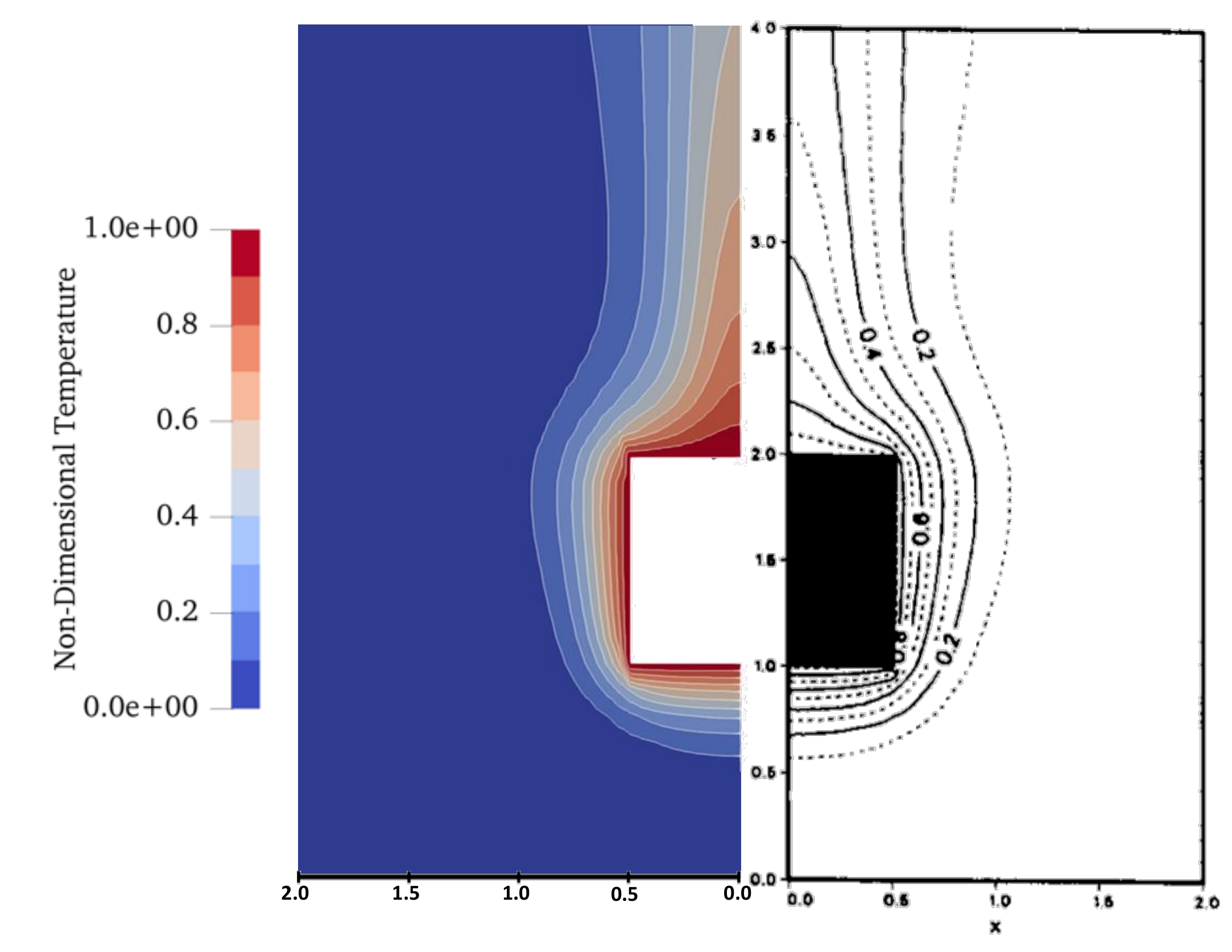}
    \caption{Non-dimensional temperature, $\theta$ in a vertical cross-section: comparison between the present simulation (left) and numerical results from Ref.~\cite{Cube_glycerine} (right).}
    \label{fig:exp_comparison}
\end{figure}

The authors of Ref.~\cite{Cube_glycerine}, instead of specifying the dimensions and temperature difference between the fluid and cube, used the non-dimensional Rayleigh and Prandtl numbers, introduced above, to describe the operating conditions. The values of Ra and Pr, used in both the experiment and our computation in OpenFOAM, are 1300 and 9840, respectively.

The size of the tank was also given in non-dimensional form in Ref.~\cite{Cube_glycerine}, using the length of the side of the cube as a reference. The length of the side of the square tank was 4 times the length of the immersed cube. In the simulation, the bottom face was assigned to be the inlet while the top face was modelled as an outlet boundary.

When setting up the simulation, fixed temperature boundary conditions were imposed on all vertical walls of the tank and on all walls of the immersed cube. The temperature of the inflow was also fixed, while at the outflow a zero-gradient boundary condition was imposed. Second-order accurate numerical schemes were used for all the quantities.

The results obtained with the OpenFOAM solver are shown in Figure~\ref{fig:exp_recr_sim}, which shows the non-dimensional temperature defined as:
\begin{equation}
    \theta = \frac{T - T_\inf}{T_w - T_\inf}.
\end{equation}
A comparison between the isotherms obtained from the CFD result and the numerical solution proposed by Cha and Cha~\cite{Cube_glycerine}, which reproduced the experiment with good accuracy, was carried out. The results are shown in Figure~\ref{fig:exp_comparison}. The results demonstrate a good agreement between the results produced by the OpenFOAM solver and the numerical solution provided by Cha and Cha~\cite{Cube_glycerine}. This validates the reliability and accuracy of the solver used in this work. Therefore, the solver will be used in the next section to evaluate the effect of the plume produced by a person inside a room. A sensitivity analysis for the values of Ra and Pr is discussed first.
    
\begin{figure}[t]
  \centering
  \includegraphics[width=0.95\linewidth]{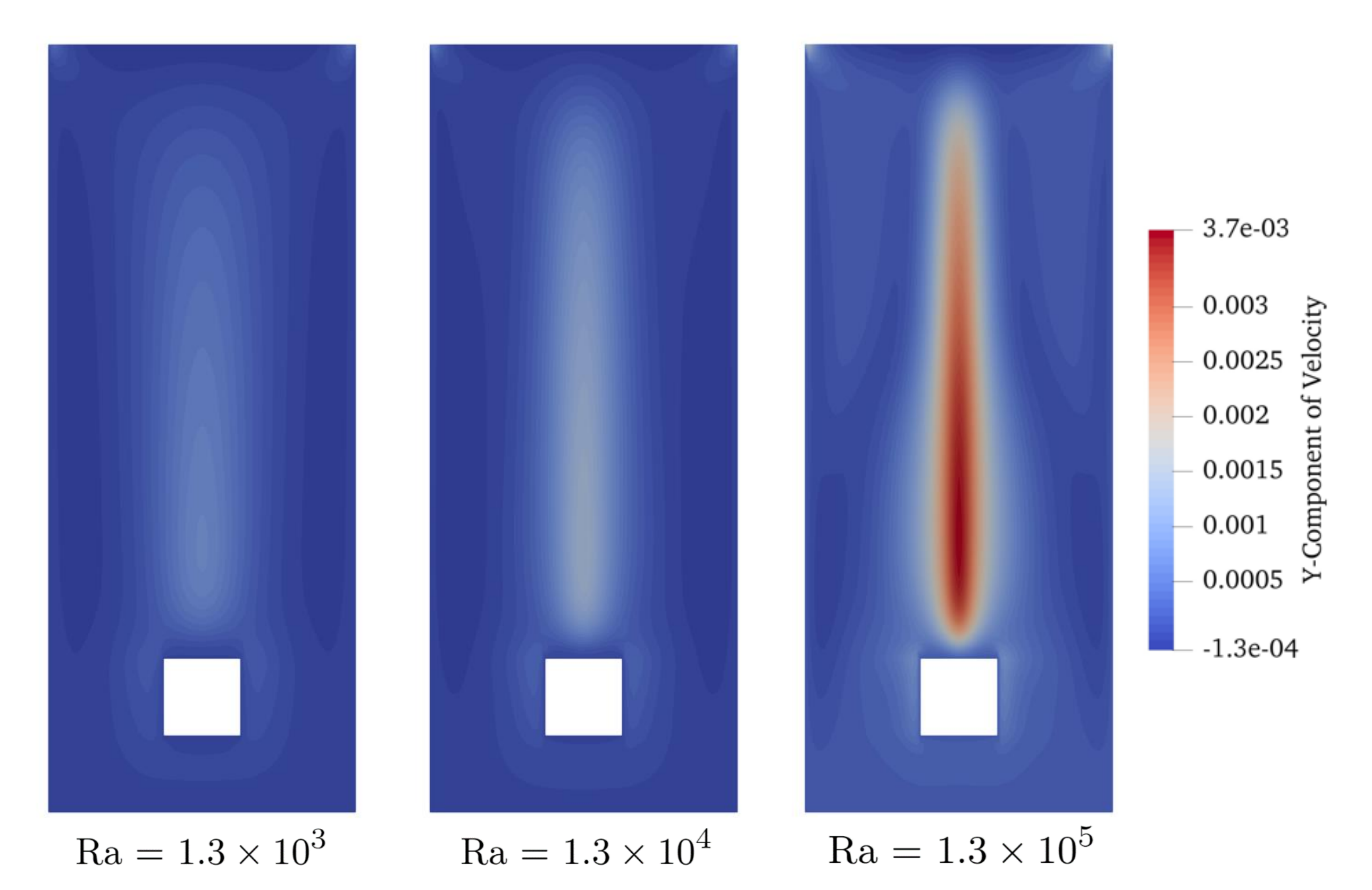}
  \caption{$y$-component of the velocity for different Rayleigh numbers in a vertical cross-section for Cases 1 to 3 (see Table~\ref{tab:Simulation Summary}).}
  \label{fig:vel_glyc}
\end{figure}

\begin{figure}[t]
  \centering
  \includegraphics[width=0.95\linewidth]{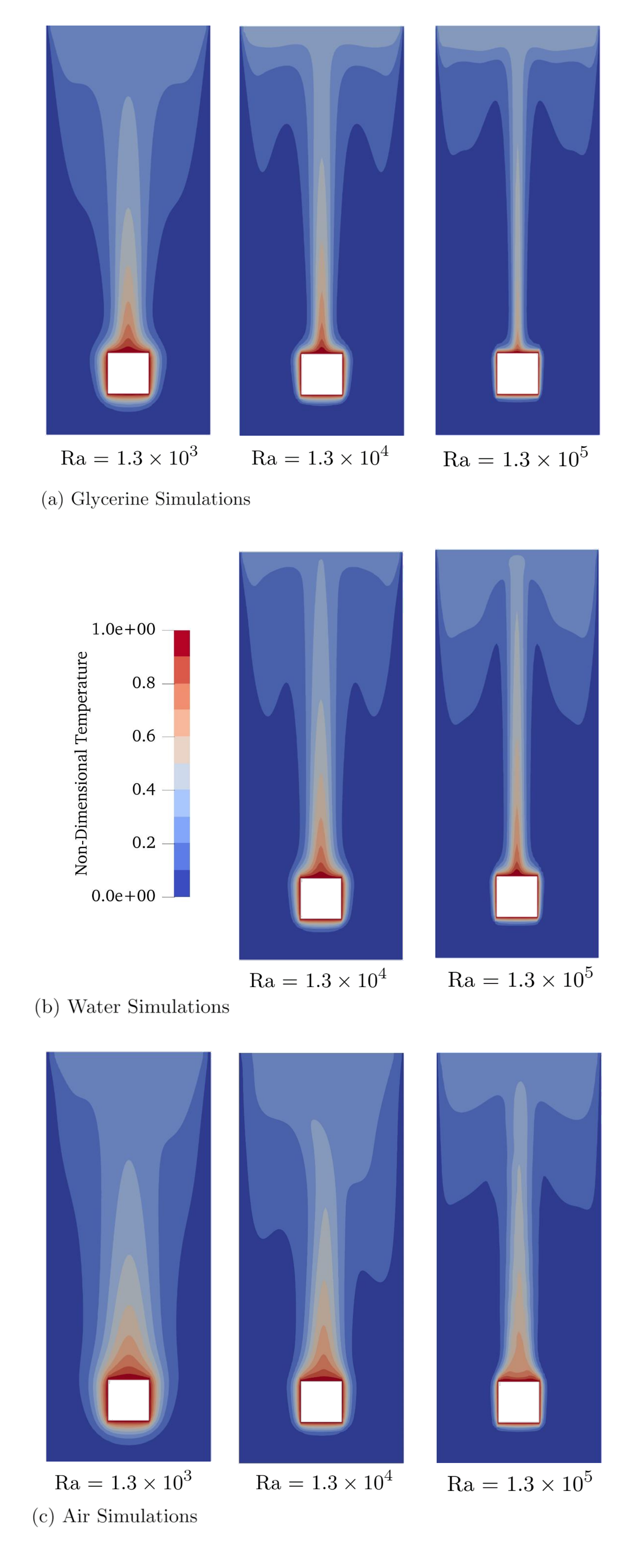}
  \caption{Non-dimensional temperature, $\theta$, in a vertical cross-section for the cases in Table~\ref{tab:Simulation Summary}.}
  \label{fig:cube_sim}
\end{figure}

\subsubsection*{Investigation of the parameters affecting the thermal plume}

As mentioned in the previous sections, the Ra and Pr numbers govern many aspects of the thermal plume, such as the thickness of the momentum and thermal boundary layers and the magnitude of the fluid velocities inside the plume. Therefore, to develop a better understanding on how these parameters influence the flow, additional simulations with the same configuration investigated in the previous section were performed for different values of Ra and Pr.

\begin{table}[b]
\caption{Parametric operating conditions.}
\label{tab:Simulation Summary}
\centering
\resizebox{0.3\textwidth}{!}{%
\begin{tabular}{llll}
\hline
\textbf{No.} & \textbf{Fluid} & \textbf{Pr} & \textbf{Ra}               \\ \hline
1  & Glycerine      & 8940        & $1.3\times10^3$ \\ 
2  & Glycerine      & 8940        & $1.3\times10^4$ \\ 
3  & Glycerine      & 8940        & $1.3\times10^5$ \\ 
4  & Water          & 5.83        & $1.3\times10^4$ \\ 
5  & Water          & 5.83        & $1.3\times10^5$ \\ 
6  & Air            & 0.705       & $1.3\times10^3$ \\ 
7  & Air            & 0.705       & $1.3\times10^4$ \\ 
8  & Air            & 0.705       & $1.3\times10^5$ \\ \hline
\end{tabular}%
}
\end{table}

In these simulations, the vertical height of the numerical domain was increased to 10 times the length of the cube, to allow more space for the plume to develop and minimise the effect of the outlet boundary condition on the plume. The Pr number was varied by changing the fluid in which the cube was submerged. Glycerine, water and air were used, in order to achieve different orders of magnitude of Pr. The Ra number was also varied by changing either the size or the temperature of the cube walls, according to Equation \ref{rayleigh}. The simulations performed for this analysis are summarised in Table~\ref{tab:Simulation Summary}. It should be noted that the simulation of water with Ra = $1.3\times10^4$ could not be conducted since it would either require a really high cube temperature or a very small cube size.

The $y$-component (vertical component) of the velocity of glycerine at different Ra numbers is shown in Figure~\ref{fig:vel_glyc}. Higher velocities are observed in the momentum boundary layer for larger Ra numbers, in agreement with the physical interpretation of the Ra number previously discussed. Also, by increasing the Ra number, the thermal diffusivity decreases relatively to the buoyancy. This leads to less heat transfer from the hot body to the fluid and therefore to a thinner boundary layer. This is evident in Figure~\ref{fig:cube_sim}, where the non-dimensional temperature for different cases is shown.

The Pr number, as defined in Equation~\ref{eq:prandtl}, is the ratio of momentum diffusivity to thermal diffusivity. Therefore, a smaller Pr number results in higher thermal diffusivity and thus thermal energy is transferred through the fluid more easily from the hot body. This leads to a thicker thermal boundary layer, as shown by the simulation results in Figure~\ref{fig:cube_sim}.
    
\clearpage

\begin{figure}[t]
  \centering
  \includegraphics[width=0.9\linewidth]{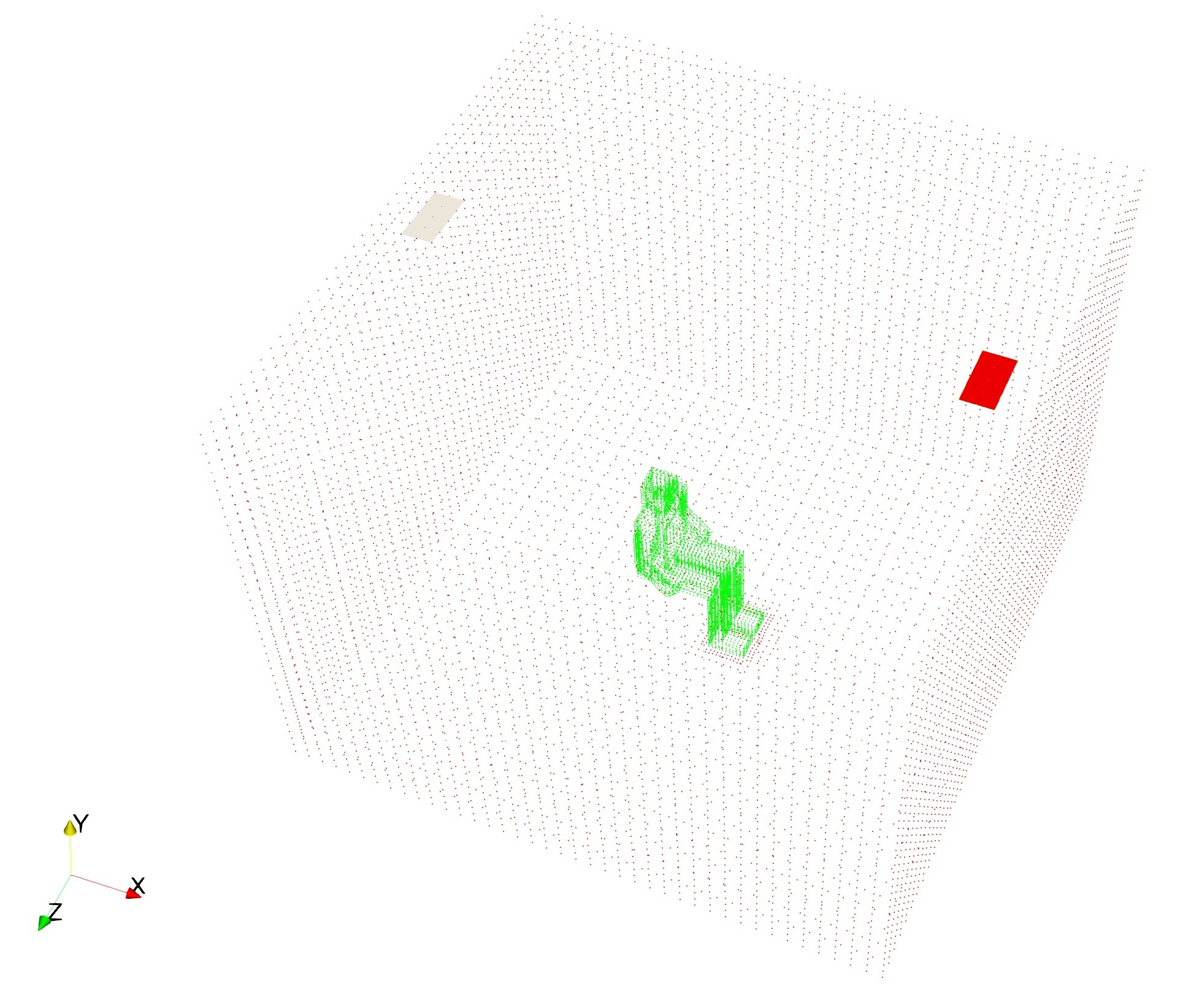}
  \caption{Positions of the ventilation ports.}
  \label{fig:mesh3d}
\end{figure}

\subsection*{Room simulations}

\subsubsection*{Room geometry and ventilation}

To determine whether the effect of the thermal plume due to the presence of a person in a room has significant effects in altering the velocity field, different simulations were performed for a room of single occupancy. The ventilation rate was varied and the velocity field at steady-state was obtained.

A square room was used, with dimensions of $4$~m $ \times~4$~m $\times~3$~m. The size of the room was selected in order to be representative of an office or a small tutorial room. A person in sitting position was added in the middle of the room and ventilation was also included. It should be noted that no chair or furniture were added, since the aim was to determine whether the effect of the plume generated by the human body was significant. Further investigation is required in the future to include the effects of furniture, which could alter the air flow, i.e., the velocity field.

\begin{table}[b]
\centering
\caption{Inflow velocity magnitudes for different ventilation rates.}
\resizebox{0.95\linewidth}{!}{%
\begin{tabular}{cc}
\hline
\textbf{Ventilation rate [vol/hour]} & \textbf{Inflow velocity [m/s]} \\ \hline
1                                    & 0.167                               \\ 
5                                    & 0.833                               \\ 
12                                   & 2.00                                \\ \hline
\end{tabular}
\label{table:velocities}
}
\end{table}

An appropriate size for the ventilation inlet and outlet was calculated by assuming a maximum velocity of the air flow and by imposing a volume flow rate based on the desired air changes per hour. Specifically, the maximum air velocity was set to 2 m/s at the maximum ventilation rate of 12 volume changes per hour~\cite{Air_changes}. The resulting size of the inlet was then calculated to be 0.08~m\textsuperscript{2}. Arbitrary dimensions of $40$~cm~$\times~20$~cm were chosen for the inlet and outlet ports. Both ports were placed on the ceiling at opposite sides of the room in order to avoid a ventilation short circuit and to ensure that the air flow follows an intended path. The positions of the inlet and outlet are shown in Figure~\ref{fig:mesh3d}, where a grey square represents the inlet and a red square the outlet. The ventilation rates were set to 1, 5 and 12 volume changes per hour, so that a wide operating range could be covered. The inflow air velocities corresponding to these rates are summarised in Table~\ref{table:velocities}.

\subsubsection*{Boundary Conditions}

The initial temperature of the air inside the room was set to $20^{\circ}$C. The boundary condition for the temperature of the room's walls was also set to $20^{\circ}$C, whereas the temperature of the body was assumed equal to $37^{\circ}$C \cite{Body_Temperature}. The temperature of the inflowing air was set to $20^{\circ}$C, assuming that the inlet operates for ventilation purposes only and not for cooling.

\begin{figure}[t]
  \centering
  \includegraphics[width=\linewidth]{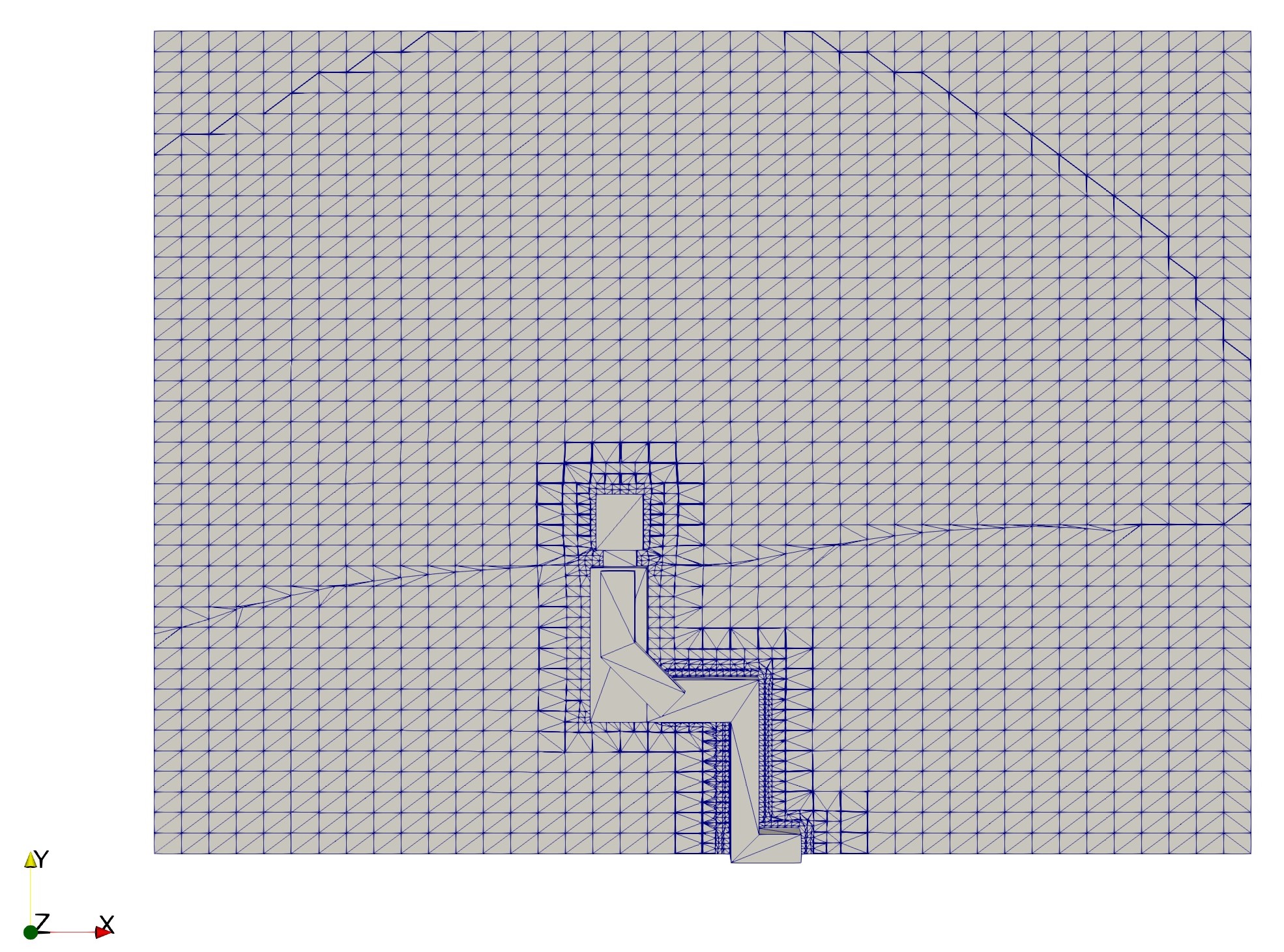}
  \caption{Overview of the mesh in the $x$-$y$ plane.}
  \label{fig:mesh1}
\end{figure}

\begin{figure}[t]
  \centering
  \includegraphics[width=\linewidth]{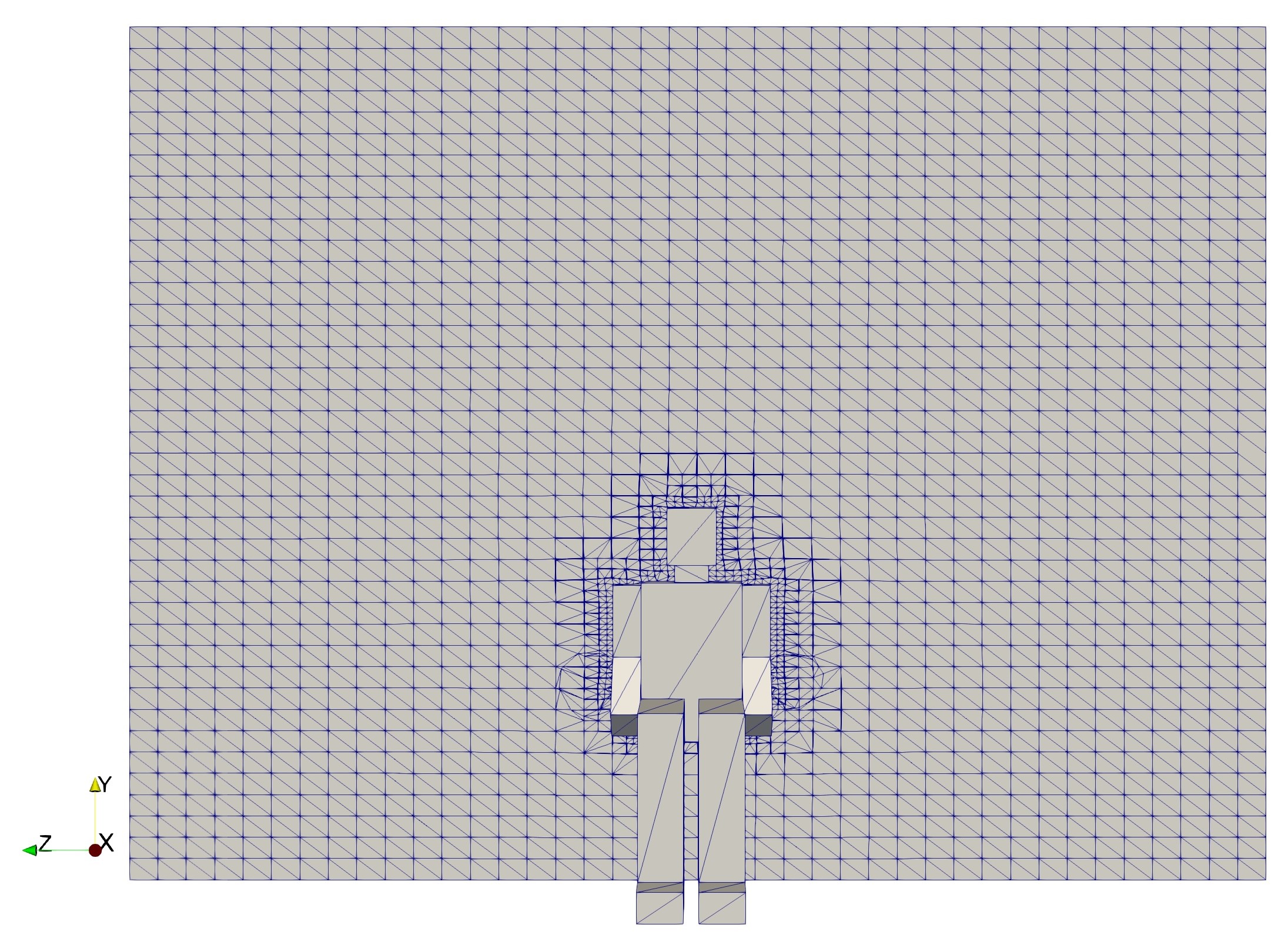}
  \caption{Overview of the mesh in the $y$-$z$ plane.}
  \label{fig:mesh2}
\end{figure}

\subsubsection*{Computational Mesh}

A hexahedral mesh with uniform distribution of nodes along the three sides ($40\times40\times30$) of the room was first created in OpenFOAM. Then, a 3-D `Stereolithography' (.stl) model representing the person was created, imported to OpenFOAM and meshed using the `snappyHexMesh' function. The sitting person was placed in the middle of the room. An overview of the mesh used for the simulations is shown in Figures~\ref{fig:mesh1} and~\ref{fig:mesh2}. A mesh refinement close to the body surface was applied to allow for a better representation of the geometry. The final mesh consists of 84,956 control volumes.
    
\subsubsection*{Results}

\begin{table}[b]
\centering
\caption{Summary of all room simulations.}
\begin{tabular}{cccc}
\hline
& \textbf{Ventilation}& \textbf{Heated /} & \textbf{Inflow angle} \\
\textbf{Case} & \textbf{rate} & \textbf{unheated} & \textbf{from negative} \\
\textbf{no.} & \textbf{[vol/hour]} & \textbf{body} & \textbf{y-axis [$^{\circ}$]} \\ \hline
1           & 0                                    & Heated                         & N/a                                                                   \\ 
2           & 1                                    & Unheated                         & 35                                                                    \\ 
3           & 1                                    & Heated                       & 35                                                                    \\ 
4           & 5                                    & Unheated                         & 35                                                                    \\ 
5           & 5                                    & Heated                       & 35                                                                    \\ 
6           & 5                                    & Unheated                         & 20                                                                    \\ 
7           & 5                                    & Heated                       & 20                                                                    \\ 
8           & 12                                   & Unheated                         & 35                                                                    \\ 
9           & 12                                   & Heated                       & 35                                                                    \\ \hline
\end{tabular}
\label{table:person_sim}
\end{table}

\begin{figure}[t]
  \centering
  \includegraphics[width=\linewidth]{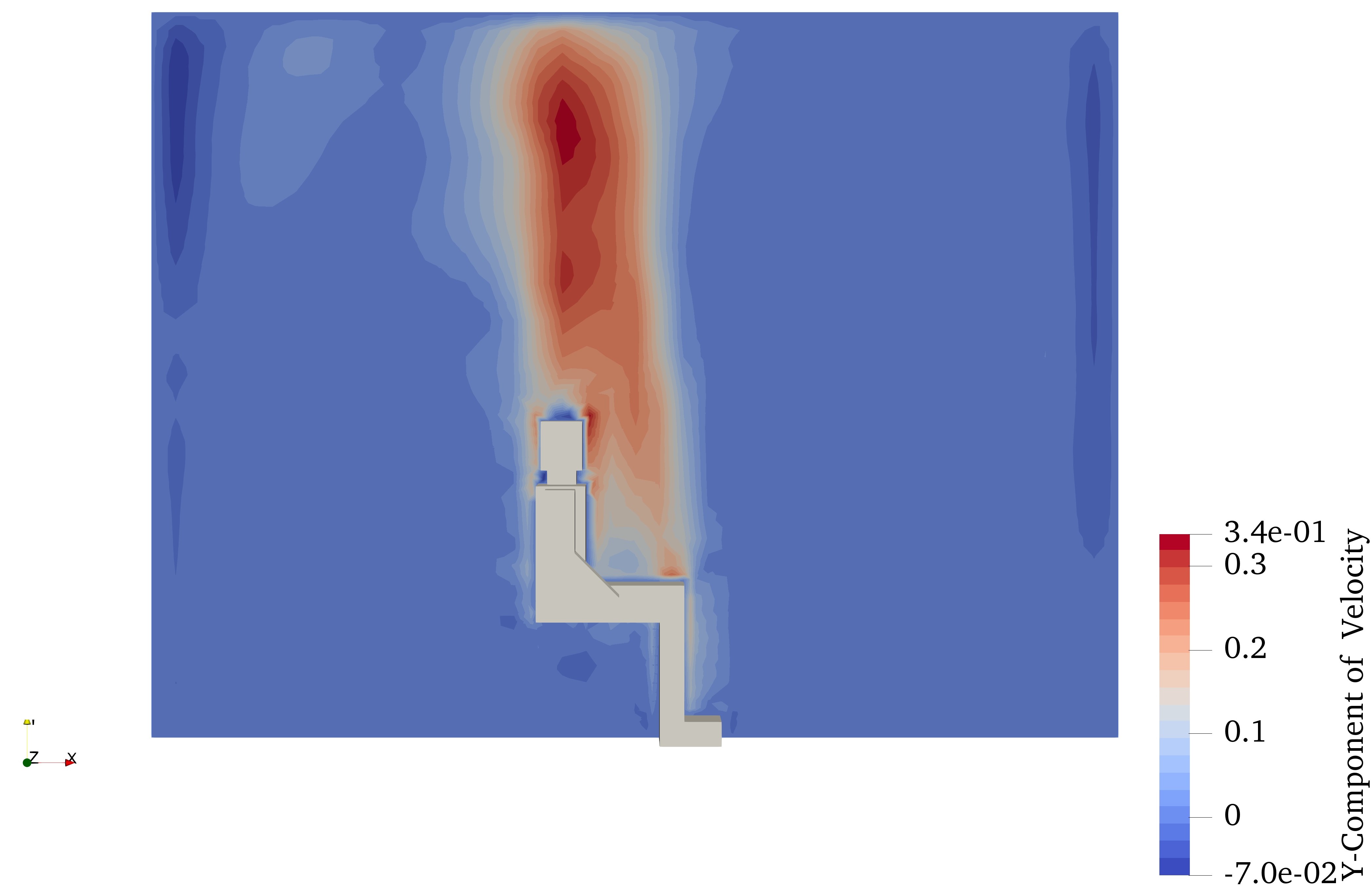}
  \caption{$y$-component of the velocity in a vertical cross-section for the case with no ventilation and a heated body (Case~1 in Table~\ref{table:person_sim}).}
  \label{fig:U_Y_sim_1}
\end{figure}

\begin{figure*}[p]
\centering
  \includegraphics[width=0.83\linewidth]{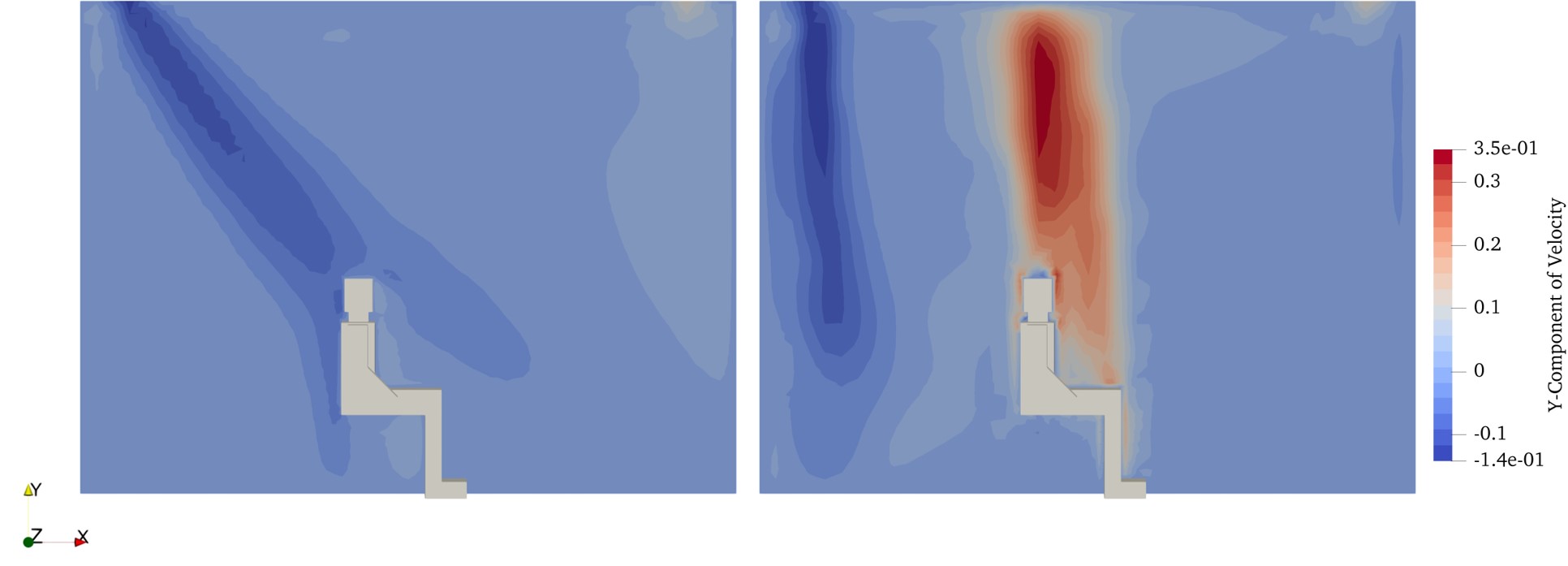}
  \caption{$y$-component of the velocity in a vertical cross-section for the simulations with 1 volume change per hour ventilation angled at $35^\circ{}$ with a heated body (left) and an unheated body (right) -- Cases~2 and 3 in Table~\ref{table:person_sim}.}
  \label{fig:U_Y_sim_23}
\end{figure*}

\begin{figure*}[p]
\centering
  \includegraphics[width=0.83\linewidth]{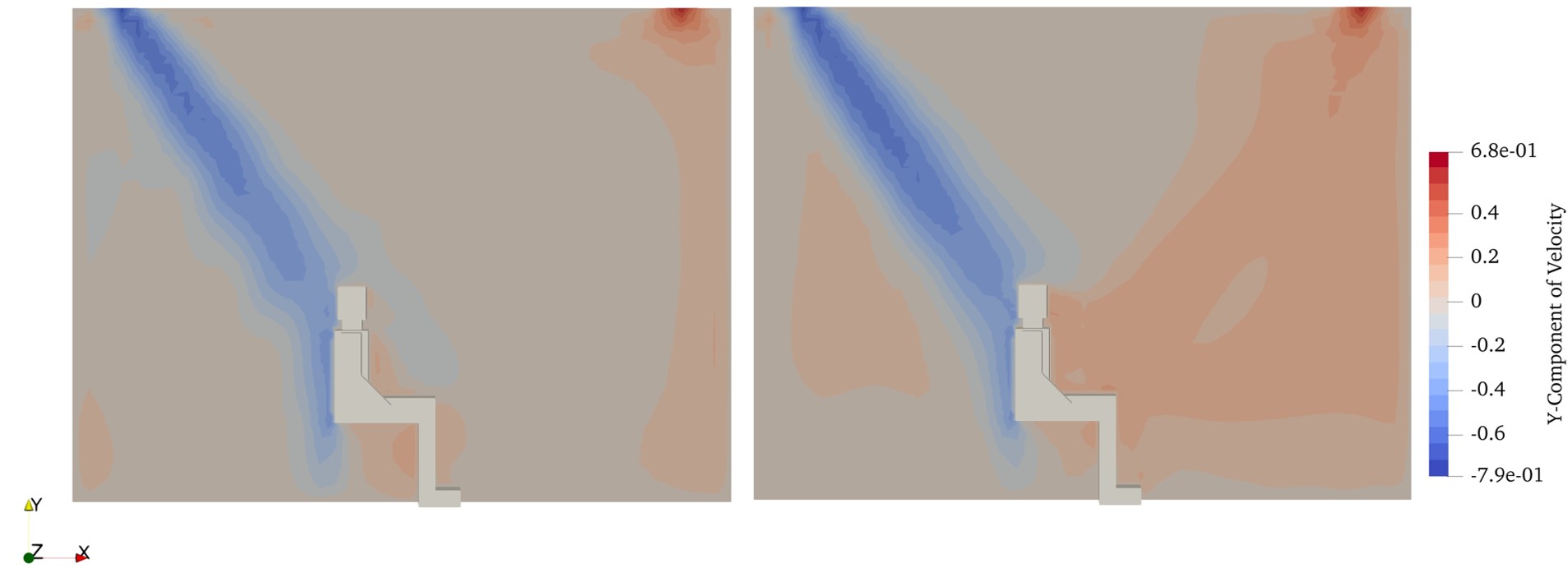}
  \caption{$y$-component of the velocity in a vertical cross-section for the simulations with 5 volume changes per hour ventilation angled at $35^\circ{}$ with a heated body (left) and an unheated body (right) -- Cases~4 and 5 in Table~\ref{table:person_sim}.}
  \label{fig:U_Y_sim_45}
\end{figure*}

\begin{figure*}[p]
  \centering
  \includegraphics[width=0.83\linewidth]{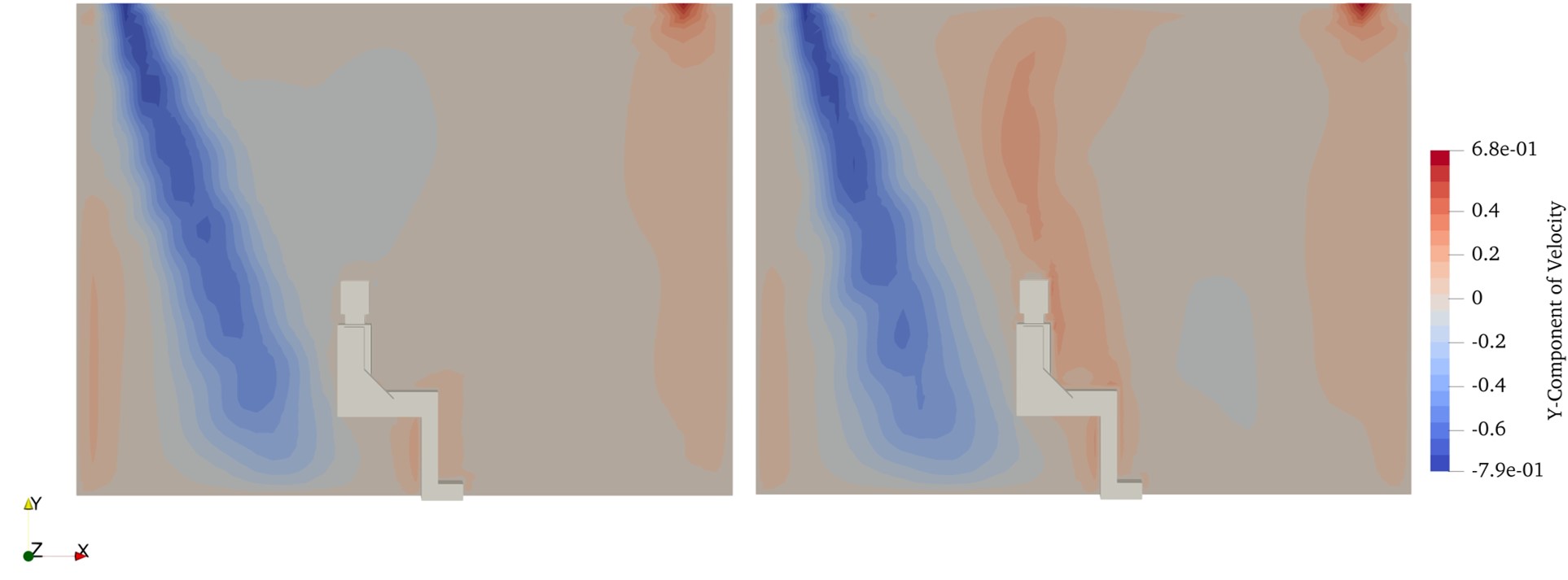}
  \caption{$y$-component of the velocity in a vertical cross-section for simulations with 5 volume changes per hour ventilation angled at $20^\circ{}$ with a heated body (left) and an unheated body (right) -- Cases~6 and 7 in Table~\ref{table:person_sim}.}
  \label{fig:U_Y_sim_67}
\end{figure*}

\begin{figure*}[t]
 \centering
  \includegraphics[width=0.83\linewidth]{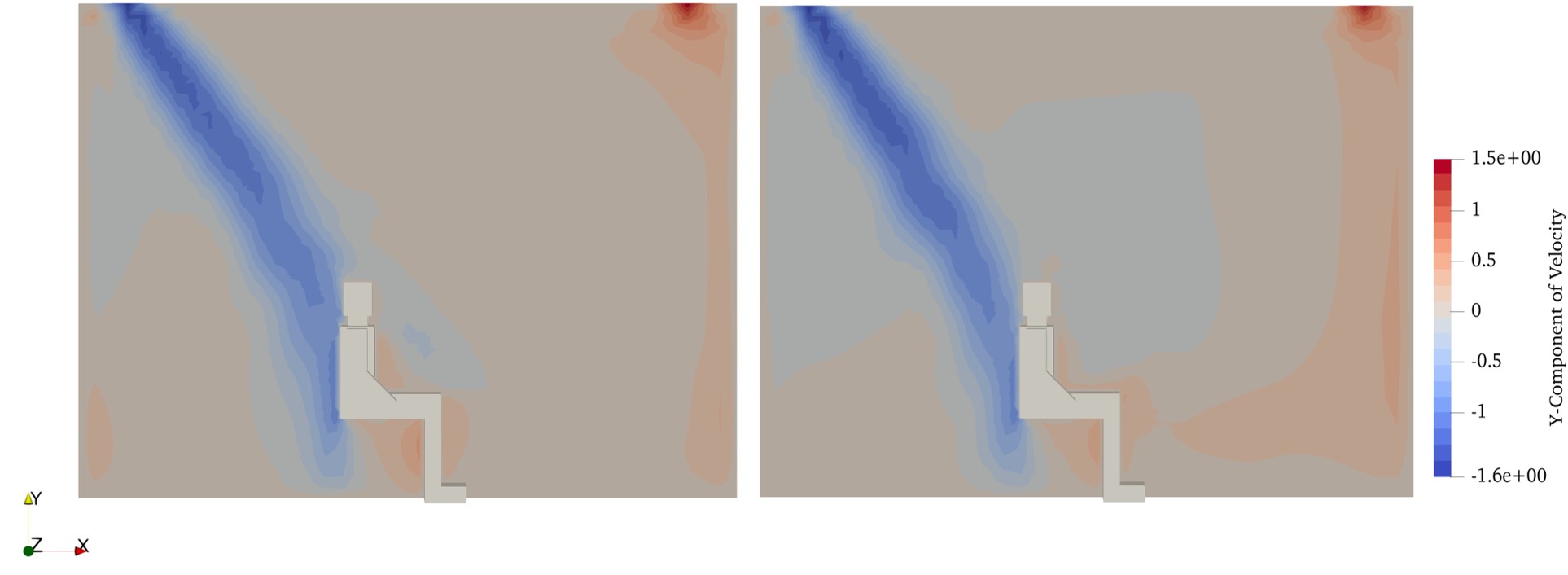}
  \caption{$y$-component of velocity in a vertical cross-section for the simulations with 12 volume changes per hour ventilation angled at $35^\circ{}$ with a heated body (left) and an unheated body (right) -- Cases~8 and 9 in Table~\ref{table:person_sim}.}
  \label{fig:U_Y_sim_89}
\end{figure*}

In order to evaluate the effect of the thermal plume on the ventilation flow field, several simulations were conducted with heated ($37^{\circ}$C) and unheated ($20 ^{\circ}$C) bodies. The effect of the ventilation rate and inflow angle on the thermal plume was also investigated. The flow field results obtained here will also be used in Section~\ref{sec:combo_plumes} to evaluate the related impact on the dispersion of droplets emitted from a person's mouth.

The effect of the inflow angle on the thermal plume was investigated by testing two different angles for the inflow. These angles were $35^{\circ}$ and $20^{\circ}$ anticlockwise from the negative vertical direction. The above angles were selected so that the air flows directly onto the person at $35^{\circ}$ and behind the person at $20^{\circ}$. The effect of the angle was only investigated at the intermediate ventilation rate of 5 volume changes per hour. It should be noted that the vertical inflow was avoided, since it is not typically used in real-life applications. Further investigation into effect of the inflow angle should be conducted in the future to develop a more detailed understanding. The main parameters of the simulations performed in this study are summarised in Table~\ref{table:person_sim}. The simulations were ran for 400,000 iterations in order to achieve sufficient numerical convergence of the solution. The (vertical) $y$-component of the velocity field was analysed and processed, as well as the non-dimensional temperature distribution in the domain.
Both parameters are visualised in a vertical cross-section.
The Rayleigh number associated with the conditions imposed in the simulations, calculated using Equation~\ref{rayleigh}, is $\mathrm{Ra}=68.69\times10^3$. Note that the characteristic length of the person used to compute Ra was evaluated as $L$ = volume / surface area = 0.096~m$^3$ / 2.720~m$^2$ = 0.035~m. 

\begin{figure}[t]
  \centering
  \includegraphics[width=\linewidth]{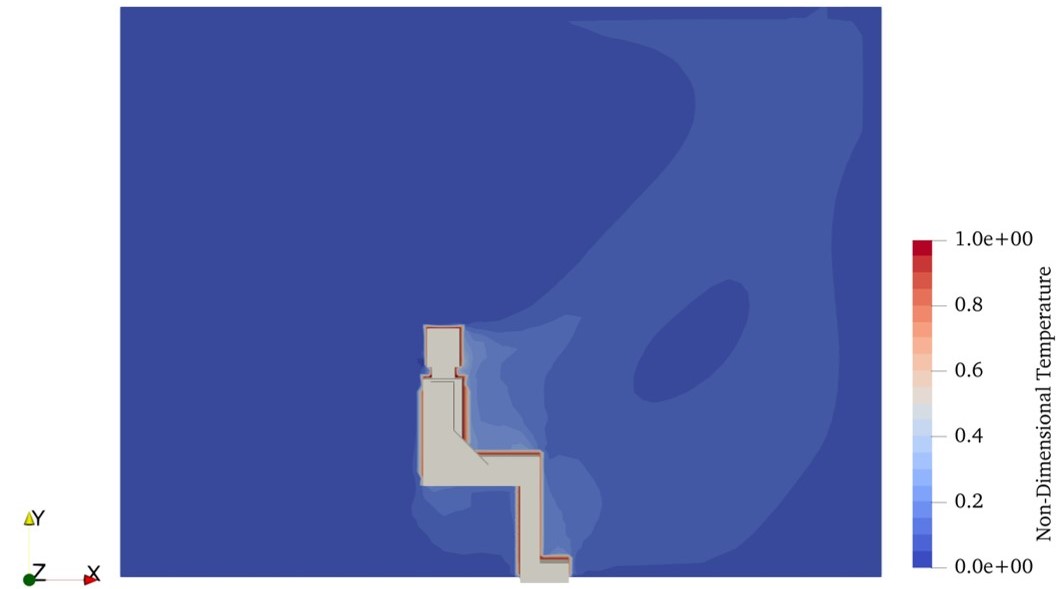}
  \caption{Non-dimensional temperature, $\theta$, in a vertical cross-section for the simulation with a 5 volume changes per hour ventilation angled at $35^\circ{}$ with a heated body -- Case~5 in Table~\ref{table:person_sim}.}
  \label{fig:non_dim_T_sim_5}
\end{figure}

In Figure~\ref{fig:U_Y_sim_1}, the vertical component of the velocity field for a heated body in a room without ventilation is shown. A significant thermal plume is observed above the human body with a maximum velocity of up to 0.34 $\mathrm{ms^{-1}}$. This indicates that without ventilation, the thermal plume has a significant effect on the velocity field. The value obtained is similar to values determined by other studies \cite{Computational_investigation_of_plumes,Numerical_investigation_plumes} and could be significant for the dispersion of droplets. In Figure~\ref{fig:U_Y_sim_23}, which shows the results for a ventilation rate of 1 volume change per hour, the thermal plume deflects the inflow stream from $35^{\circ}$ to being almost vertical. Under these conditions, a large plume forms with vertical velocities reaching the same magnitude as for the non-ventilated case. Therefore, under poor ventilation conditions, the effect of the plume is still significant. Figure \ref{fig:U_Y_sim_45} shows the results of the simulation with a ventilation rate of 5 volume changes per hour and an inlet angle of $35^{\circ}$. There is no inflow stream deflection observed, since the velocity of the inlet ($0.83\:\mathrm{ms^{-1}}$) is much larger in magnitude than the velocity inside the thermal plume. The inflow stream reaches the person directly, causing the plume to shift to the right, as also seen in Figure~\ref{fig:non_dim_T_sim_5}. This leads to larger velocities, up to about $0.20\:\mathrm{ms^{-1}}$ on the right-hand side of the person. Hence, the effect of the thermal plume at 5 volume changes per hour is still significant. In Figure~\ref{fig:U_Y_sim_67}, the ventilation rate is set to 5 volume changes per hour and the inflow is angled at $20^{\circ}$ from the negative vertical direction. The results indicate that the thermal plume forms normally, like in Figure~\ref{fig:U_Y_sim_1}, since the inflow is not directed onto the person. Similar to the Cases~4 and 5, the effect of the thermal plume is considerable. Figure~\ref{fig:U_Y_sim_89} shows the results for the maximum ventilation rate of 12 volume changes per hour. Under these conditions, the effect of the thermal plume is not as significant as for the lower ventilation rates, because the velocity of the plume is much lower than the velocity of the incoming ventilation stream.

\begin{figure}[t]
  \centering
  \includegraphics[width=\linewidth]{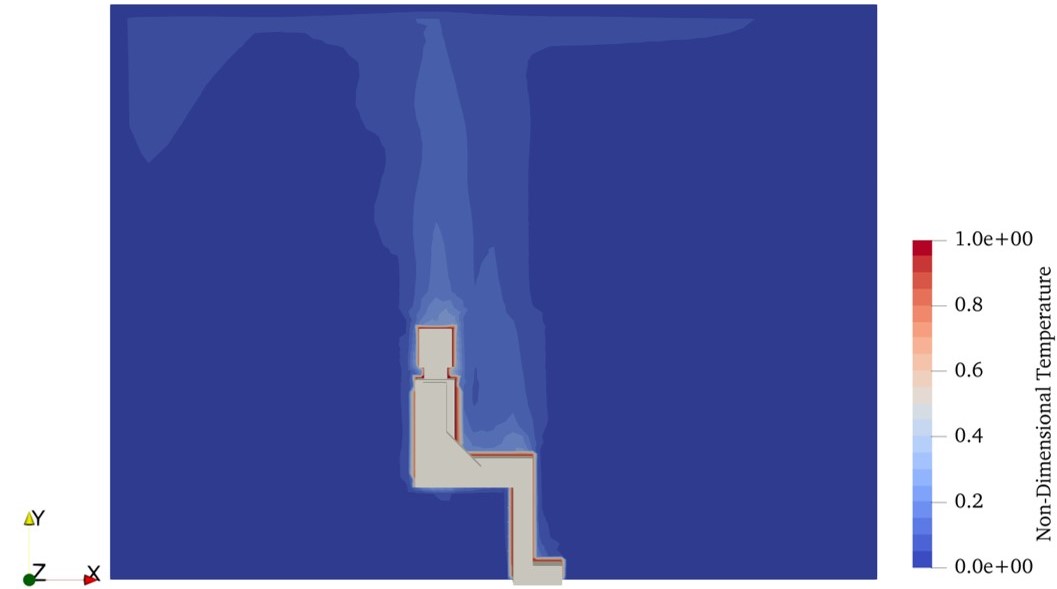}
  \caption{Non-dimensional temperature, $\theta$, in a vertical cross-section for the simulation with no ventilation and with a heated body -- Case~1 in Table~\ref{table:person_sim}.}
  \label{fig:non_dim_T_sim_1}
\end{figure}

\begin{figure}[t]
  \centering
  \includegraphics[width=\linewidth]{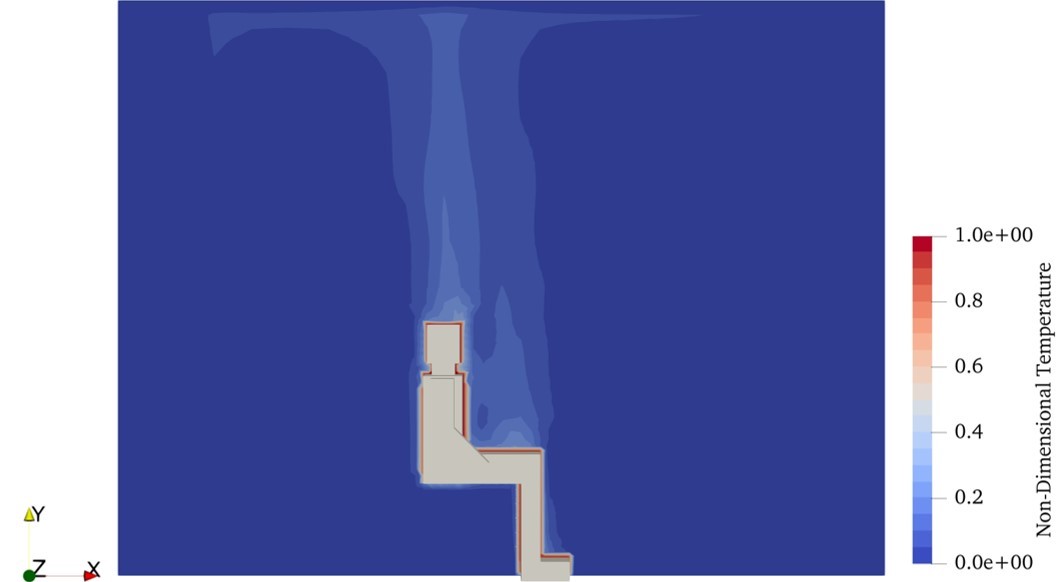}
  \caption{Non-dimensional temperature, $\theta$, in a vertical cross-section for the simulation with a 1 volume change per hour ventilation angled at $35^\circ{}$ with a heated body -- Case~3 in Table~\ref{table:person_sim}.}
  \label{fig:non_dim_T_sim_3}
\end{figure}

\begin{figure}[t]
  \centering
  \includegraphics[width=\linewidth]{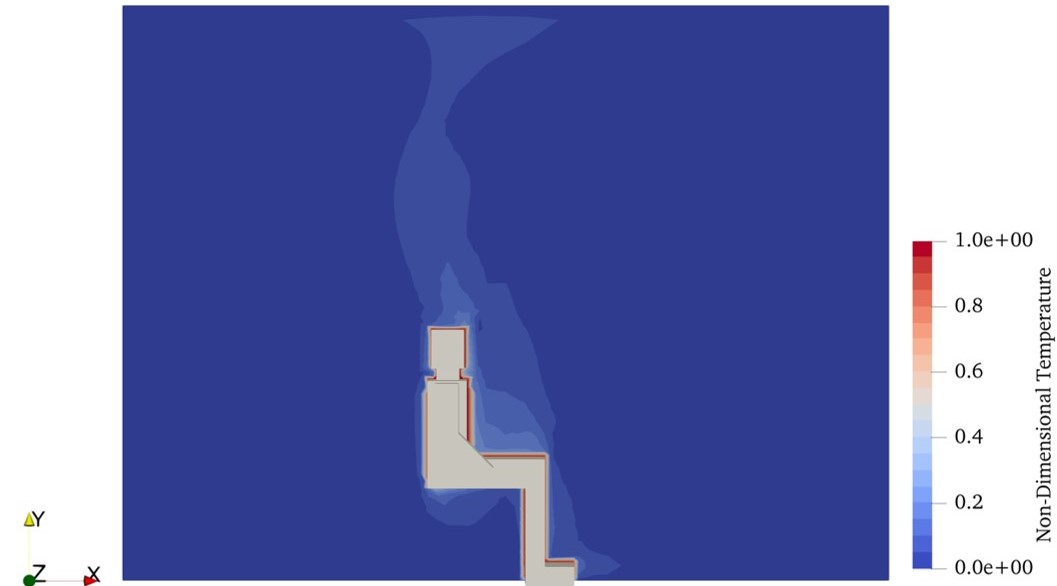}
  \caption{Non-dimensional temperature, $\theta$, in a vertical cross-section for the simulation with a 5 volume changes per hour ventilation angled at $20^\circ{}$ with a heated body -- Case~7 in Table~\ref{table:person_sim}.}
  \label{fig:non_dim_T_sim_7}
\end{figure}

\begin{figure}[t]
  \centering
  \includegraphics[width=\linewidth]{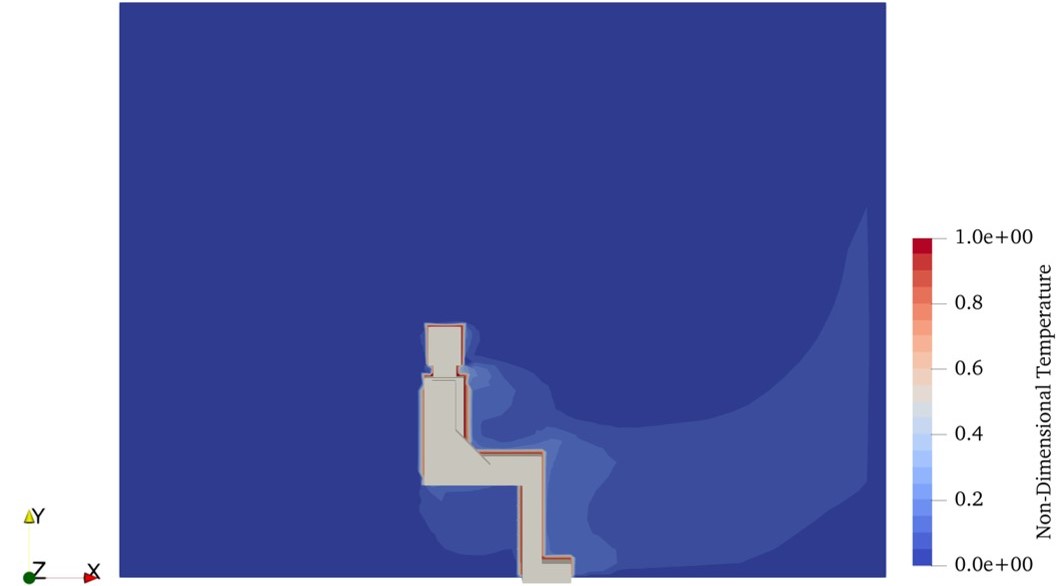}
  \caption{Non-dimensional temperature, $\theta$, for the simulation with a 12 volume changes per hour ventilation angled at $35^\circ{}$ with a heated body -- Case~9 in Table~\ref{table:person_sim}.}
  \label{fig:non_dim_T_sim_9}
\end{figure}

The interaction and coupling between the temperature field and the velocity field is demonstrated through the results shown in Figures~\ref{fig:non_dim_T_sim_1} to \ref{fig:non_dim_T_sim_9}, where the non-dimensional temperature in a vertical cross-section for the various cases with a heated body is shown. The non-dimensional temperature also allows us to better identify the location of the thermal plume. In the case of poor ventilation (less or equal to 1 volume change per hour), the region with a high temperature is located straight above the body, determining the high vertical velocities there. A similar behaviour is observed for the case with the ventilation flow not directed towards the body (Figure~\ref{fig:non_dim_T_sim_7}). For moderate or high ventilation rates and a flow directed towards the body, see Figures~\ref{fig:non_dim_T_sim_7} and~\ref{fig:non_dim_T_sim_9}, the high temperature region is significantly deflected by the ventilation flow. Interestingly, in the case of very strong ventilation (Figure~\ref{fig:non_dim_T_sim_9}), the thermal plume is located in front of the body, below the head.

\subsubsection*{Conclusions}

The effect of the thermal plume generated by a human body on the velocity field of a ventilated room has been investigated. For this purpose, the ventilation flow rates and the direction of the ventilation flow were varied. Results show that the effect of the thermal plume is significant for rooms with poor to moderate ventilation. The effect becomes insignificant for large ventilation rates. Therefore, it is expected that the thermal plume could affect the dispersion of saliva droplets in case of poor or moderate room ventilation. This is further investigated in Section~\ref{sec:combo_plumes}.

\clearpage

\section{Modelling the dispersion of breath}\label{sec:Breath}

\vspace{3mm}

\noindent By \emph{A.M. Akbar, A. Giusti and D. Fredrich}

\vspace{3mm}

\noindent The breathing and related emissions of carbon dioxide (CO\textsubscript{2}) from the mouth locally modify the background ventilation flow field. In order to improve the accuracy of the computation of spray and aerosol dispersion, it is important to include into the model the local jets emitted from the mouth. In addition, even in the case of no viral transmission (i.e., saliva droplets with no viral content), the dispersion of CO\textsubscript{2} in the room is an important aspect of air quality that must be monitored. Therefore, this section aims at selecting appropriate models to evaluate the velocity field in the near-mouth region and compute the dispersion of CO\textsubscript{2} in a closed environment.

As a first approximation, the breath from our mouth can be considered a continuous jet. In this approximation, a breath plume is defined as a body of exhaled fluid moving through the surrounding air fluid. In the field of environmental science, many models have already been developed to compute the dispersion of pollutants exiting factory stacks. Such models will be considered first to reproduce the dispersion of CO\textsubscript{2} exiting mouths in indoor environments. Hence, ventilation conditions have been used as opposed to atmospheric conditions to determine the direction of the jet. Differently to outdoors, mixing ventilation generally leads to an average steady-state CO\textsubscript{2} content in the room. This is accounted for by imposing a uniform background CO\textsubscript{2} field. In addition, once a plume reaches a wall, ceiling, or floor boundary, the plume must be, to some extent, `reflected'. Reflections will thus be taken into account as well. Technically, this is achieved by expanding the domain dimensions of the room and calculating the CO\textsubscript{2} concentration should the plume continue to grow beyond the boundary, and then by projecting that concentration value onto the reflected coordinate inside the room. The plume approximation can also be used to estimate the gas velocity close to the mouth, and its decay at greater distance. A jet velocity profile can be produced at every mouth exit to improve the accuracy of the computation of spray and aerosol dispersion discussed in Section~\ref{sec:dispersion}.

It should be noted that, in reality, CO\textsubscript{2} is not released continuously from the mouth, as described by the plume model, but instead as pulsed puffs. In order to further improve the accuracy of the computation of CO\textsubscript{2} dispersion, a puff model, where the puffs are released at specified time intervals, is implemented. The pulsed puff model also allows for a more realistic coupling with the background ventilation pattern. The puffs follow a trajectory based on the ventilation velocity, and the expansion of the puff is determined by the dispersion coefficients in the ventilation, normal, and vertical directions \cite{Silva2013} (described by dispersion parameters, $\sigma_{x'}$, $\sigma_{y'}$, and $\sigma_{z'}$, respectively, where ($x',y',z'$) are the coordinates representing the location of a given point in a frame of reference with the $x'$-axis aligned with the `wind' direction and the origin located at the puff's centre of mass -- this will be discussed in the following).

The specific objectives of this study can be summarised as: (i) choose a plume model for the dispersion of CO\textsubscript{2} pollutants; (ii) select an approximation for the jet velocity profiles that form at each mouth exit; (iii) choose a puff model for the dispersion of CO\textsubscript{2} pollutants; (iv) implement the plume, the jet velocity profile, and the puff models, each into Python; (v) calculate the mixing value, which indicates the average CO\textsubscript{2} in the room; (vi) analyse a case study to provide more insight into the performance of each model.

The evaluation of the models has been performed by considering a model tutorial room ($15.0~ $m$~\times~8.5~$m$~\times~3.0~$m) occupied by 10 people. The distribution of tables is representative of a tutorial room at Imperial College London. The developed models and their coupling with the tool designed in this project will provide an insight into the behaviour of breath plumes and their impact as a function of room ventilation, the spatial distribution of occupants in the room, and reflections from the walls, ceiling, and floor. The model is conceived to be flexible and inputs corresponding to each of the occupants can be set independently to represent different exhalation activities, such as speaking and coughing.

\subsection*{Methods}

Gaussian models were chosen for the plume, jet, and puff representations. The Gaussian plume model, which is commonly used in environmental sciences, provides an approximation close to experimental results on the dispersion of pollutants from stacks \cite{Miller1987}, hence it was selected as a reliable method. The jet velocity profiles are well approximated by the Gaussian curve for distances away from the potential core region of the jet \cite{Aziz2008}. Additionally, the Gaussian distribution provides a good model for puff concentrations \cite{Cao2011}. 

\subsubsection*{Modelling plumes}

\begin{figure}[t]
  \centering
  \includegraphics[width=1\linewidth]{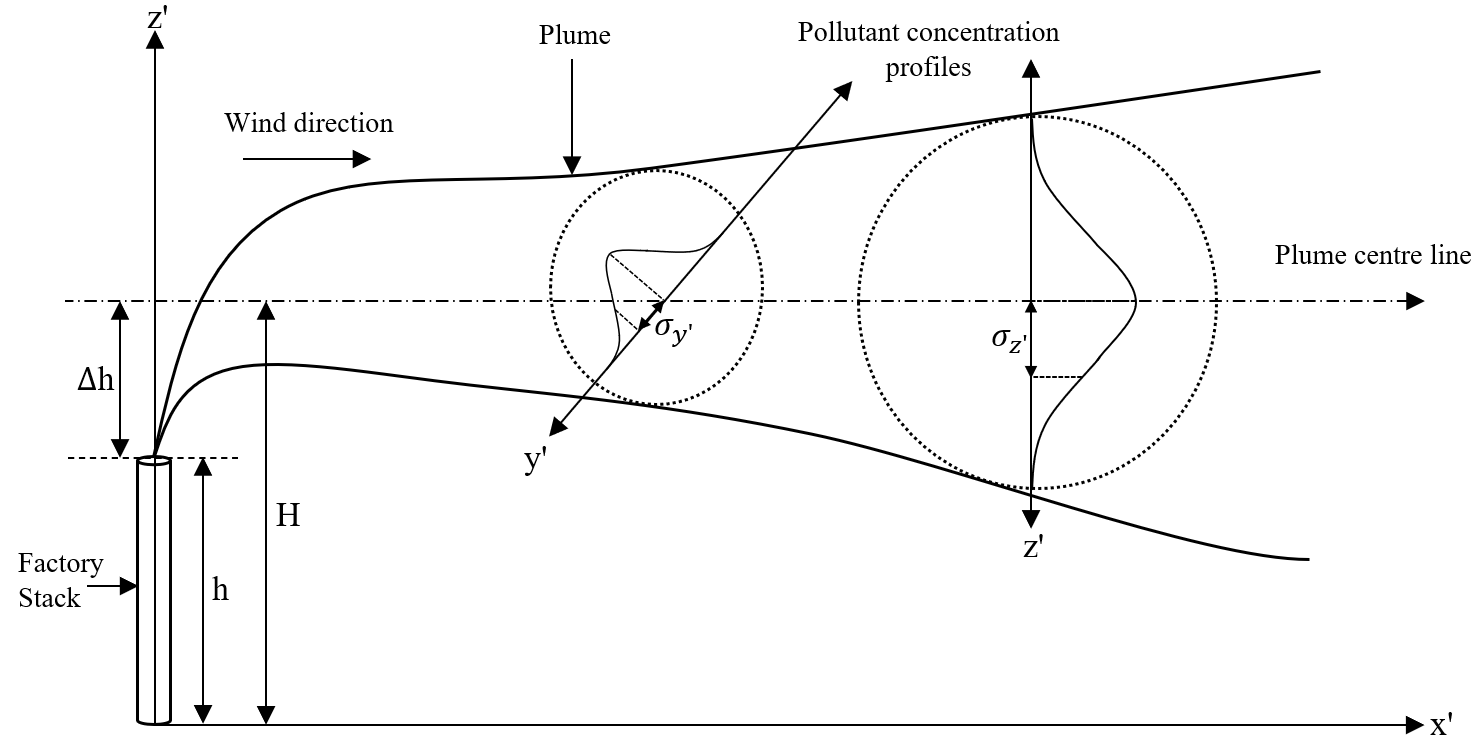}
  \caption{Schematic of the plume and relevant quantities for the Gaussian plume model.}
  \label{fig:Plume}
\end{figure}

The Gaussian plume model follows a normal statistical distribution. Note that in the following, the coordinates ($x',y',z'$) will be used to indicate the location of a given point in a frame of reference with the $x'$-axis aligned with the `wind' direction and the origin located at the stack position, as shown in Figure~\ref{fig:Plume}. The coordinates ($x,y,z$) will indicate the position with respect to a frame of reference with the axes aligned with the sides of the room (a cuboid in all the cases presented here - the origin typically coincides with one of the corners). For the sake of simplicity, a horizontal mean ventilation velocity is considered. Two dispersion coefficients are used, $\sigma_{y'}$ and $\sigma_{z'}$, representing the dispersion in the direction normal to the page and in the vertical direction, respectively \cite{Abdel2008}. In other words, $\sigma_{y'}$ and $\sigma_{z'}$ represent the dispersion of pollutant concentration along two Cartesian coordinates in a plane normal to the plume centre line.

In the Gaussian plume model, the concentration of the pollutant exiting the orifice follows the equation~\cite{Abdel2008}:
\begin{multline}\label{eq:Plume}
C \left(x',y',z'\right)=\frac{Q}{2\pi\sigma_{y'}\sigma_{z'}U}\mathrm{exp}\left[-\frac{1}{2}\left(\frac{y'}{\sigma_{y'}}\right)^2\right] \times \\ \left[\mathrm{exp}\left(-\frac{(z'-H)^2}{2\sigma_{z'}^2}\right) + \mathrm{exp}\left(-\frac{(z'+H)^2}{2\sigma_{z'}^2}\right)\right],
\end{multline}
where $C$ is the plume concentration at a given coordinate ($x',y',z'$), $Q$ is the pollutant emission rate from the source, $U$ is the magnitude of the mean `wind' velocity which determines the $x'$-direction, $y'$ is the lateral distance normal to the `wind' direction, $z'$ is the vertical distance and $H$ is the height of the plume; $H = h + \Delta h$, where $h$ is the height of the source and $\Delta h$ is the plume rise height.

The Gaussian plume model is based on the following assumptions: (i) the conditions are steady-state; (ii) the emission of pollutants from the source is continuous; (iii) there is negligible diffusion in the $x'$-direction; (iv) the $U$ value has a constant magnitude and direction with time and varying height; (v) the dispersion coefficients are functions of $x'$ \cite{Abdel2008}. 

For this study, the atmospheric conditions have been replaced by ventilation conditions, assumed to be uniform in the room (note that to relax this assumption, the pulsed puff model described below has been implemented), and the source is the mouth of the occupant in sitting position. Thus, in Equation~\ref{eq:Plume}, $U$ represents the ventilation speed. Additionally, $\Delta h=0$~m because the plume rise due to the hot air from the mouth is assumed to be negligible, therefore the plume will be injected into the $x$-$y$ plane of the room. Moreover, in all the examples shown in this section, $h=1.2$~m, as this was assumed to be the average height of an occupant in sitting position. Hence, $H=h=1.2$~m. The room of cuboidal shape has been discretised with a Cartesian grid and the value of $C$ in each grid point has been evaluated using Equation~\ref{eq:Plume}.

The $Q$ value is calculated by using an average breathing rate of 12 breaths per minute~\cite{BBCBitesize}, an average tidal volume of 0.5~L~\cite{BBCBitesize}, and the density, $\rho$, of CO\textsubscript{2} at normal temperature and pressure, which is 1.842 kg/m$^3$ \cite{EngTool2019}. In general, the mass flow rate from a given source can be expressed as:
\begin{equation}\label{eq:mdot}
    \dot{m} = \rho \dot{V},
\end{equation}
where $\dot{m}$ is the mass flow rate of CO\textsubscript{2}, and $\dot{V}$ is the volume flow rate of CO\textsubscript{2}. $Q$ can be expressed as either the mass flow rate or volume flow rate. In this work, $Q$ is expressed as the mass flow rate ($Q=\dot{m}$) and Equation~\ref{eq:mdot} is used to calculate the related value.

The dispersion coefficients, $\sigma_{y'}$ and $\sigma_{z'}$, were calculated using formulae modified from the Brigg’s model~\cite{Mohan1997}. The atmospheric stability was categorised according to Pasquill and Gifford stability classes~\cite{Woodward2010}. As the investigation was performed for an indoor environment, the most stable class, F, was chosen. Additionally, an open-country environment was selected rather than an urban environment as it was assumed that there were no significant obstacles in the room to impact the plume dispersion. Class F corresponded to the following formulae for calculating $\sigma_{y'}$ and $\sigma_{z'}$~\cite{Mohan1997}:
\begin{equation}\label{eq:sigy}
    \sigma_{y'}=\frac{0.04x'}{\sqrt{1+0.0001x'}},
\end{equation}

\begin{equation}\label{eq:sigz}
    \sigma_{z'}=\frac{0.016x'}{1+0.0003x'},
\end{equation}
where $x'$ is the distance from the source in the `wind' direction.

The distance $x'$ was calculated using the scalar product with the unit ventilation velocity vector. The distance $y'$ was calculated by finding the magnitude of the vector normal to the ventilation direction. Moreover, the height $z'$ was computed by finding the difference in heights between the two coordinates. From Equations \ref{eq:sigy} and \ref{eq:sigz}, $\sigma_{y'}$ and $\sigma_{z'}$ vary with $x'$, thus, both values were calculated for every coordinate in the room. Then, the corresponding $C$ values at every coordinate were stored in a 3D matrix using Python. The overlapping plumes from different occupants contributed to the $C$ value at that coordinate, and the results were displayed as $z$-contour plots of the room.

\begin{figure}[t]
  \centering
  \includegraphics[width=1\linewidth]{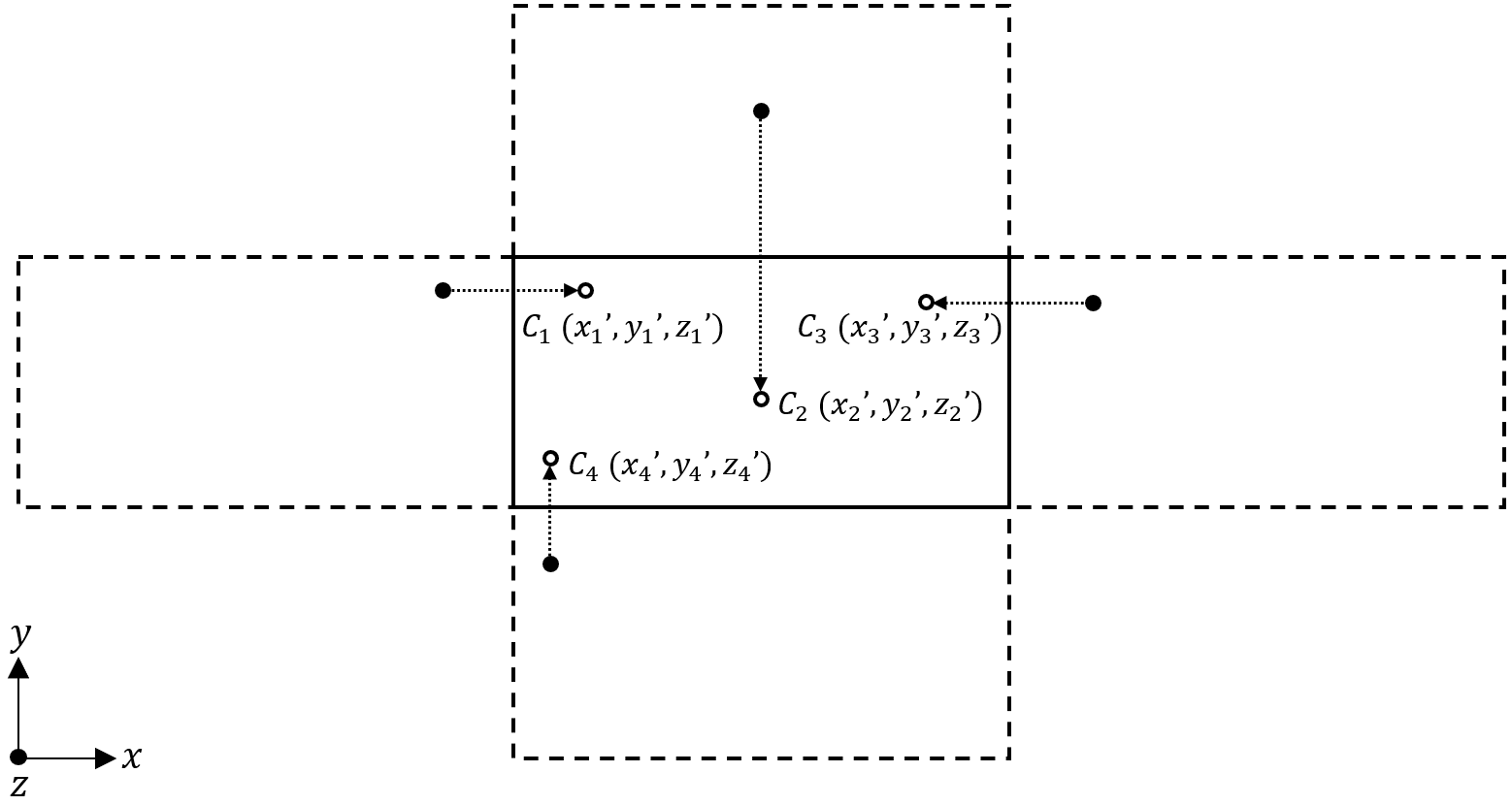}
  \caption{Schematic of the plume reflections from boundaries displaying the projection of $C$ values for singular coordinates (expansion shown here in the positive and negative $x$ and $y$-directions, but in the calculations, expansion along the $z$-axis was also included). The solid line rectangle represents the $x$-$y$ plane of the room ($15.0~$m$~\times~8.5~$m$~\times~3.0~$m) and the dashed line rectangles indicate the expanded domains, each with dimensions equivalent to the room.}
  \label{fig:RoomReflections}
\end{figure}

The plume reflections (see Figure~\ref{fig:RoomReflections} for reference) were included in the Python code and $z$-contour plots were produced. To incorporate the reflections, the domain was expanded in the positive and negative directions along the three coordinate axes. In particular, the room domain was duplicated at each of the six faces. The effect of reflections at the corners were neglected as well as the reflection contribution for distances higher than the length of the room along each side. Depending on which boundary the plume had reached, $C$ was computed for a point in the expanded domain. Then, this value was projected onto the reflected coordinate within the room domain. The reflections were applied to portray the realistic behaviour of plumes when reaching a boundary.

In order to verify the plume reflection Python code, (i) the ventilation velocity was varied and (ii) the coordinates of occupants were distributed uniformly within the room. Verification was performed through direct inspection of the results. In particular, the direction of the plumes and the decay of their concentration were used as key parameters to check the implementation.

\subsubsection*{Jet velocity profile}

\begin{figure}[t]
  \centering
  \includegraphics[width=1\linewidth]{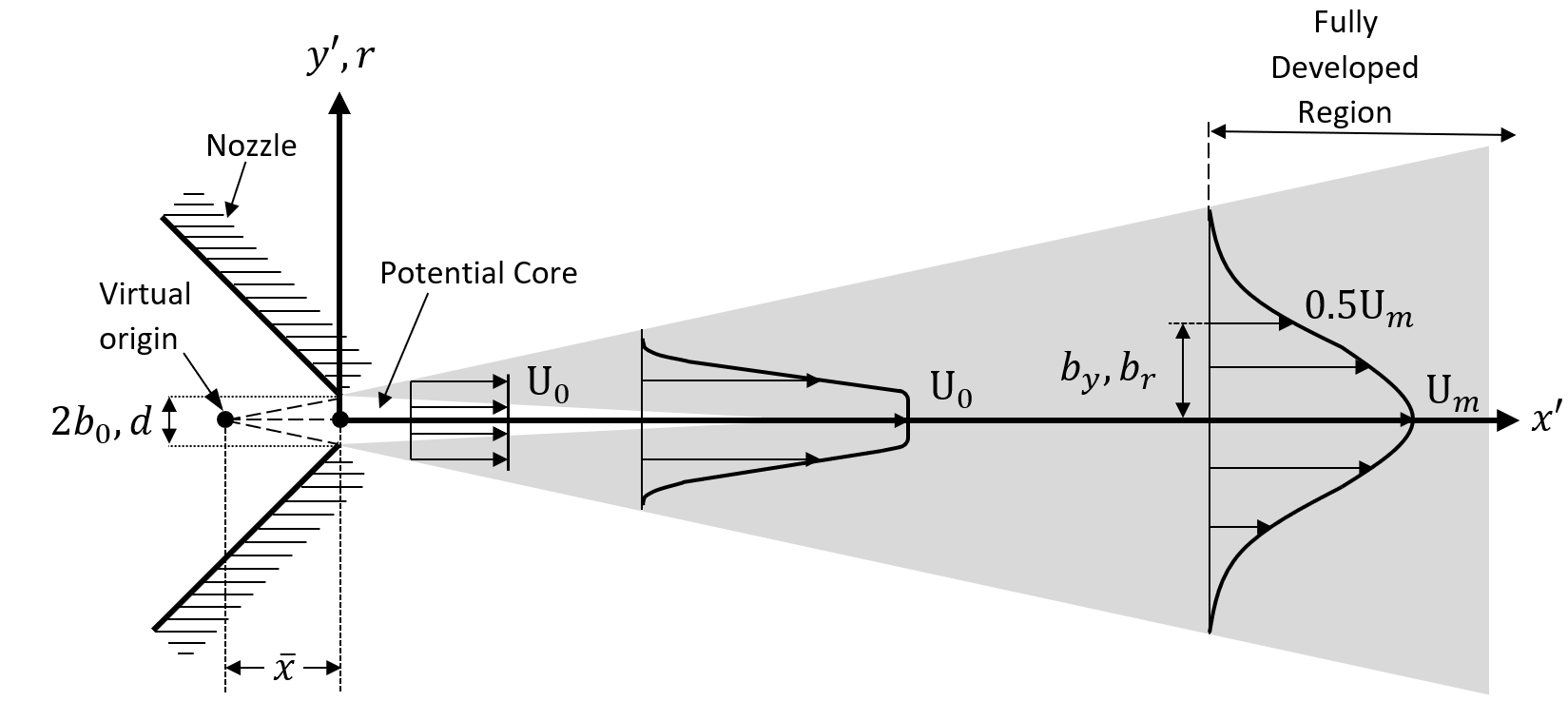}
  \caption{Jet velocity profile diagram, where $2b_0$, $b_y$ and $x'$-$y'$ coordinates are used for planar jets and $d$, $b_r$ and $x'$-$r$ coordinates are used for circular jets.}
  \label{fig:Jet}
\end{figure}

Jet velocity profiles originating from the mouths were modelled by a Gaussian curve approximation. This approximation is valid since the mouth exit size is small, resulting in a short initial potential jet region. Hence, the velocity profile is mainly in the fully developed region which fits the Gaussian distribution curve well~\cite{Lee2011}. This jet velocity profile is shown in Figure~\ref{fig:Jet}. Note that, in the following, the coordinates ($x',y',z'$) will be used to indicate the location of a given point in a frame of reference with the $x'$-axis aligned with the `wind' direction (and the jet centre line) and the origin located at the nozzle exit, i.e., the mouth. 

Two types of incompressible submerged jet flows were modelled and compared: planar ($x'$-$y'$ coordinates) and circular ($x'$-$r$ coordinates). The equations representing planar jets are written as~\cite{Aziz2008}:
\begin{equation}
    b_{y}=A_{1}\left(x'+\bar{x}\right),
\end{equation}
\begin{equation}
    U_{m}=\frac{A_{3}U_{0}}{\sqrt{x'/b_{0}+\alpha_{1}}},\label{eq:by}
\end{equation}
\begin{equation}\label{eq:ujP}
    U_{j}=U_{m}\times\mathrm{exp}\left[-0.693\left(\frac{y'}{b_{y}}\right)^2\right],
\end{equation}
where $b_{y}$ is the distance in the $y'$-direction, $A_{1}$ and $A_{3}$ are planar jet coefficients, $x'$ is the distance along the jet centre line, $\bar{x}$ is the distance between the nozzle and virtual origin, $U_{m}$ is the velocity at any point along the centre line, $U_{0}$ is the initial uniform nozzle exit velocity, $b_{0}$ is half the nozzle width, $\alpha_{1}$ is the planar jet correction value for the virtual origin, $U_{j}$ is the jet velocity given by the Gaussian distribution and $y'$ is the distance in the direction normal to the jet centre line.

The equations representing circular jets are as follows~\cite{Aziz2008}:
\begin{equation}\label{eq:br}
    b_{r}=A_{2}\left(x'+\bar{x}\right),
\end{equation}
\begin{equation}\label{eq:umr}
    U_{m}=\frac{A_{4}U_{0}}{x'/d+\alpha_{2}},
\end{equation}
\begin{equation}\label{eq:ujC}
    U_{j}=U_{m}\times\mathrm{exp}\left[-0.693\left(\frac{r}{b_{r}}\right)^2\right],
\end{equation}
where $b_{r}$ is the distance in the $r$-direction, $A_{2}$ and $A_{4}$ are circular jet coefficients, $d$ is the nozzle diameter, $\alpha_{2}$ is the circular jet correction value for the virtual origin and $r$ is the radial distance in the direction normal to the jet centre line. All the other quantities are defined as in Equations~\ref{eq:by}--\ref{eq:ujP}.

The values for $A_{1}$, $A_{2}$, $A_{3}$ and $A_{4}$ were set to 0.097, 0.097, 3.5 and 6.3, respectively, as presented in Aziz et al.~\cite{Aziz2008}. The $U_{0}$ values used were 3.9 m/s and 11.7 m/s, as these are the average expiration air velocities for speaking and coughing, respectively~\cite{Chao2009}. Additionally, $\bar{x}=0$ m since the mouth is the origin and exit for the air, hence $\alpha_{1}=\alpha_{2}=0$. The average mouth width is 50~mm~\cite{Christensen2010}, thus $d=0.05$~m and $b_{0}=0.025$~m.

Equations~\ref{eq:by}--\ref{eq:ujC} were implemented into Python where a 3D matrix was formed for each type of jet to store the velocities at every coordinate for every occupant in the room. Then, $z$-contour plots were generated for each of the speaking and coughing scenarios.    

\subsubsection*{Modelling puffs}

\begin{figure}[t]
  \centering
  \includegraphics[width=1\linewidth]{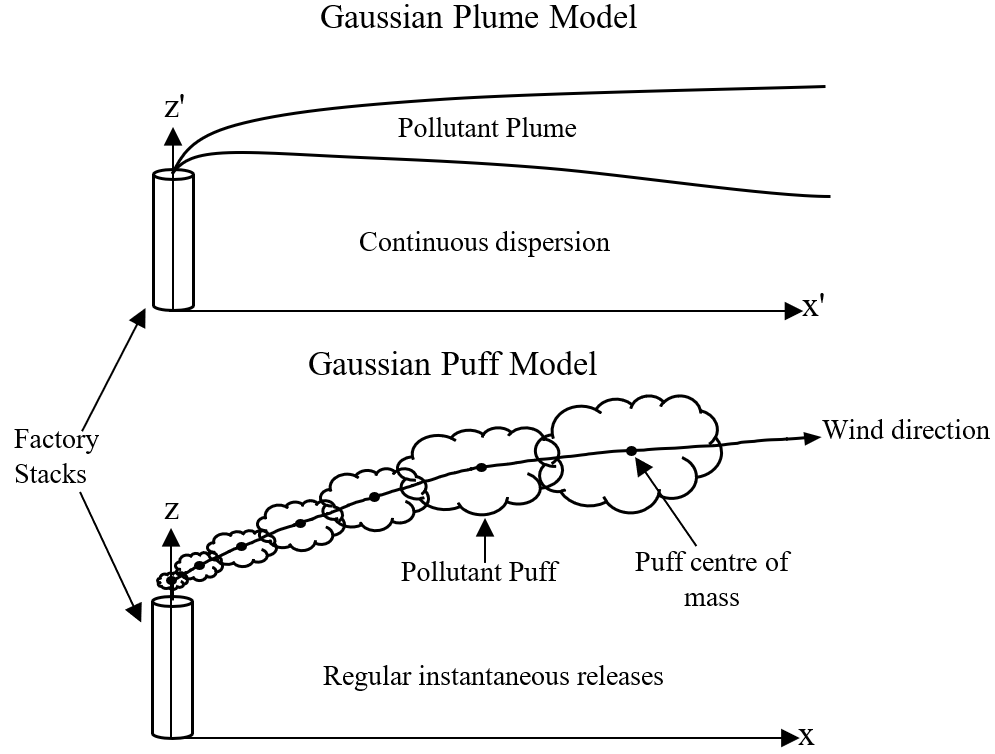}
  \caption{Comparison between the Gaussian plume and puff models.}
  \label{fig:PlumeVsPuff}
\end{figure}

The Gaussian puff model was used to demonstrate a more realistic dispersion of the CO\textsubscript{2} pollutants by injecting puffs into the room at timed intervals, unlike the Gaussian plume model, which represented a continuous exhalation of breath as shown in Figure~\ref{fig:PlumeVsPuff}. Note that the coordinates ($x',y',z'$) will be used here to indicate the location of a given point in a frame of reference with the $x'$-axis aligned with the `wind' direction and the origin located at the puff's centre of mass.

A Lagrangian particle tracking model was implemented into Python, in which the injected Lagrangian particle represented the puff's centre of mass. It was assumed that the particle velocity was the same as the background ventilation field. For the sake of simplicity, in the examples shown in this section, a uniform velocity value of 0.1~m/s in the $x$ and $y$-directions was imposed. However, it should be noted that when the model is applied to real cases, the velocity value is directly taken from the computational fluid dynamics simulations of the time-averaged ventilation field. 

Since on average, people breathe about 12 times per minute \cite{BBCBitesize}, particles were injected into the room every 5 seconds. Three dispersion coefficients in the ventilation, normal, and vertical directions, $\sigma_{x'},~\sigma_{y'},~\sigma_{z'}$, respectively, were used for the Gaussian puff model to depict the enlargement of the puff due to diffusion \cite{Silva2013}. The age of a puff, $t$, is the total time elapsed since the puff was injected into the room. The $\sigma_{x'},~\sigma_{y'},~\sigma_{z'}$ values varied with $t$ rather than distance, to provide a more accurate result. The Gaussian puff model equation is as follows \cite{Cao2011}:
\begin{multline}\label{eq:Puff}
    c (x',y',z',t)=\frac{Q_{m}}{\left(2\pi\right)^{3/2}\sigma_{x'}\sigma_{y'}\sigma_{z'}}\times \\ \mathrm{exp}\left[-\frac{1}{2}\left(\frac{x'}{\sigma_{x'}}\right)^2-\frac{1}{2}\left(\frac{y'}{\sigma_{y'}}\right)^2-\frac{1}{2}\left(\frac{z'}{\sigma_{z'}}\right)^2\right],
\end{multline}
where $c$ is the puff concentration at a given coordinate and time $\left(x',y',z',t\right)$, $Q_{m}$ is the initial mass released, and $x'$, $y'$ and $z'$ are, respectively, the distance in the `wind', normal to the `wind' and vertical direction at time, $t$. This model, originally developed in the context of atmospheric dispersion of pulsed emissions, will be applied in the present study to investigate indoor dispersion due to ventilation.

The dispersion coefficients were assumed to be isotropic to form spherical puffs. They are determined using an intermediate-field approximation, assuming the viscosity and initial puff size are minor factors~\cite{DeHaan1998}: 
\begin{equation}\label{eq:sigif}
    \sigma_{if}^2(t)=C_{if}\varepsilon t^3,
\end{equation}
where $\sigma_{if}$ is the dispersion coefficient for a given time instant ($\sigma_{if}=\sigma_{x'}=\sigma_{y'}=\sigma_{z'}$), $C_{if}$ is a constant that will be equal to unity for this investigation, and $\varepsilon$ is the turbulence kinetic energy dissipation rate. The value of $\varepsilon$ was set to 0.001~$\mathrm{m^2/s^3}$. The puff dispersion for different values of $\varepsilon$, namely $\varepsilon=0.002$~$\mathrm{m^2/s^3}$ and $\varepsilon=0.006$~$\mathrm{m^2/s^3}$ was also evaluated and compared. Note again that $\varepsilon$ is directly computed by the computational fluid dynamics simulations of the room ventilation. Therefore, in the evaluation of real cases, the local value of this quantity will be directly computed from the solution of the ventilation field. The calculation of $Q_{m}$ uses the following expression:
\begin{equation}\label{eq:m}
    Q_m = m = \rho V,
\end{equation}
where $m$, $\rho$ and $V$ represent, respectively, the mass, density and volume of CO\textsubscript{2} released per breath.

Using Equations~\ref{eq:Puff}--\ref{eq:m}, the $c$ values in the room domain were stored in a 3D matrix in Python and $z$-contour plots were produced for the different $\varepsilon$ values. 

\subsubsection*{Mixing Value}

The mixing value in the model tutorial room at steady-state conditions can be determined via the following equation:
\begin{equation}\label{eq:mix}
    Y_{CO_{2}mix}=\frac{N\, \dot{m}_{CO_{2}}}{\dot{m}_{mix}}=\frac{N\, \dot{m}_{CO_{2}}}{N\, \dot{m}_{CO_{2}}+\dot{m}_{air}},
\end{equation}
where $Y_{CO_{2}mix}$ is the mixing value, $N$ is the number of occupants in the room, $\dot{m}_{CO_{2}}$ is the mass flow rate of CO\textsubscript{2} exiting the mouth, $\dot{m}_{mix}$ is the mass flow rate of the mixture of exhaled CO\textsubscript{2} and air exiting the room at steady-state conditions, $\dot{m}_{air}$ is the mass flow rate of the air from the ventilation. The value $Y_{CO_{2}mix}$ can be used to set the background time-averaged value of CO\textsubscript{2} in the room in the case of perfect mixing.

\subsection*{Results and discussion}

\begin{figure}[t]
  \centering
  \includegraphics[width=0.95\linewidth]{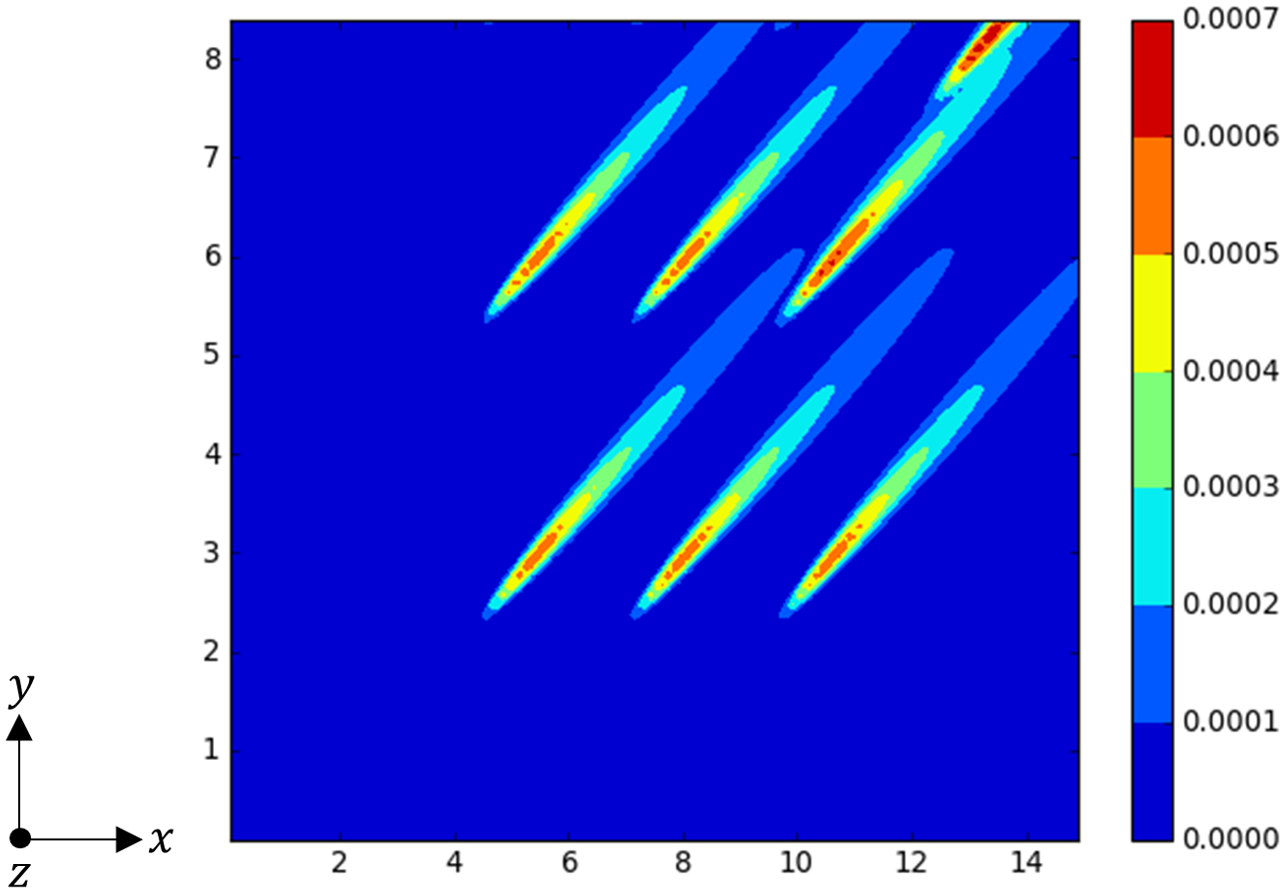}
  \caption{Gaussian plume model $z$-contour plot. The $x$-$y$ axes represent the room lengths [m], the colour bar displays the CO\textsubscript{2} plume concentrations [kg/m$^3$] within the room ($15.0~$m$~\times~8.5~$m$~\times~3.0~$m), at a height of 1.24~m. The ventilation velocity components are 6~m/s and 4~m/s in the $x$ and $y$-directions, respectively.}
  \label{fig:PlumeResults}
\end{figure}

\begin{figure}[t]
  \centering
  \includegraphics[width=0.95\linewidth]{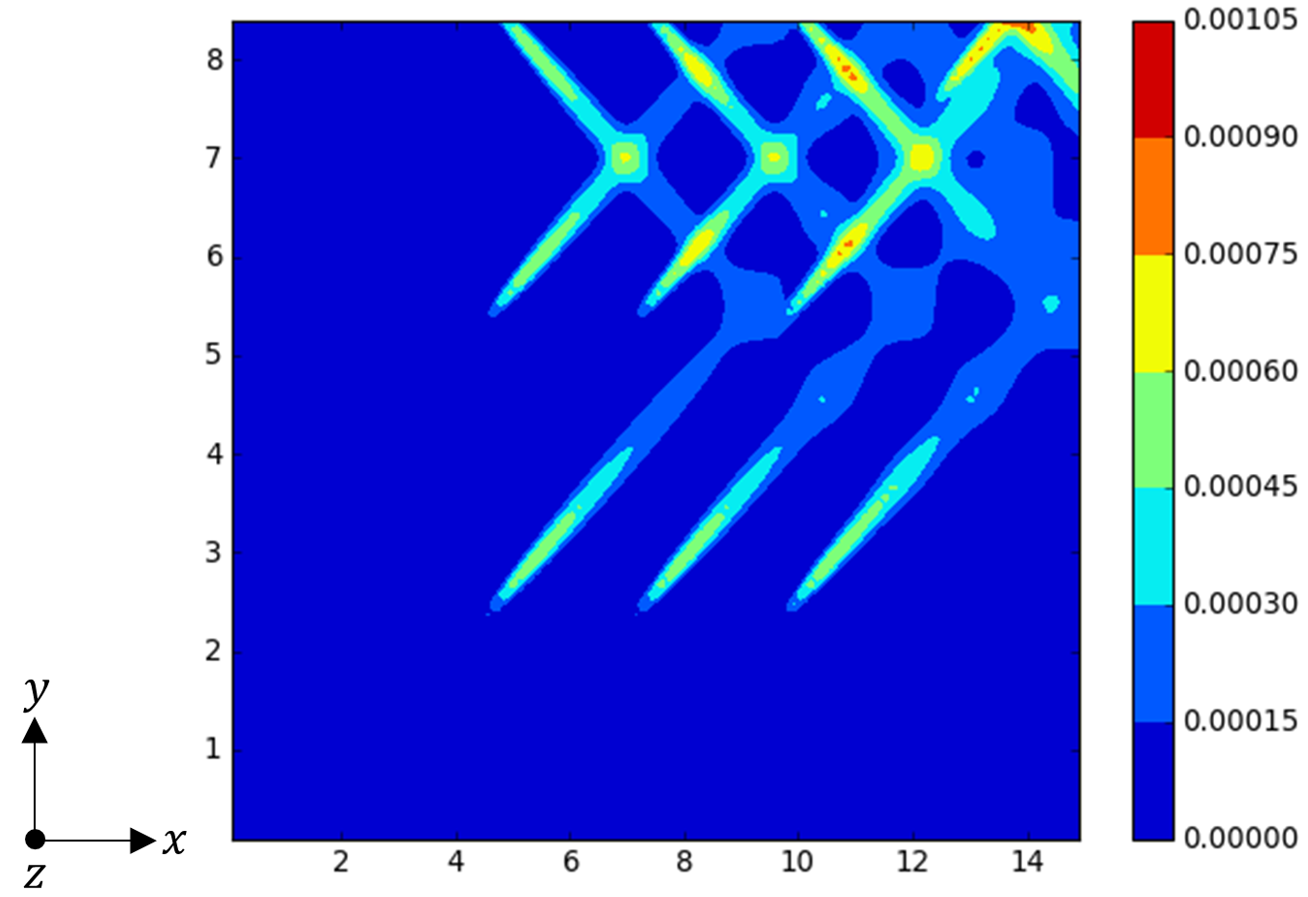}
  \caption{Gaussian plume model $z$-contour plot including reflections. The $x$-$y$ axes represent the room lengths [m], the colour bar displays the CO\textsubscript{2} plume concentrations [kg/m$^3$] within the room ($15.0~$m$~\times~8.5~$m$~\times~3.0~$m), at a height of 1.24~m. The ventilation velocity components are 6~m/s and 4~m/s in the $x$ and $y$~directions, respectively.}
  \label{fig:ReflectionResults}
\end{figure}

Figure~\ref{fig:PlumeResults} demonstrates the CO\textsubscript{2} plume concentration dispersion approximately at the height of the occupants' mouths in sitting position. It is expected that at this height, the plume concentrations would be the highest. Furthermore, the plumes are spread in the direction of the ventilation velocity. The reflected plumes and their overlap with the original plumes, producing regions of high CO\textsubscript{2} concentrations, are shown in Figure~\ref{fig:ReflectionResults}.

\begin{figure}[t]
  \centering
  \includegraphics[width=0.95\linewidth]{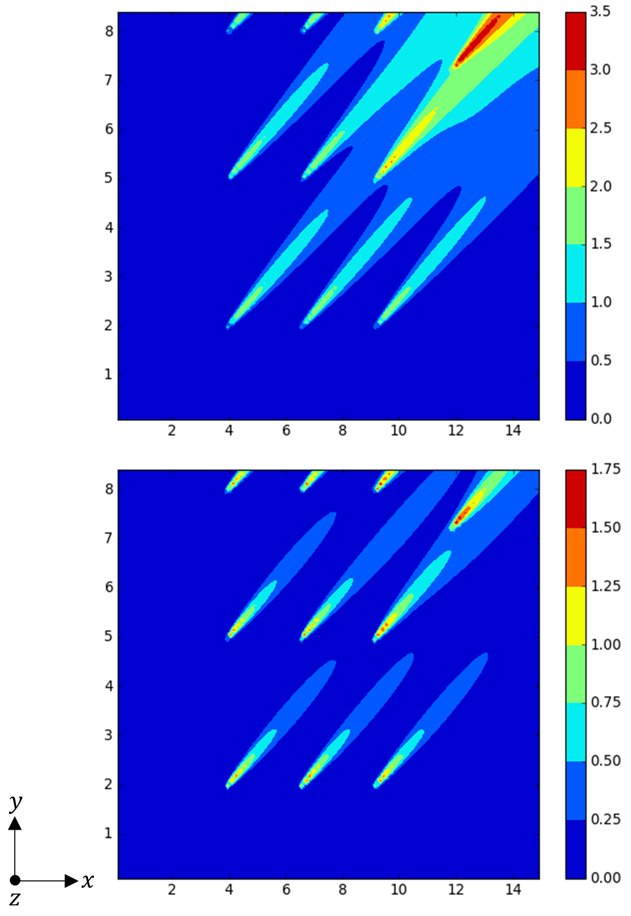}
  \caption{Planar (top) and circular (bottom) jet velocity profile $z$-contour plots. The $x$-$y$ axes represent the room lengths [m], the colour bars display the air velocities [m/s] within the room ($15.0~$m$~\times~8.5~$m$~\times~3.0~$m), at a height of 1.24~m. The ventilation velocity components are 6~m/s and 4~m/s in the $x$ and $y$~directions, respectively, where $U_{0} = 3.9$~m/s for speaking. Note that a different colour bar scaling is used for each contour plot.}
  \label{fig:JetResults}
\end{figure}

In Figure~\ref{fig:JetResults}, two different jet velocity profiles are shown, in which the observed maximum velocity is double for the planar jet compared to the circular jet. Hence, there is a greater decay from $U_0$ for the circular jet.

\begin{figure}[t]
  \centering
  \includegraphics[width=0.95\linewidth]{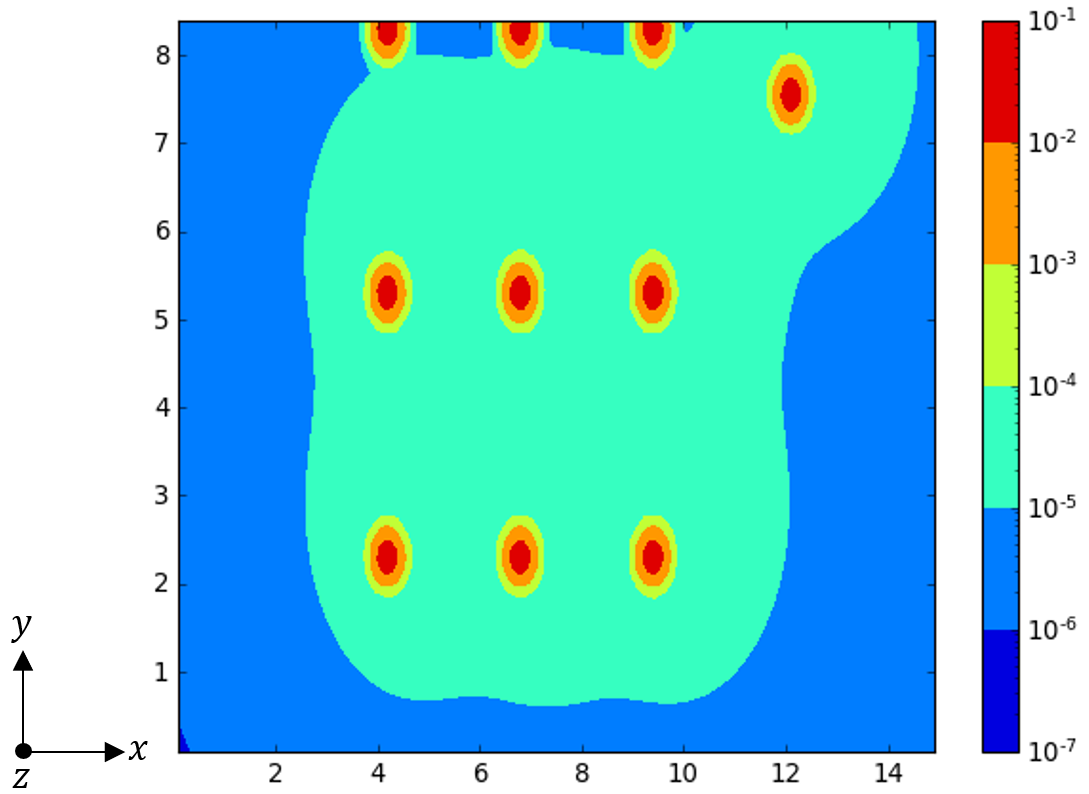}
  \caption{Gaussian puff model $z$-contour plot with $\varepsilon~=~0.001$. The $x$-$y$ axes represent the room lengths [m], the colour bar displays the CO\textsubscript{2} puff concentrations [kg/m$^3$] within the room ($15.0~$m$~\times~8.5~$m$~\times~3.0~$m), at a height of 1.24~m, for a total time of 120~s. The puff velocity components are 0.1~m/s in both the $x$ and $y$-directions. Note that the colour bar has a logarithmic scale.}
  \label{fig:PuffResults}
\end{figure}

\begin{figure}[t]
  \centering
  \includegraphics[width=0.95\linewidth]{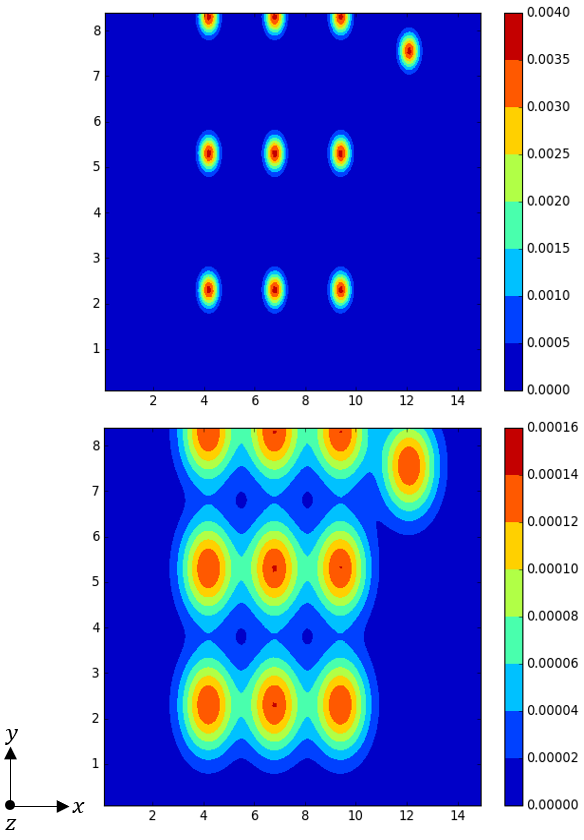}
  \caption{Gaussian puff model $z$-contour plots with $\varepsilon=0.002$ (top) and $\varepsilon=0.006$ (bottom). The $x$-$y$ axes represent the room lengths [m], the colour bars display the CO\textsubscript{2} puff concentrations [kg/m$^3$] within the room ($15.0~$m$~\times~8.5~$m$~\times~3.0~$m), at a height of 1.24~m, for a total time of 120~s. The puff velocity components are 0.1~m/s in both the $x$ and $y$-directions. Note that a different colour bar scaling is used for each contour plot.}
  \label{fig:PuffComparisonResults}
\end{figure}

Results of the puff model are shown in Figure~\ref{fig:PuffResults}, where a logarithmic scale is used to display the entire puff dispersion over the time period. The results indicate that the puff CO\textsubscript{2} concentration decays rapidly. Thus, only the puffs with the shortest age are shown.
In addition, the plots in Figure~\ref{fig:PuffComparisonResults} show a direct comparison between the puff sizes when $\varepsilon$ is varied. Larger $\varepsilon$ values resulted in a stronger dispersion due to higher turbulence, which produced larger puffs. A logarithmic scale was not used for these results to provide a clearer comparison between the puffs.

\subsubsection*{Discussion}

In Figure~\ref{fig:PlumeResults}, the plume CO\textsubscript{2} concentrations decrease as the distance from the occupants increases. This physically represents the diffusion of CO\textsubscript{2} in the surrounding air (or equivalently the entrainment of air in the plume). However, the following assumptions of the Gaussian plume model \cite{Abdel2008} may not be accurate for  modelling breathing activity. Firstly, steady-state conditions and dispersion coefficients that are functions of $x'$ were used. If the dispersion coefficients varied with time, then the results may be more accurate and representative of the CO\textsubscript{2} dispersion in the room. Secondly, a continuous pollutant emission does not accurately represent the dispersion of pollutants exhaled since breaths are emitted as short discrete puffs. Thirdly, $U$ is constant in its magnitude and direction with time and varying height. However, the air ventilation velocity would vary when moving away from the mouth. To address these drawbacks and provide more realistic results, the Gaussian puff model was selected and implemented. In this model the ventilation velocity field from computational fluid dynamics simulations can be used to model the velocities at every coordinate in the room and to provide more realistic information about the background turbulence.

Several of the values used in the calculations were averages, e.g., a sitting height of 1.2~m, an average breathing rate of 12 breaths per minute \cite{BBCBitesize} (therefore 1 breath every 5 seconds), an average tidal volume of 0.5~L \cite{BBCBitesize}, an average mouth width of 50~mm \cite{Christensen2010}, and the average expiration air velocities for speaking and coughing were 3.9~m/s and 11.7~m/s, respectively \cite{Chao2009}. In principle, the model could be improved by using a statistical distribution for the values of each parameter. However, it is expected that a statistical representation of such parameters is unlikely to significantly affect the results.

Formulae modified from the Brigg’s Model~\cite{Mohan1997} were designed to represent outdoor environments -- open-country and urban. Selecting an open-country environment to represent the room may not be accurate. Likewise, the $A_{1},~ A_{2},~A_{3}$ and $A_4$ values were assumed equal to the values in the report by Aziz et al.~\cite{Aziz2008}, and those values were selected according to their experimental data. Similarly, $C_{if}~=~1$ was used in the report by De Haan et al.~\cite{DeHaan1998}. Hence, these values may not be suitable for use in this study, especially for small rooms or injection very close to the walls. Future work should therefore focus on experiments (or high-fidelity simulations) to evaluate such parameters for ventilated indoor environments.

Moreover, $\sigma_{x'},~\sigma_{y'},$ and $\sigma_{z'}$ were assumed to be isotropic to produce spherical puffs, where ($x',y',z'$) are the coordinates representing the location of a given point in a frame of reference with the $x'$-~axis aligned with the `wind' direction and the origin located at the puff's centre of mass. However, $\sigma_{x'}$ may be larger than $\sigma_{y'}$ due to shear effects from the ventilation, which may also have a bearing on the results \cite{Cao2011}. In addition, the intermediate-field approximation selected for calculating $\sigma_{x'}, ~\sigma_{y'},$ and $\sigma_{z'}$ assumes that viscosity and initial puff size are insignificant factors, which may require further investigation for the specific case of emissions from the mouth.

The maximum CO\textsubscript{2} concentration is greater in Figure~\ref{fig:ReflectionResults} (10.5 $\times 10^{-4}$~kg/m$^3$) than in Figure~\ref{fig:PlumeResults} (7 $\times 10^{-4}$~kg/m$^3$). This is expected because Figure \ref{fig:ReflectionResults} includes the CO\textsubscript{2} concentrations from the plume reflections, unlike Figure \ref{fig:PlumeResults}. Therefore, Figure \ref{fig:ReflectionResults} is a more accurate display of plumes. However, the plume representation could have been further improved if other boundaries, such as tables, were included. Such developments should be the focus of future studies.
The performed verifications (not all shown here) produced the expected results with varying ventilation velocities as the plumes were directed in the velocity direction. Similarly, expected results were produced for the uniform placement of occupants in the room as the plumes were uniformly distributed. These verifications were useful to assess the reliability of the utilised models and code.

By comparing the two plots shown in Figure~\ref{fig:JetResults}, the circular jet plot appears to provide a more realistic display than the planar jet, since a jet of breath is not expected to disperse far across the room. The Gaussian approximation for the jets are valid due to the small size of the mouth exit. As a consequence, the velocity profile was predominantly in the fully developed region of the jet, which is approximated well by the Gaussian distribution. This velocity profile will be incorporated into the model described in Section~\ref{sec:dispersion} to improve the computation of the dispersion of spray and aerosols in the vicinity of the mouth.

In terms of the pulsed puff model, due to the rapid decay of the CO\textsubscript{2} concentrations, only the puffs with the shortest age are distinguishable in Figure~\ref{fig:PuffResults}. This rapid decay may be due to the swift enlargement of the puffs which is determined by $\sigma_{x'},~\sigma_{y'},$ and $\sigma_{z'}$. Hence, it is of crucial importance to reliably compute these values. The puff size comparison in Figure~\ref{fig:PuffComparisonResults} highlights the significance of the $\varepsilon$ value. Since $\varepsilon$ represents the turbulence kinetic energy dissipation rate, it is expected that a larger $\varepsilon$ value would result in a larger puff, which supports the results shown in Figure~\ref{fig:PuffComparisonResults}. The control of the $\sigma_{x'},~\sigma_{y'},$ and $\sigma_{z'}$ values through the local turbulence (i.e., $\varepsilon$) could provide an additional path to manage the dilution of breath in the room.

\subsubsection*{Conclusions}

Models for the velocity profile in the vicinity of the mouth and for the dispersion of CO\textsubscript{2} emitted by occupants have been evaluated.
Regarding the dispersion of CO\textsubscript{2}, a continuous Gaussian plume model~\cite{Abdel2008} was first evaluated by using the Pasquill and Gifford stability classes \cite{Woodward2010} and formulae modified from the Brigg’s model \cite{Mohan1997} to calculate $\sigma_{y'}$ and $\sigma_{z'}$. The model was implemented in Python and verified through the analysis of a model tutorial room. The model was complemented with plume reflections at the walls, ceiling, and floor to give a more realistic representation of dispersion in a closed environment. Verification was conducted by varying the ventilation velocities and position of occupants in the room. Results indicate the correct implementation of the models.

To make the evaluation of CO\textsubscript{2} dispersion more physically consistent, a Gaussian pulsed puff model~\cite{Cao2011} was implemented. This model uses an intermediate-field approximation to calculate $\sigma_{x'},~\sigma_{y'}$ and $\sigma_{z'}$ \cite{DeHaan1998}, where ($x',y',z'$) are the coordinates representing the location of a given point in a frame of reference with origin located at the puff's centre of mass. All the quantities required by the model, such as the local velocity of the air and the dissipation rate of the turbulence kinetic energy, can be directly obtained from computational fluid dynamics simulations. Therefore, it is an ideal candidate to be coupled with the framework developed in this project where numerical simulations are used to compute the background ventilation field.
A Gaussian model~\cite{Aziz2008} was also implemented to compute the velocity field in the vicinity of the mouth. Both planar and circular jets were evaluated. Results obtained with the circular jet approximation were considered more realistic. This is the model recommended to be used to locally modify the background flow field in the computation of spray and aerosol dispersion.

Overall, the developed code could provide useful insight into the dispersion of breath and therefore the distribution of CO\textsubscript{2} concentrations within the room. This may be used as a tool to evaluate the air quality in indoor environments.

\clearpage

\section{Ventilation field in indoor environments}\label{sec:cfd}

\vspace{3mm}

\noindent By \emph{M.F. bin Mohd Fadzil, A. Giusti and D. Fredrich}

\vspace{3mm}

\noindent The development of a tool to predict the time-averaged ventilation pattern in indoor environments is discussed in this section. The focus here lies on tutorial and lecture rooms, and their peculiar characteristics (e.g., tables and lecterns) based on their specific design at Imperial College London. For the sake of simplicity, the geometry is created without considering occupants. The presence of occupants will be taken into account in the computation of spray and aerosol dispersion (Section~\ref{sec:dispersion}) by injecting droplets at the mouth locations and by modifying the predicted ventilation field to include the specific characteristics of breathing (e.g., near-field jet issued by the mouth, Section~\ref{sec:Breath}).

To predict the ventilation field, computational fluid dynamics (CFD) simulations are performed using OpenFOAM. Equations for incompressible turbulent flows are solved, neglecting the effects of buoyancy and gravity on the flow. The mean flow is also assumed statistically stationary, given that the boundary conditions do not change in time (also, the effects due to movement of people and the pulsed nature of the breath are neglected). Hence, the steady-state `simpleFoam' solver was chosen to determine the ventilation field in the rooms. This solver computes the time-averaged incompressible Navier-Stokes equations using the SIMPLE algorithm for the pressure-velocity coupling. The solver requires the properties of the fluid as an input, together with a turbulence model to reproduce the effects of the turbulence on the mean flow. In this study, analysis of the sensitivity of the solution to the turbulence model has been performed. In addition, a mesh (i.e., the space discretisation used to solve the flow equations) sensitivity analysis has been conducted to provide indications on the required mesh refinements.

The data from this pre-processing tool is combined in post-processing with tailored models to account for the effects of thermal and exhaled plumes generated by the occupants of a room.
This combined approach enables a `real-time' estimation of droplet dispersion and the related spatially-resolved infection risk from airborne diseases.

\subsection*{Automated CFD workflow} \label{subsection: IV - Automated CFD Workflow}

An OpenFOAM simulation requires a minimum number of case files to run. Essentially, these files define the parameters of the problem that is being studied, including the thermophysical properties of the fluid, the control parameters of the simulation and solvers, and the initial and boundary conditions for the various quantities computed by the model. The files have to be present in three separate directories: \textit{constant}, \textit{system} and \textit{0} (which is the identified initial time step). Table~\ref{tab:IV - Directory Function} describes the main function of the files contained in each directory.

\begin{table}[t]
\centering
\caption{Directories and their function in an OpenFOAM case.}
\label{tab:IV - Directory Function}
\begin{tabular}{cc}
\hline
\textbf{Directory} & \textbf{Files}                                                                                            \\ \hline
\textit{0}         & \begin{tabular}[c]{@{}c@{}}Specify initial values\\ and boundary conditions\\ of the problem\end{tabular}    \\ \hline
\textit{constant}  & \begin{tabular}[c]{@{}c@{}}Define the mesh and \\ physical properties of\\ the fluid\end{tabular} \\ \hline
\textit{system}    & \begin{tabular}[c]{@{}c@{}}Set parameters involved\\ in solving the problem\end{tabular}                 \\ \hline
\end{tabular}
\end{table}

\usetikzlibrary{shapes.geometric, arrows, decorations.markings}
\tikzstyle{process} =  [rectangle, rounded corners, semithick, minimum width = 6 cm, minimum height= 1 cm, text centered, text width = 6cm, draw = black, fill = white, line width = 1.4pt]
\tikzstyle{arrow} = [thick, decoration={markings,mark=at position
   1 with {\arrow[semithick]{open triangle 60}}},
   double distance=1.4pt, shorten >= 5.5pt,
   preaction = {decorate},
   postaction = {draw,line width=1.4pt, white,shorten >= 4.5pt}]
\tikzstyle{method} = [semithick, text centered, text width = 1 cm,  fill = white]

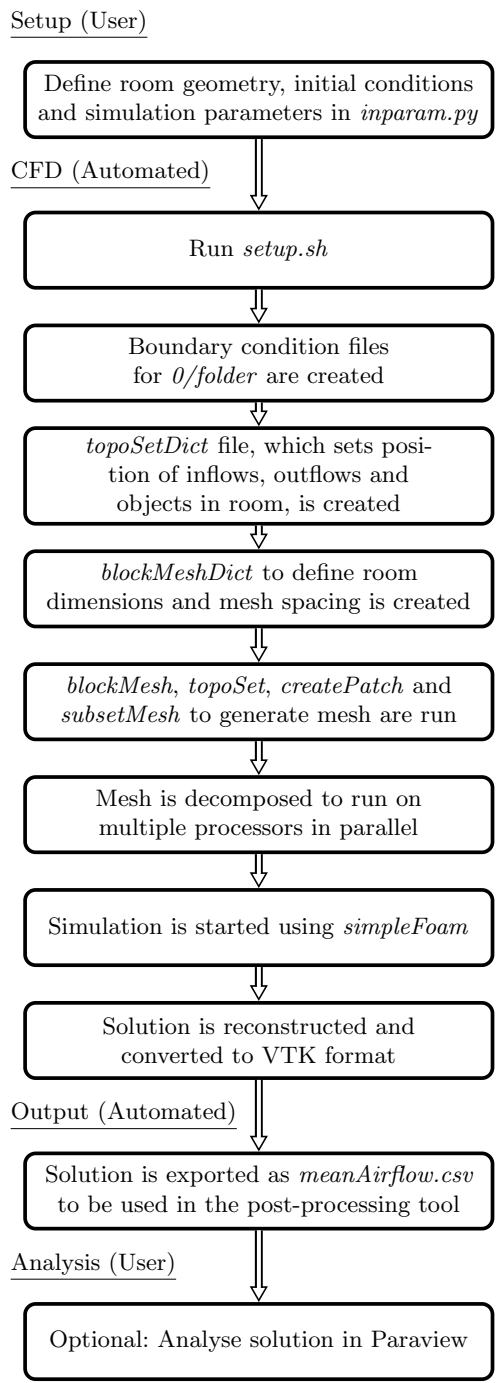
\begin{figure}[t]
    \centering
    \begin{tikzpicture}[node distance = 3cm, thick]
    \node (Geometry) [process] {Define room geometry, initial conditions and simulation parameters in \textit{inparam.py}};
    \node (Category: Setup) [method, above of = Geometry, yshift = -2 cm, xshift = -2.8 cm] {\underline{Setup (User)}};
    \node (Category: CFD) [method, below of = Geometry, yshift = 2 cm, xshift = -2.8 cm] {\underline{CFD (Automated)}};
    \node (Setup) [process, below of = Geometry, yshift = 1 cm] {Run \textit{setup.sh}};
    \node (Boundary Condition) [process , below of = Setup, yshift = 1.5cm] {Boundary condition files for \textit{0/folder} are created};
    \node (topoSetDict) [process , below of = Boundary Condition, yshift = 1.5cm] {\textit{topoSetDict} file, which sets position of inflows, outflows and objects in room, is created};
    \node (blockMeshDict) [process , below of = topoSetDict, yshift = 1.5cm] {\textit{blockMeshDict} to define room dimensions and mesh spacing is created};
    \node (blockMesh) [process , below of = blockMeshDict, yshift = 1.5cm] {\textit{blockMesh}, \textit{topoSet}, \textit{createPatch} and \textit{subsetMesh} to generate mesh are run};
    \node (Decompose) [process , below of = blockMesh, yshift = 1.5cm] {Mesh is decomposed to run on multiple processors in parallel};
    \node (Start) [process , below of = Decompose, yshift = 1.5cm] {Simulation is started using \textit{simpleFoam}};
    \node (Reconstruct) [process , below of = Start, yshift = 1.5 cm] {Solution is reconstructed and converted to VTK format};
    \node (Category: Output) [method, below of = Reconstruct, yshift = 2 cm, xshift = -2.8 cm] {\underline{Output (Automated)}};
    \node (Export) [process , below of = Reconstruct, yshift = 1 cm] {Solution is exported as \textit{meanAirflow.csv} to be used in the post-processing tool};
    \node (Category: Analysis) [method, below of = Export, yshift = 2 cm, xshift = -2.8 cm] {\underline{Analysis (User)}};
    \node (Analyse) [process , below of = Export, yshift = 1cm] {Optional: Analyse solution in Paraview};
    \draw[arrow](Geometry) -- (Setup);
    \draw[arrow](Setup) -- (Boundary Condition);
    \draw[arrow](Boundary Condition) -- (topoSetDict);
    \draw[arrow](topoSetDict) -- (blockMeshDict);
    \draw[arrow](blockMeshDict) -- (blockMesh);
    \draw[arrow](blockMesh) -- (Decompose);
    \draw[arrow](Decompose) -- (Start);
    \draw[arrow](Start) -- (Reconstruct);
    \draw[arrow](Reconstruct) -- (Export);
    \draw[arrow](Export) -- (Analyse);
    \end{tikzpicture}
    \caption{Workflow for the first stage (pre-processor) of the developed model.}
    \label{fig:IV - Flowchart}
\end{figure}

The workflow to obtain the solution of the flow field in a ventilated room can be divided into 4 distinct processes, as shown in Figure \ref{fig:IV - Flowchart}. First, the geometry of the room and the control parameters of the simulation (e.g., number of cores used for the computation) should be set and given as inputs to the CFD solver. Then, the input files for the simulation, including the mesh and all the files contained in the folders listed in Table~\ref{tab:IV - Directory Function}, need to be generated. When all the files are ready, the simulation is run. This is followed by the analysis of the results (note that this step is necessary for the evaluation of the solver and obtained results, but it can be skipped if the setup is judged reliable). Finally, the output file containing the solution in terms of the velocity and turbulence fields, to be given as input to the post-processing tool for the computation of spray and aerosol dispersion (Section~\ref{sec:dispersion}), is generated. A method of automating this tedious workflow was devised in the present work and will be explained in the following.

\subsubsection*{Setup}

To initialise the OpenFOAM simulation, the room geometry first needs to be defined. Traditionally, the method for defining the geometry would require manually editing the \textit{topoSetDict} and \textit{blockMeshDict} files to reflect the room being studied. However, this method is relatively inefficient and difficult to execute for users who are not familiar with the OpenFOAM suite. Several scripts were therefore written in Python to automate this procedure and allow any user to easily set up a case.

First, a Python script called \textit{inparam.py} was written, where the dimensions of the room and its objects such as tables, chairs, steps, etc. (represented by parallelepipeds) can be entered.
Additionally, the inlet and outlet conditions are defined in this file. The inlet was assumed to have a steady inflow velocity. This can be calculated, e.g., by considering the number of air changes per hour (ACH) required for a particular room. For example, from Ref.~\cite{Air_changes}, a tutorial room is recommended to have an ACH value of 6. For the tutorial room investigated in the present work, this assumption results an inflow velocity of 0.459 m/s.
The outlets, i.e., any outflow ventilation ports, windows or doors, are assumed to have a simple zero-gradient boundary condition.

\begin{table}[t]
\centering
\caption{Python files and their respective functions.}
\label{tab:IV - Python File Function}
\begin{tabular}{cc}
\hline
\textbf{File}           & \textbf{Function}                                                                                                                                  \\ \hline
\textit{bmd\_write.py}  & \begin{tabular}[c]{@{}c@{}}Contains function that creates \\ the \textit{blockMeshDict} file\end{tabular}                                                       \\ \hline
\textit{cpd\_write.py}  & \begin{tabular}[c]{@{}c@{}}Contains function that creates \\ the \textit{createPatchDict} file\end{tabular}                                                     \\ \hline
\textit{inparam.py}     & \begin{tabular}[c]{@{}c@{}}Defines the room geometry,\\ inlet and outlet conditions, \\ and simulation parameters \end{tabular}                                                  \\ \hline
\textit{start.py}       & \begin{tabular}[c]{@{}c@{}}Changes \textit{inparam.py} input values\\ to a suitable format and calls the \\ functions from the other files\end{tabular}     \\ \hline
\textit{tsd\_write.py}  & \begin{tabular}[c]{@{}c@{}}Contains function that creates \\ the \textit{topoSetDict} file\end{tabular}                                                         \\ \hline
\textit{zero\_write.py} & \begin{tabular}[c]{@{}c@{}}Contains functions that create\\ the \textit{pressure}, \textit{omega}, \textit{k}, \textit{epsilon}, \\ \textit{nut} and \textit{velocity} files for \\ the \textit{0} directory\end{tabular} \\ \hline
\end{tabular}
\end{table}

The parameters to control the simulation and define the number of partitions (i.e., cores/processors) used in the computation could also be set in the \textit{inparam.py} file. These parameters are then passed on to the \textit{controlDict} and \textit{decomposeParDict} files. The \textit{controlDict} file defines the number of steps a simulation will run for. For cases with constant inflow velocity (as studied here), it is important to run the simulation for a sufficient number of steps to ensure a converged solution is obtained. This number is typically dependent on the complexity of the flow field (e.g., it has been observed that with smaller rooms it takes a shorter number of steps to reach convergence).
In the study of the tutorial room and lecture room presented here, each simulation was run up to 50,000 steps. The ventilation field was analysed in 5000-step intervals to check for convergence, as explained later.
Meanwhile, the \textit{decomposeParDict} file is used to allow the simulation to be performed in parallel across multiple processors. This reduces the computing time required for a simulation to run. This is very useful especially in complex cases with multiple inlets and outlets.

A number of files (i.e., \textit{bmd\_write.py}, \textit{cpd\_write.py}, \textit{tsd\_write.py} and \textit{zero\_write.py}) were also written, which contain functions to create the OpenFOAM files needed for the simulation. Another Python file, \textit{start.py} takes the values defined in \textit{inparam.py} and converts them into `string' format, a necessary step to ensure the values are written in the form required by OpenFOAM. Any additional calculations required for the case setup are also performed in this file. Lastly, the \textit{start.py} file calls the various functions to create the OpenFOAM files. A summary of the function of each file is given in Table~\ref{tab:IV - Python File Function}.

The devised approach is user-friendly as only the values in the file \textit{inparam.py} need to be changed for specific room geometries and inflow conditions.
It should be noted that a graphical user interface (GUI) was also developed, in which the user can `draw' the geometry of the room, objects and ventilation ports and input the respective ventilation rates, etc. The GUI then directly feeds the required information into \textit{inparam.py}. 
The rest of the process occurs automatically and does not require any input from the user. This allows users to perform CFD simulations of any indoor environment with no occupants (or with occupants who are stationary for a period of time).

\subsubsection*{Computational fluid dynamics}

Once the case setup has been defined, the user simply needs to execute the script \emph{setup.sh} (e.g., in a Linux environment, the command \textit{./setup.sh} is used) to run the OpenFoam simulation. This script essentially chains a few shell commands together in a file for ease of use. Hence, when the terminal executes that file, it will execute all the specified commands in the script in the specified order. A brief summary of the procedure contained in \emph{setup.sh} is provided below.

The \textit{setup.sh} script first removes any previous mesh geometry. Then, it creates and moves the necessary files defining the current indoor environment to their respective directories by running the \textit{start.py} file. The \textit{blockMesh}, \textit{topoSet}, \textit{createPatch} and \textit{subsetMesh} commands are then executed to generate the mesh. The simulation is decomposed into multiple processors using the \textit{decomposePar} command. Finally, the OpenFOAM simulation is run in the background. A \emph{log} file is also generated to monitor the simulation. After the simulation is completed, \textit{setup.sh} runs the \textit{reconstructPar} command to reconstruct the ventilation field from the decomposed solution, since each individual processor solved a given part of the domain. The command \textit{simpleFoam --postProcess --func yPlus} could also be run to compute additional outputs to be exported with the solution (in this specific case the quantity $y^+$). The $y^+$ value, which represents the non-dimensional wall distance of the cells adjacent to a wall, is useful for the evaluation of the mesh quality and turbulence modelling analysis. However, it is not necessary for the computation of the dispersion of sprays and aerosols done in post-processing, hence this command can be commented out in fully-automated operation. Lastly, the solution is converted into VTK format for optional analysis and sanity checks performed by the user. Listing~\ref{lst:IV - setup.sh} shows all the commands included in the \textit{setup.sh} file.

\subsubsection*{Output}

The command \textit{pvpython paraview\_export.py} is automatically executed, which runs the script \textit{paraview\_export.py} converting the VTK file into .csv file format. The .csv file is then imported into the droplet dispersion post-processing code in Python (see Section~\ref{sec:dispersion}) to be run in `real time'.

\definecolor{backcolour}{rgb}{0.95,0.95,0.92}
\definecolor{codegreen}{rgb}{0,0.6,0}
\definecolor{codegray}{rgb}{0.5,0.5,0.5}
\definecolor{codepurple}{rgb}{0.58,0,0.82}
\definecolor{ballblue}{rgb}{0.13, 0.67, 0.8}

\begin{lstlisting}[
    language=sh, 
    basicstyle=\footnotesize,
    backgroundcolor=\color{backcolour},   
    commentstyle=\color{ballblue},
    morekeywords={mv,rm, mkdir, cd, echo},
    otherkeywords={>, -},
    keywordstyle=\color{magenta},
    stringstyle=\color{red},
    emph={fr, overwrite, patch, noFields, np, parallel, postProcess, func},
    emphstyle=\color{codepurple},
    basicstyle=\ttfamily\footnotesize,
    breakatwhitespace=false,         
    breaklines=true,                 
    captionpos=b,                    
    keepspaces=true,
    showspaces=false,                
    showstringspaces=false,
    showtabs=false,                  
    tabsize=2,
    caption = Contents of setup.sh file.,
    label={lst:IV - setup.sh},
    float
]
#!/bin/sh

rm -fr ../constant/polyMesh 
python start.py 
rm *.pyc 
mv U ../0/ 
mv epsilon ../0/ 
mv k ../0/ 
mv p ../0/ 
mv nut ../0/ 
mv omega ../0/ 
mkdir ../constant/polyMesh 
mv blockMeshDict ../constant/polyMesh/ 
mv topoSetDict ../system/ 
mv createPatchDict ../system/ 
cd ../ 
blockMesh 
topoSet 
createPatch -overwrite 
subsetMesh -overwrite table -patch walls -noFields 
decomposePar -force 
echo "Simulation in progress - This may take a few minutes"
mpirun -np 8 simpleFoam -parallel > log &  
reconstructPar
simpleFoam  -postProcess  -func  yPlus
foamToVTK
cd setup/
pvpython paraview_export.py
\end{lstlisting}

\subsection*{Assessment of the utilised methods}

\subsubsection*{Investigated geometries}

\begin{figure*}[t]
    \centering
    \includegraphics[width=\linewidth]{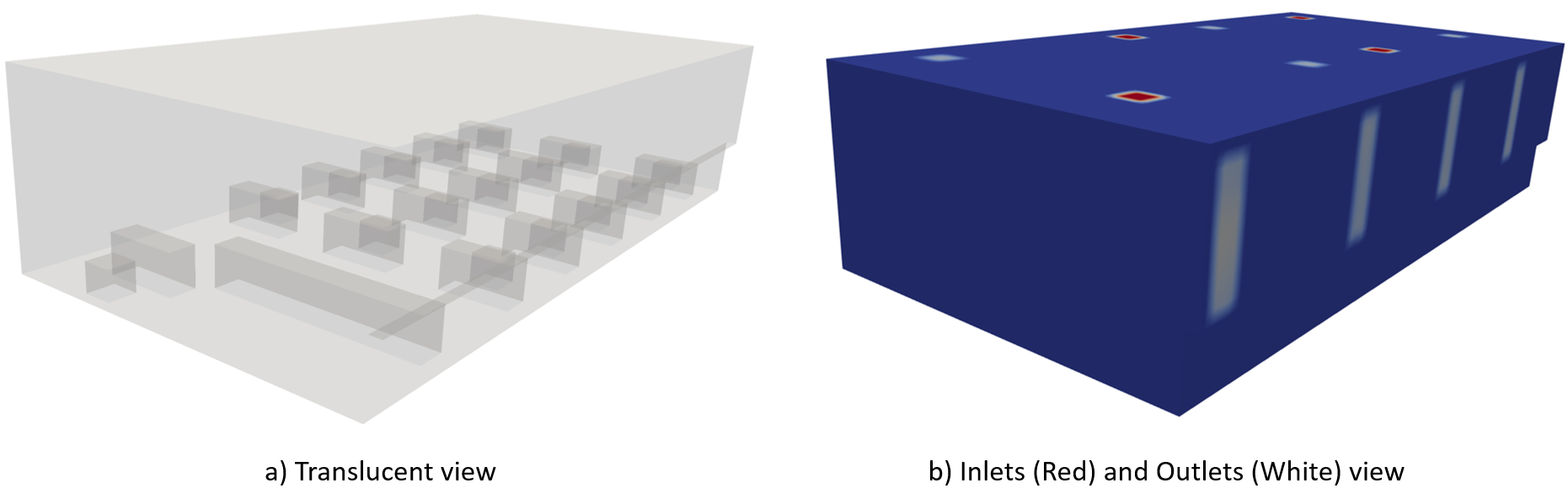}
    \caption{View of the tutorial room. The geometry reproduces the room CAGB 649/650 at Imperial College London.}
    \label{fig:IV - TR}
\end{figure*}

\begin{figure*}[t]
    \centering
    \includegraphics[width=\linewidth]{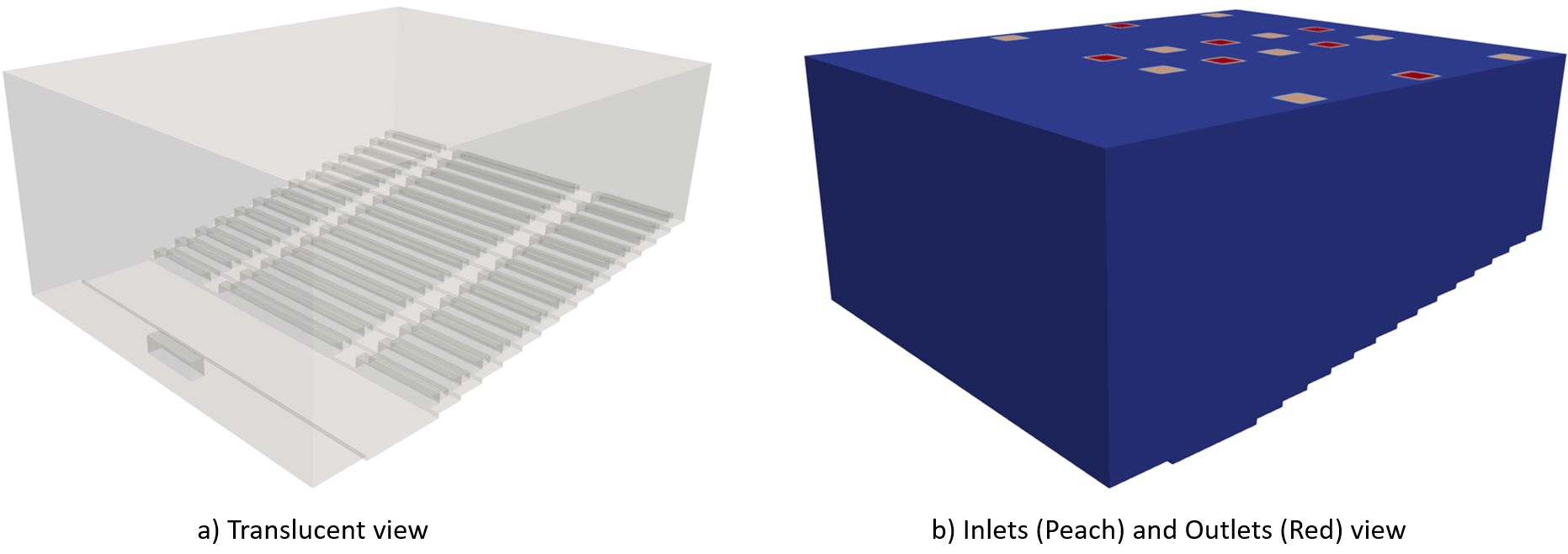}
    \caption{View of the lecture theatre. The geometry reproduces the lecture theatre in room CAGB 500 at Imperial College London.}
    \label{fig:IV - LR}
\end{figure*}

Two teaching rooms in the City and Guilds building (CAGB) at the South Kensington Campus, Imperial College London, have been investigated. Figure~\ref{fig:IV - TR} shows a tutorial room, comprising a lectern and a number of desks arranged in such a way that social distancing between students is established. This arrangement follows the guidelines developed by the College for the academic year 2020-2021. This is the first teaching room investigated here to assess the numerical tools. The second room simulated in this study is the lecture theatre shown in Figure~\ref{fig:IV - LR}. The room is characterised by a lectern and a series of rows consisting of desks and benches at different heights.

\subsubsection*{Ventilation pattern analysis}

In the analyses below, Paraview was used to visualise the solutions obtained from the automated CFD simulations of the tutorial room and lecture theatre. These analyses were necessary and crucial to ensure that the teaching rooms were correctly dimensioned, the inlet and outlet conditions were accurate, and to ensure the results are physically consistent. For example, as shown in the sections below, using the $k$-$\omega$ turbulence model resulted in a ventilation field of the tutorial room with velocities in excess of 30~m/s, which was judged not physically possible.

\subsubsection*{Mesh sensitivity analysis} \label{subsection: IV - Mesh Sensitivity Analysis}

Generally, in CFD simulations, the mesh discretisation is known to have a direct influence on the accuracy of the solution, with coarser meshes possibly yielding inaccurate results. For a sufficiently fine mesh, the ventilation flow field should converge towards the same solution if the number of mesh points is further increased. However, the computational time to run a simulation drastically increases with increasing mesh size. Hence, a compromise is needed between the computational cost and accuracy. To ensure that the ventilation field is sufficiently accurate and takes a reasonable amount of time to run, a mesh sensitivity analysis is performed here. The objective is to find the minimum mesh size that guarantees converged results, i.e.~results independent of the mesh size.

The tutorial room, with geometry as shown in Figure~\ref{fig:IV - TR}, was used to investigate the effect of the chosen mesh size on the ventilation field. The tutorial room has 4 inlets, each with an inflow velocity of 0.459~m/s. There are 4 outlets on the ceiling and an additional 4 outlets representing open windows on the left-hand side wall of the room. The room was configured to have an ACH of 6. The ventilation field was expected to develop some degree of turbulence and the turbulence model $k$-$\omega$ SST was used to model the turbulence effects. It should be noted that this investigation assumed that the mesh sensitivity analysis conducted is independent of the turbulence model. The effect of the turbulence model on the mesh size required to have mesh independence should be addressed in future work. The influence of turbulence models on the ventilation field will be discussed in the next section. 

\begin{table}[b]
\centering
\caption{Mesh sizes used for the sensitivity analysis.}
\label{tab:IV - Mesh}
\begin{tabular}{ccc}
\hline
\textbf{Mesh size} & \textbf{Number} & \textbf{Simulation runtime} \\
\textbf{[m]} & \textbf{of cells} & \textbf{[hours]} \\ \hline
0.1 & 267 832 & 3.1 \\ 
0.05 & 2 166 976 & 14.2 \\ \hline
\end{tabular}
\end{table}

The strategy used to build the mesh is based on the use of a uniform mesh (i.e., the geometry is discretised using tetrahedral cells of equal size). The alternative is to generate local refinements by means of e.g., the OpenFOAM `snappyHexMesh' utility, which was used in Section~\ref{sec:tPlume} to study the thermal plume generated by a person. When a uniform mesh is used, the maximum cell size should be chosen in such a way that all the geometry details can be properly reproduced. This means that the geometry should be simplified as much as possible to avoid the necessity of very small cells. The mesh generation process follows a `sculptor' approach. First, the mesh of a cuboid that consists of the outside room dimensions is created. Then, cells are removed to create the features of the room (e.g.,~cells defining parallelepipeds are removed to generate table surfaces, etc.). This process has been automated for the user by controlling OpenFOAM with dedicated scripts, as described above.

It should be noted that having a mesh that reproduces all the geometry features is not sufficient to have physically consistent results. The mesh (i.e., spatial discretisation) should also guarantee a proper spatial resolution to accurately solve all the flow features, especially when strong gradients are present in the flow. The mesh sensitivity analysis allows us to identify the maximum possible cell size that still maintains a mesh independent solution, which is independent of the mesh resolution. For this purpose, two mesh sizes were compared with uniform cell edge lengths of 0.1~m and 0.05~m, respectively. Each simulation was run on 8 processors in parallel up to 50,000 steps. The time taken to run each simulation and the total number of cells are reported in Table~\ref{tab:IV - Mesh}.

\begin{figure*}[t]
    \centering
    \includegraphics[width =\linewidth]{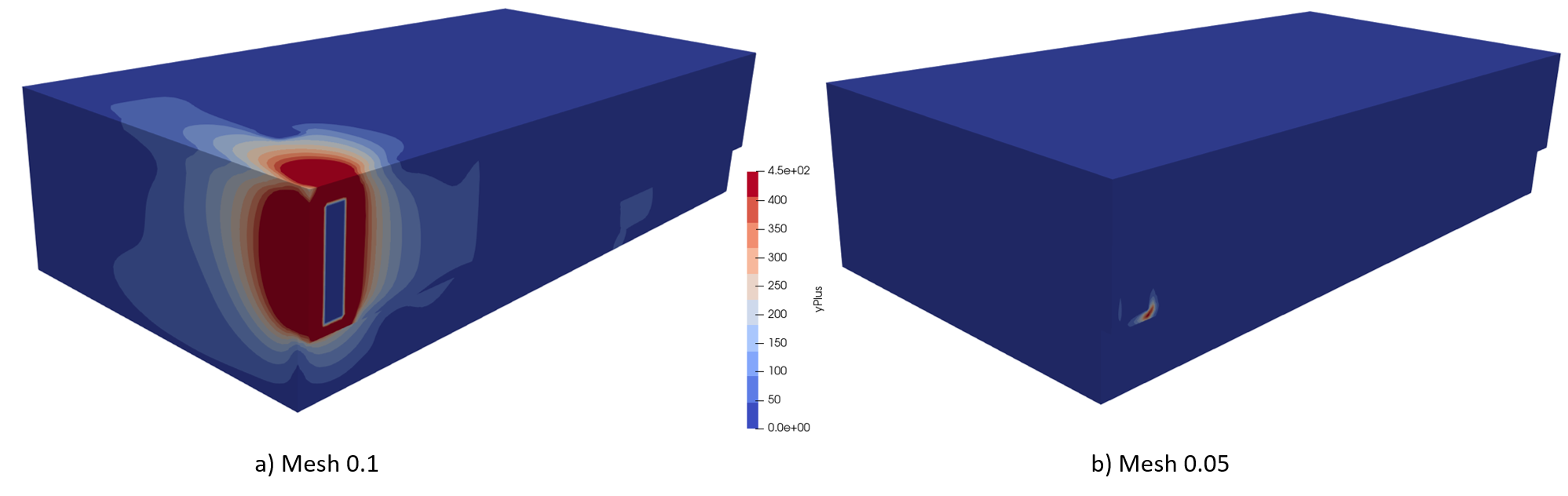}
    \caption{View of the $y^+$ values for two different mesh sizes after 50,000 steps.}
    \label{fig:IV - yPlus View}
\end{figure*}

\begin{figure*}[t]
    \centering
    \includegraphics[width =\linewidth]{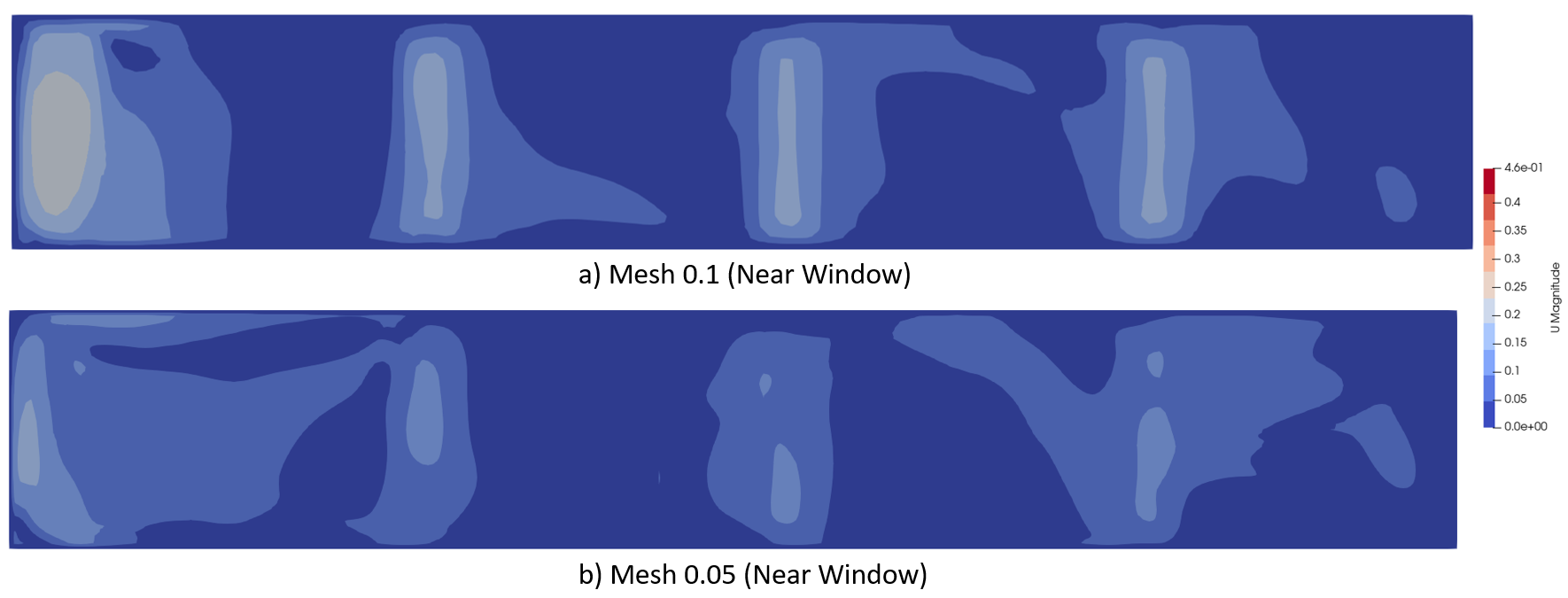}
    \caption{Velocity magnitude, $U$, in the vicinity of the wall containing the windows for two different mesh sizes after 50,000 steps.}
    \label{fig:IV - Near Window}
\end{figure*}

In CFD simulations, it is computationally expensive to resolve the velocity field near walls. The velocity gradients near the walls are steep, which would require an extremely fine mesh to properly resolve the flow. Hence, in these regions, a wall function is used, which allows us to use a mesh with lower resolution by directly imposing the solution given by the log-law of the wall. The dimensionless number, $y^+$, which represents the non-dimensional distance normal to the wall, is useful in determining whether the wall function has been applied correctly. The $y^+$ value should be sufficiently high that the first grid point is far from the wall to ensure the application of a wall function is valid (a value lower than 20 would place the first node in the laminar sub-layer where the log-law does not hold). However, the location of the first node must not be too far away, as this would lead to an incorrect resolution of the boundary layer (the log-law is not valid anymore for $y^+$ values greater than 200). A $y^+$ value of greater than 30 and lower than 200 was assumed to be of sufficient accuracy for the present analysis.

Figure~\ref{fig:IV - yPlus View} shows the $y^+$ field for the two different mesh sizes. With the coarser mesh, there is an extended region of high $y^+$ values around the left window. The window is located very close to one of the ventilation ports. It is therefore speculated that a possible reason for the high $y^+$ values in this region are the relatively high flow velocities induced by the presence of walls close to the ventilation port and the left window. The values of $y^+$ can be reduced in this region by applying the mesh with higher resolution (Figure~\ref{fig:IV - yPlus View}, right). It should be noted that when $y^+$ is outside of the range where the wall function is valid, then the prediction of the velocity field may also be directly affected. Figure~\ref{fig:IV - Near Window} shows the ventilation pattern near the windows. The velocity field predicted with the two meshes is different. The difference in velocity was estimated to be in the range of $\pm$0.05 m/s. The more refined mesh is recommended, however, it is important to note that it is difficult to give general recommendations on the mesh size since the mesh resolution necessary to properly resolve the flow field strongly depends of the features of the flow. More work should be carried out in the future to develop solid best practice guidelines. From Table~\ref{tab:IV - Mesh}, it is noted that decreasing the uniform cell edge length by half increased the number of cells by about a factor of 8 and the simulation took approximately 4 times longer to run. This made the finer mesh computationally expensive. However, since the computation of the flow field is done in pre-processing and only has to be conducted once per room geometry and ventilation pattern, this does not affect the `real-time' performance of the developed model.

\subsubsection*{Turbulence model analysis} \label{subsection: IV - Turbulence Model Analysis}

There are three common approaches to account for turbulence in the simulation of fluid flows: direct numerical simulation (DNS), large eddy simulation (LES), and Reynolds-averaged Navier-Stokes (RANS) simulations. The first two methods grant better accuracy at the cost of computational time. For the relatively simple case of rooms with steady inflows, the RANS method was chosen in the present work.

In the RANS method, flow variables (e.g.~velocity, pressure) are separated into mean and fluctuating components, which are then substituted into the original form of the Navier-Stokes equations. The final result is a set of equations for the time-averaged quantities, with additional terms to include the effect of turbulence on the mean flow. Such terms are known as the Reynolds stress terms. To close these equations, the Reynolds stresses must be expressed as a function of the mean quantities. This is done using so-called `turbulence models'. Every turbulence model has its own peculiarities. Since the large scales are typically not isotropic, different predictions of the flow field depending on the specific turbulence model used in the simulation may be expected.

An investigation was conducted into how three different types of turbulence models affect the predicted ventilation pattern in the tutorial room. The turbulence models chosen are the $k$--$\varepsilon$, $k$--$\omega$ and $k$--$\omega$ SST models. The conditions chosen for the tutorial room were similar to the ones used to investigate the mesh sensitivity. Each simulation used 8 processors, a mesh size with an edge cell length of 0.1~m, and the ventilation field was captured every 5,000 steps up to 50,000 steps. All three models required a similar amount of time for the simulation to complete, as shown in Table~\ref{tab:IV - Turbulence}.

\begin{table}[t]
\centering
\caption{Turbulence models assessed in this work.}
\label{tab:IV - Turbulence}
\begin{tabular}{cc}
\hline
\textbf{Turbulence} & \textbf{Simulation runtime} \\
\textbf{model} & \textbf{[hours]} \\ \hline
kEpsilon & 3.22 \\ 
kOmega & 3.04 \\ 
kOmegaSST & 3.1 \\ \hline                 
\end{tabular}
\end{table}

\begin{figure}[t]
    \centering
    \includegraphics[width=\linewidth]{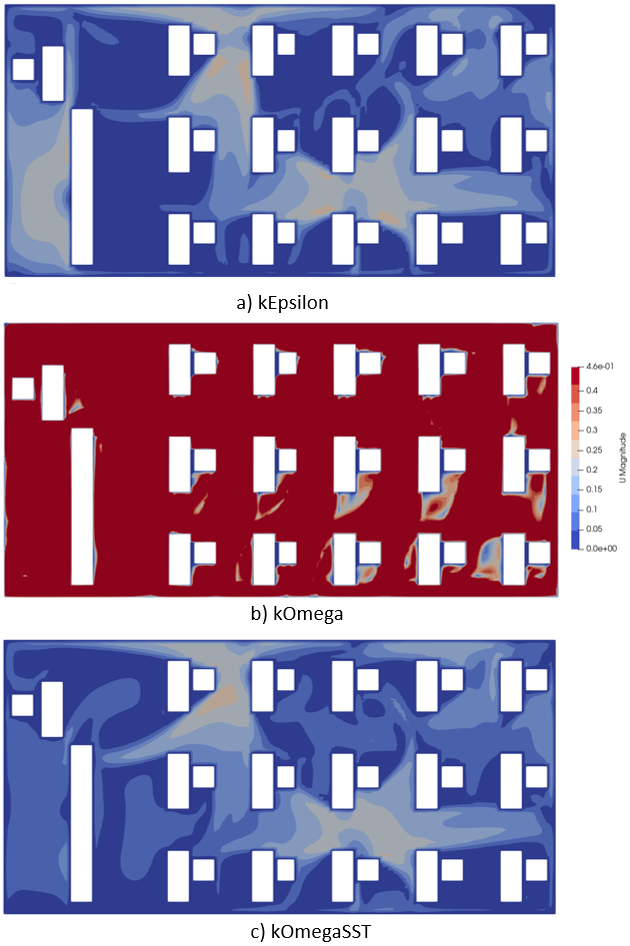}
    \caption{Velocity magnitude in a horizontal cross-section of the tutorial room at $z = 0.1$~m for three different turbulence models.}
    \label{fig:IV - Turbulence Floor Comparison}
\end{figure}

\begin{figure}[t]
    \centering
    \includegraphics[width=\linewidth]{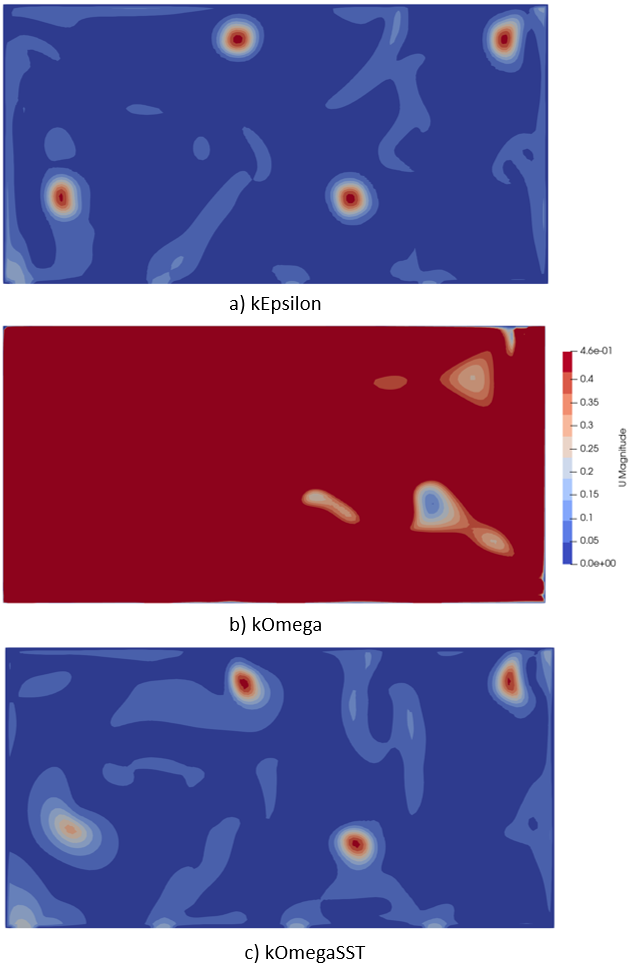}
    \caption{Velocity magnitude in a horizontal cross-section of the tutorial room at $z = 1.62$~m for three different turbulence models.}
    \label{fig:IV - Turbulence Midway Floor Comparison}
\end{figure}

The next step in this study was to compare the ventilation field predicted by each model. A horizontal slice was taken of the tutorial room at $z = 0.1$~m and $z = 1.62$~m for each model. Note that the $z$-axis represents the vertical direction of the tutorial room with a total height of $3.24$~m. Figures~\ref{fig:IV - Turbulence Floor Comparison} and \ref{fig:IV - Turbulence Midway Floor Comparison} show the magnitude of the velocity field for all three models at $z = 0.1$~m and $z = 1.62$~m, respectively after 50,000 steps.

From the results, it was clear that the $k$--$\omega$ model was not suitable for this application. The velocity field it produced was in excess of 30~m/s. This is considered not physically possible, given the configuration of the room. Note that there are more outlets than inlets and that the inlets have a constant inflow velocity of 0.459~m/s each, which means the velocity field is off by a factor of about 100. A possible explanation for this is $k$-$\omega$ being a turbulence model which is sensitive to the free-stream values of $\omega$ \cite{kOmega1,kOmega2}. In this simulation, the turbulent kinetic energy at the inlet was automatically calculated by an OpenFOAM function (type \textit{turbulentMixingLengthFrequencyInlet}) by imposing a turbulence intensity 0.05 and a turbulence length scale equal to 0.25~m. Further investigation is needed to determine the exact cause of the $k$--$\omega$ model not producing a physically valid velocity field.

The ventilation fields generated by the $k$--$\varepsilon$ and $k$--$\omega$ SST models are similar and the time taken for the simulations to complete showed a negligible difference. Based on Ref.~\cite{kEpsilon1}, the $k$--$\varepsilon$ model \cite{Jones1972} might struggle near boundaries and under strong adverse pressure gradients. This is not a problem for the $k$--$\omega$ SST model as it uses a blending function to combine the $k$--$\varepsilon$ and $k$--$\omega$ models, depending on the distance of the nearest grid point from the wall. Therefore, it was concluded that the $k$--$\omega$ SST model was a safer choice to model ventilation flows, especially as this workflow is intended to be used for other rooms, which may have adverse pressure gradients. 

\subsubsection*{Results} \label{subsection: IV - Results}

\begin{figure}[t]
    \centering
    \includegraphics[width=\linewidth]{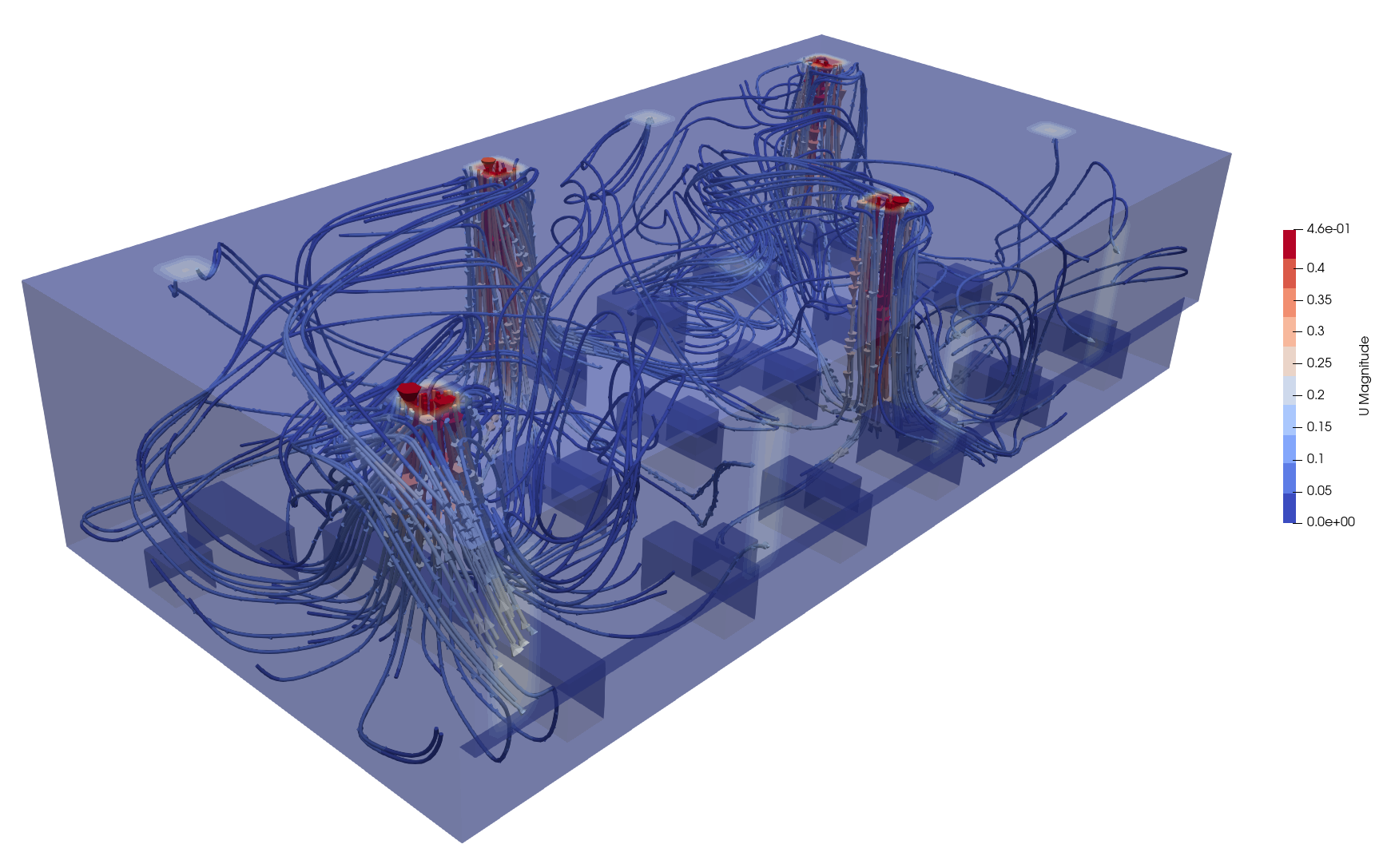}
    \caption{Flow streamlines inside the tutorial room.}
    \label{fig:IV - TR Streamlines}
\end{figure}

\begin{figure}[t]
    \centering
    \includegraphics[width=\linewidth]{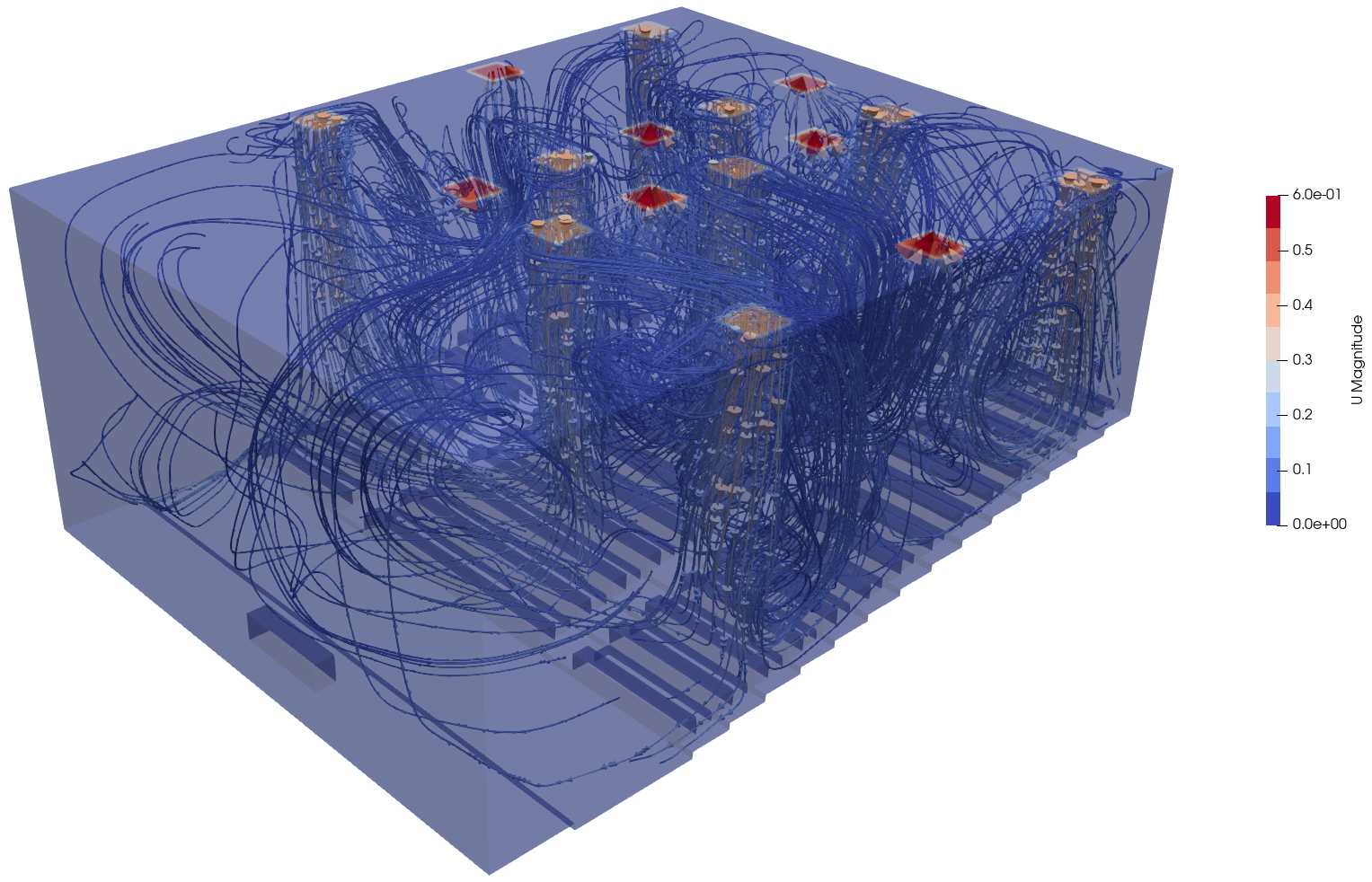}
    \caption{Flow streamlines inside the lecture theatre.}
    \label{fig:IV - LR Streamlines}
\end{figure}

This section presents a short summary of the results of the CFD simulation of ventilation in indoor environments, specifically, in a tutorial room and a lecture theatre. The streamline function in Paraview was used to visualise how the velocity vectors would flow from the inlets. Figures \ref{fig:IV - TR Streamlines} and \ref{fig:IV - LR Streamlines} show the streamlines in the tutorial room and lecture theatre, respectively.

The velocity vectors have the highest magnitudes near the inlets. As the flow penetrates further into the room, the velocity magnitudes decrease. The velocity is relatively low in the vicinity of the floor and at heights comparable with the location of the head of a seated student. Hence, this emphasises why the background ventilation simulation needs to be combined with additional models in post-processing to account for the effects of thermal plumes and breath jets produced by the occupants of a room, which could be quite significant considering the relatively low velocity magnitudes of the ventilation field.

Interestingly, from Figure~\ref{fig:IV - LR Streamlines} it becomes clear that the inlets of the lecture theatre have a lower velocity magnitude compared to the outlets despite the absence of any open windows or doors. This outcome is possible as there are more inlets compared to outlets (9 to 7) in the room. It is suggested that further investigation is performed to understand any impact this could potentially have on the ventilation pattern. In general, analysis of the flow distribution between inlets and outlets, as well as their placement in the room, coupled with an estimate of the infection risk, could make the developed framework an important tool to assist the design of ventilated indoor environments.

\clearpage
    
\section{Droplet dispersion under thermal plumes}\label{sec:combo_plumes}

\vspace{3mm}

\noindent By \emph{A.M. Akbar, A. Giorgallis, A. Kruse, N. Liniger, \\ 
L. Papachristodoulou, A. Giusti and D. Fredrich}

\vspace{3mm}

\noindent This section describes the results obtained by combining the circular jet model derived in Section~\ref{sec:Breath}, for the velocity profile due to breathing, with the flow field obtained by simulating the thermal plume from a body in Section~\ref{sec:tPlume}, and applying them to the study of saliva droplet dispersion in the single person occupation of a small room using the methods discussed in Section~\ref{sec:dispersion}. The flow chart in Figure~\ref{fig:PostProcFlow} indicates the steps taken to combine Sections~\ref{sec:dispersion}, \ref{sec:tPlume} and \ref{sec:Breath}. The main aim is to study the effect of the thermal plume on saliva droplet dispersion. This section also provides an example of the application of the full model to a realistic case study.

It should be noted that the pre-processing and post-processing tools are coupled in a way that allows the user to execute the workflow described in Figure~\ref{fig:PostProcFlow} by simply running the developed Python code. The only input that needs to be manually entered are the approximate locations of the occupants' mouths within the room.

\subsection*{Post-processing}

\subsubsection*{Mesh initialisation from CFD}

The spray dispersion simulation is based on a Lagrangian tracking method utilising numerical parcels (see Section~\ref{sec:dispersion}). Since in the computation of the dispersion of droplets the gas-phase is assumed to remain unaffected by the droplet motion, the number of grid nodes used to import (interpolate) the CFD solution into the post-processing tool does not significantly affect the computational cost (an increase in the number of grid nodes mainly affects the amount of RAM necessary to store the variables). This also means that, in general, a very refined grid can be used without affecting the `real-time' performance of the model. Note, however, that the computational time required for the pre-processed CFD simulations (see Section~\ref{sec:tPlume}) is strongly affected by the mesh size. Grid independence studies should therefore be performed (see Section~\ref{sec:cfd}) and the number of cells must be kept at a minimum.

Objects within a room will have a significant effect on the velocity field. Hence, it was important to run the CFD with the main obstacles (in this case the person) included in the domain. To be able to use the results from the CFD for post-processing, the velocity field was interpolated onto a finer hexahedral mesh of rectangular shape (based on the cuboid describing the outer boundaries of the room). Objects within the domain in the CFD simulation were also transferred to the post-processor by deactivating individual nodes of the grid located outside of the fluid domain. It is therefore possible for any objects within the domain of the CFD to be transferred to the post-processor, such as human bodies, tables, chairs, etc. To interpolate the grid, a nearest interpolation method was used, as this is the fastest and most computationally efficient method. However, if for some reason the accuracy needs to be improved, the method can be changed by simply specifying a linear interpolation scheme. In Python, the \verb|scipy.interpolate.griddata()| function was used. The interpolated quantities are the gas-phase velocity, $\vec{U}_g$, the turbulent kinetic energy, $k$, and the turbulent energy dissipation rate, $\varepsilon$. 

\usetikzlibrary{shapes.geometric, arrows, decorations.markings}
\tikzstyle{process} =  [rectangle, rounded corners, semithick, minimum width = 6 cm, minimum height= 1 cm, text centered, text width = 6cm, draw = black, fill = white, line width = 1.4pt]
\tikzstyle{arrow} = [thick, decoration={markings,mark=at position
   1 with {\arrow[semithick]{open triangle 60}}},
   double distance=1.4pt, shorten >= 5.5pt,
   preaction = {decorate},
   postaction = {draw,line width=1.4pt, white,shorten >= 4.5pt}]
\tikzstyle{method} = [semithick, text centered, text width = 5cm,  fill = white]

\begin{figure}[t]
    \centering
    \begin{tikzpicture}[node distance = 3cm, thick]
    \node (Data From CFD) [process] {Data from CFD containing: $k$, $\varepsilon$, $\vec{U}_g$ and grid node coordinates};
    \node (Interpolation) [method, below of =Data From CFD, yshift = 1.5cm , xshift = +2.8 cm ] {Interpolating the CFD data onto the mesh grid in the Python environment using a nearest neighbour method from SciPy library};
    \node (Python Environment) [process, below of = Data From CFD] {Interpolated velocity field in Python environment};
    \node (Superposition) [method, below of =Python Environment, yshift = 1.5cm , xshift = +2.8 cm ] {Superimposing the model for the circular breath jet onto the mesh grid in Python};
    \node (Breath) [process, below of = Python Environment] {Velocity field modified in vicinity of the mouth};
    \node (Dispersion) [process , below of = Breath, yshift = 1.5cm] {Initialising the dispersion model at the location of the mouth};
    \node (NonDim) [method, below of = Dispersion, yshift = 1.5cm , xshift = +2.8 cm ] {Non-dimensionalisation based on the age of each individual droplet to generate a probability density function};
    \node (Result) [process , below of = Dispersion] {Non-dimensionalised dispersion in the form of a probability density function};
    \draw[arrow](Data From CFD) -- (Python Environment);
    \draw[arrow](Python Environment) -- (Breath);
    \draw[arrow](Breath) -- (Dispersion);
    \draw[arrow](Dispersion) -- (Result);
    \end{tikzpicture}
    \caption{Flow chart of the steps taken to combine Sections~\ref{sec:dispersion}, \ref{sec:tPlume} and \ref{sec:Breath}.}\label{fig:PostProcFlow}
\end{figure}
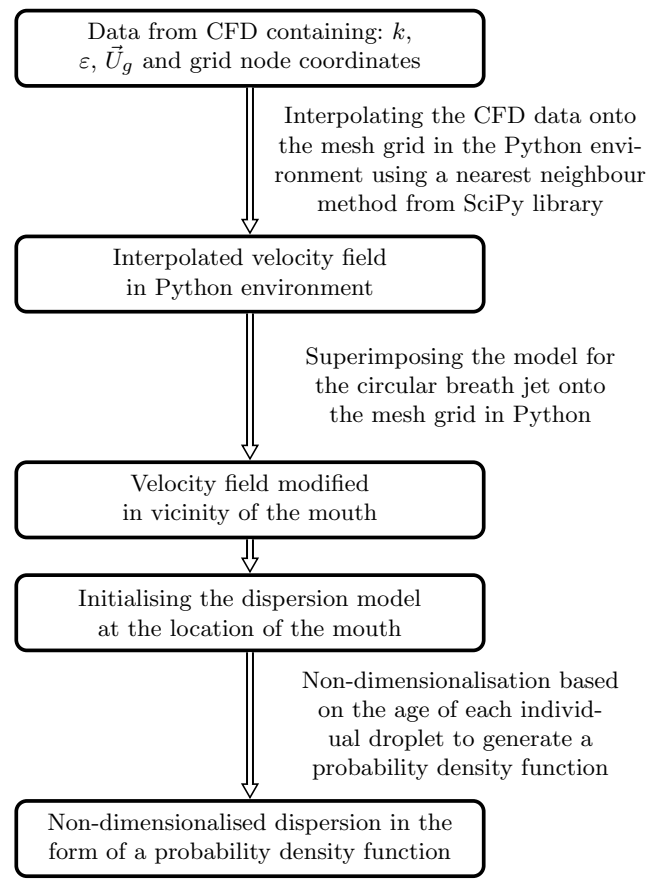

\subsubsection*{Breath jet}

\begin{figure}[t]
    \centering
    \includegraphics[width=\linewidth]{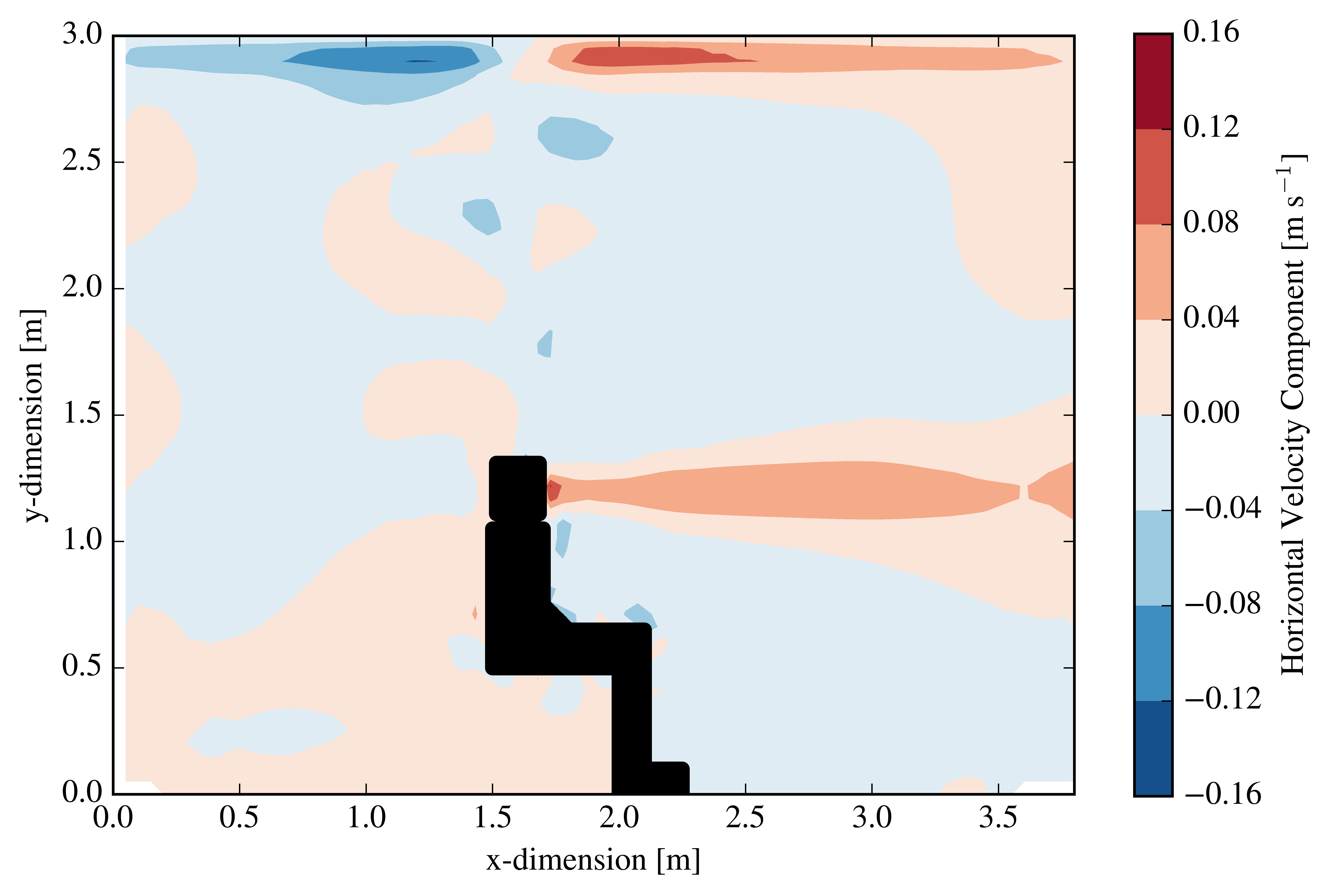}
    \caption{Contour plot of the modified horizontal component of the velocity field in a vertical cross-section. The background flow field is extracted from the CFD simulation of Case~1 in Table~\ref{table:person_sim}.}
    \label{fig:VelProf}
\end{figure}

The Gaussian model for the circular jet velocity profile described in Section~\ref{sec:Breath} was superimposed and added to the velocity field extracted from the CFD performed in Section~\ref{sec:tPlume}. It is important to note that this step is not theoretically correct, as it involves superimposing velocity fields over each other, which are described by non-linear equations (i.e.,~linear superimposition of effects is not valid). Additionally, the turbulence parameters in the region of the circular breath jet were not altered, i.e.~$\varepsilon$ and $k$ were kept the same. Furthermore, only the horizontal velocity component was altered, as the direction of injection was chosen to be fully horizontal. The superposition of the jet was done close to the area where a mouth would be located on the person in the simulation. The resulting velocity field is shown in Figure~\ref{fig:VelProf} for Case~1 in Table~\ref{table:person_sim}.

\subsubsection*{Multi-threading the dispersion simulation}

With large numbers of parcels, the simulation can become computationally heavy and therefore relatively slow. In order to speed-up the computational time, the simulation can be multi-threaded by starting the simulation in multiple threads and at the end adding the tracking mesh together. This is possible as the model does not account for interactions between individual droplets, which makes it possible to superimpose the spatial probability distributions of droplets (described in the next section) computed by each thread.
To make the computation as statistically relevant as possible, a total of 7,000 parcels were injected into each of the scenarios. This number is somewhat arbitrarily chosen, however after non-dimensionalisation, 7,000 parcels proved to be sufficient for most of the simulations to show statistical relevance, i.e. give a full probability distribution. Further analysis on the sensitivity of the results to the number of parcels should be conducted in the future. This will also provide guidelines for the set-up of dispersion computations in e.g., teaching rooms. 

\subsubsection*{Non-dimensionalisation}

In order to track the position of droplets over time, a tracking mesh is implemented in the simulation. At every time step, the location of each parcel is recorded and the parcel count at the closest tracking mesh node is incremented by one. Hence, every node in the mesh records how many times a parcel passes by. For clarification, the tracking arrays are projections of all the parcel paths onto a plane (indicated below with the indices $i$,~$j$). The floor deposition refers to the parcels that came into contact with the floor of the room. 

In a further step, this tracking mesh is non-dimensionalised to enable a more intuitive comparison between the different ventilation patterns. The non-dimensionalisation is based on the total number of time steps during the lifetime of a parcel, i.e.~the time a parcel moves within the simulation until it touches any object within the room (parcel age). In order to calculate the probability density function, as parcels may disperse at different time periods, a tracking array, $T_{i,j,k}$, was created for each parcel, $k$. The age of each injected parcel, $t_{l,k}$, is recorded, and is finally used to non-dimensionalise the tracking array, $T_{i,j,k}^*$:
\begin{equation}
    \label{eq:NonDimensionalisation}
    T_{i,j,k}^* = T_{i,j,k} \: \frac{\mathrm{d}t}{t_{l,k}} .
\end{equation}
This gives the probability of each parcel to be in a given location, conditioned on being suspended in the air. After each individual path of the parcels is non-dimensionalised, the overall probability of a droplet being suspended at a given location is found by first multiplying Equation~\ref{eq:NonDimensionalisation} by the probability of a given parcel to be suspended, i.e., $t_{l,k}/\mathrm{max}\left(t_l\right)$. Then, the values of all the paths are summed and divided by the number of injected parcels, $N$. In that way, the non-dimensional tracking mesh, $T_{i,j}^*$, can be obtained as:
\begin{equation}
    \label{eq:sumnondim}
    T_{i,j}^* = \frac{1}{N}\sum\limits_{k=0}^N T_{i,j,k}^* \frac{t_{l,k}}{\mathrm{max}\left(t_l\right)}.
\end{equation}
The resulting values may be expressed per unit volume of air. As far as the deposition on surfaces is concerned, the counting per parcel is multiplied by $\left(\mathrm{max}\left(t_l\right) - t_{l,k}\right)/\mathrm{d}t$ before proceeding with the evaluation of the related non-dimensional matrix. As the simulation was multi-threaded, the previously non-dimensionalised tracking arrays were summed up (from each thread) and divided by the total number of threads.

\begin{figure}[t]
  \centering
  \includegraphics[width=1.0\linewidth]{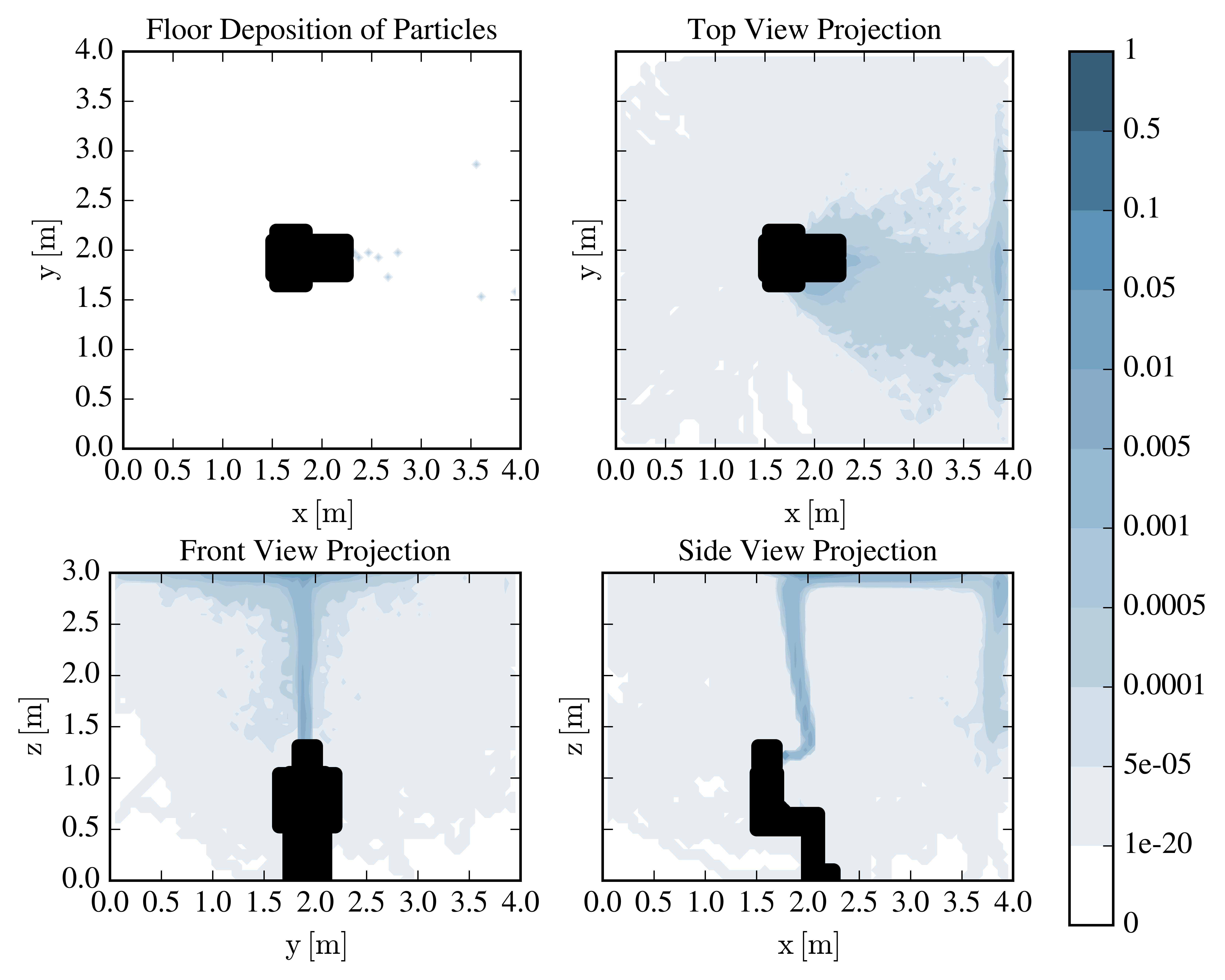}
  \caption{Probability distribution of droplet dispersion with no ventilation and a heated body -- Case~1 in Table~\ref{table:person_sim}.}
  \label{fig:S1}
\end{figure}

\subsection*{Results and conclusions}
\begin{figure}[t]
  \centering
  
  \begin{subfigure}{1.0\linewidth}
  	\includegraphics[width=1.0\linewidth]{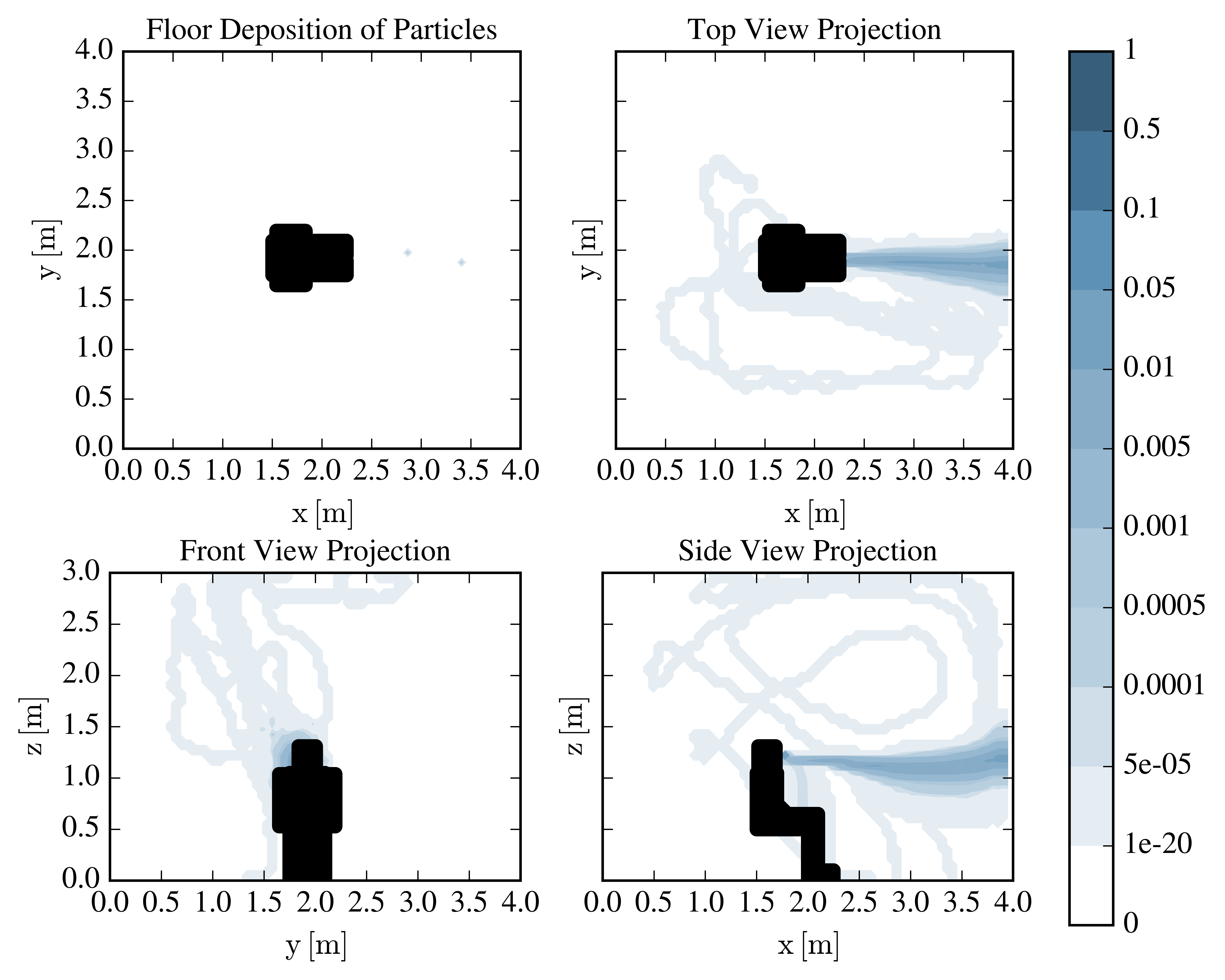}
	\caption{Unheated}
  \end{subfigure}
  
    \begin{subfigure}{1.0\linewidth}
  	\includegraphics[width=1.0\linewidth]{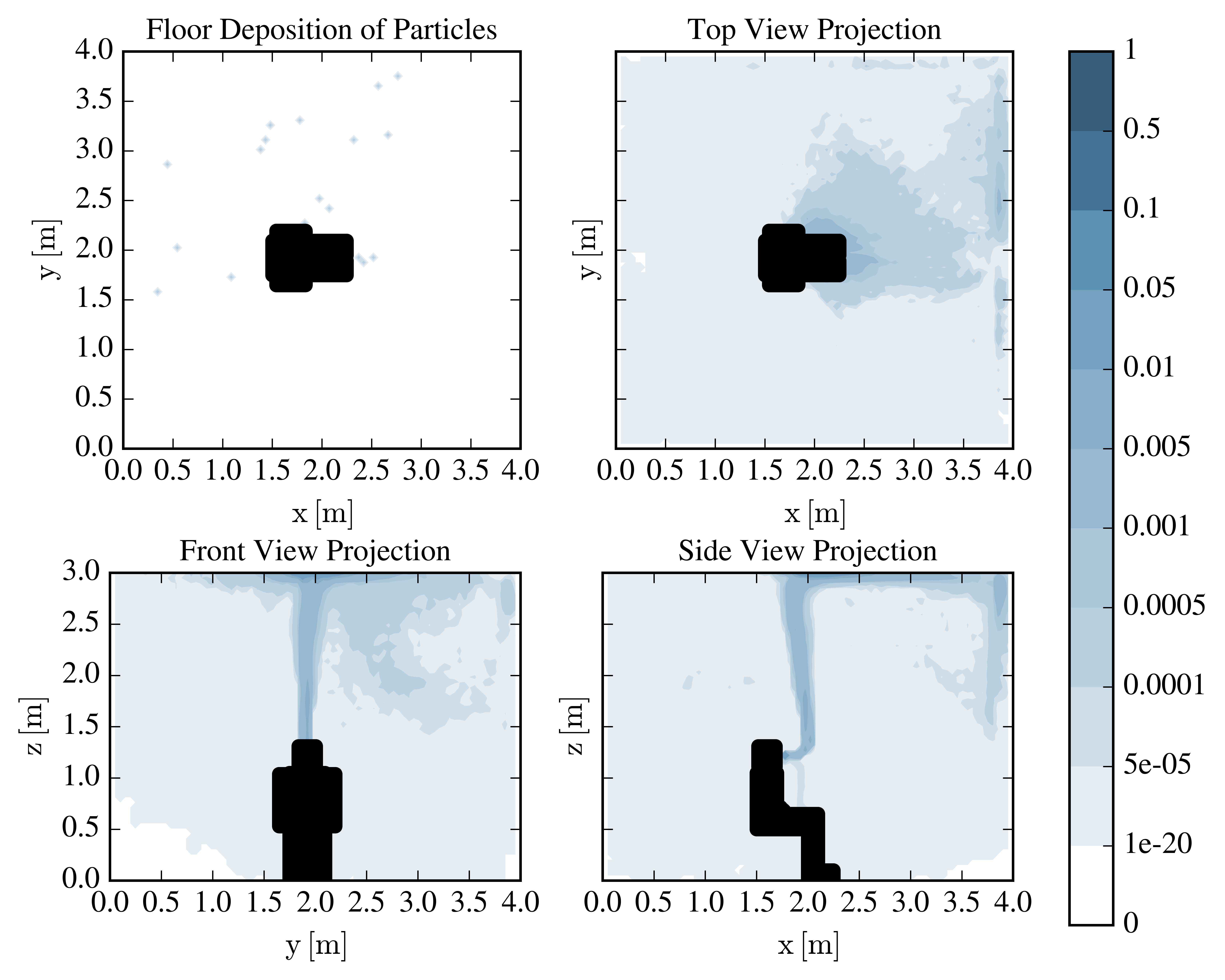}
	\caption{Heated}
  \end{subfigure}
  
  \caption{Probability distribution of droplet dispersion with 1 volume change per hour, $35^{\circ}$ inflow angle and a (a) unheated and (b) heated body -- Cases~2 and 3 in Table~\ref{table:person_sim}.}
  \label{fig:S2S3}
\end{figure}

\begin{figure}[t]
  \centering
  
  \begin{subfigure}{1.0\linewidth}
  	\includegraphics[width=1.0\linewidth]{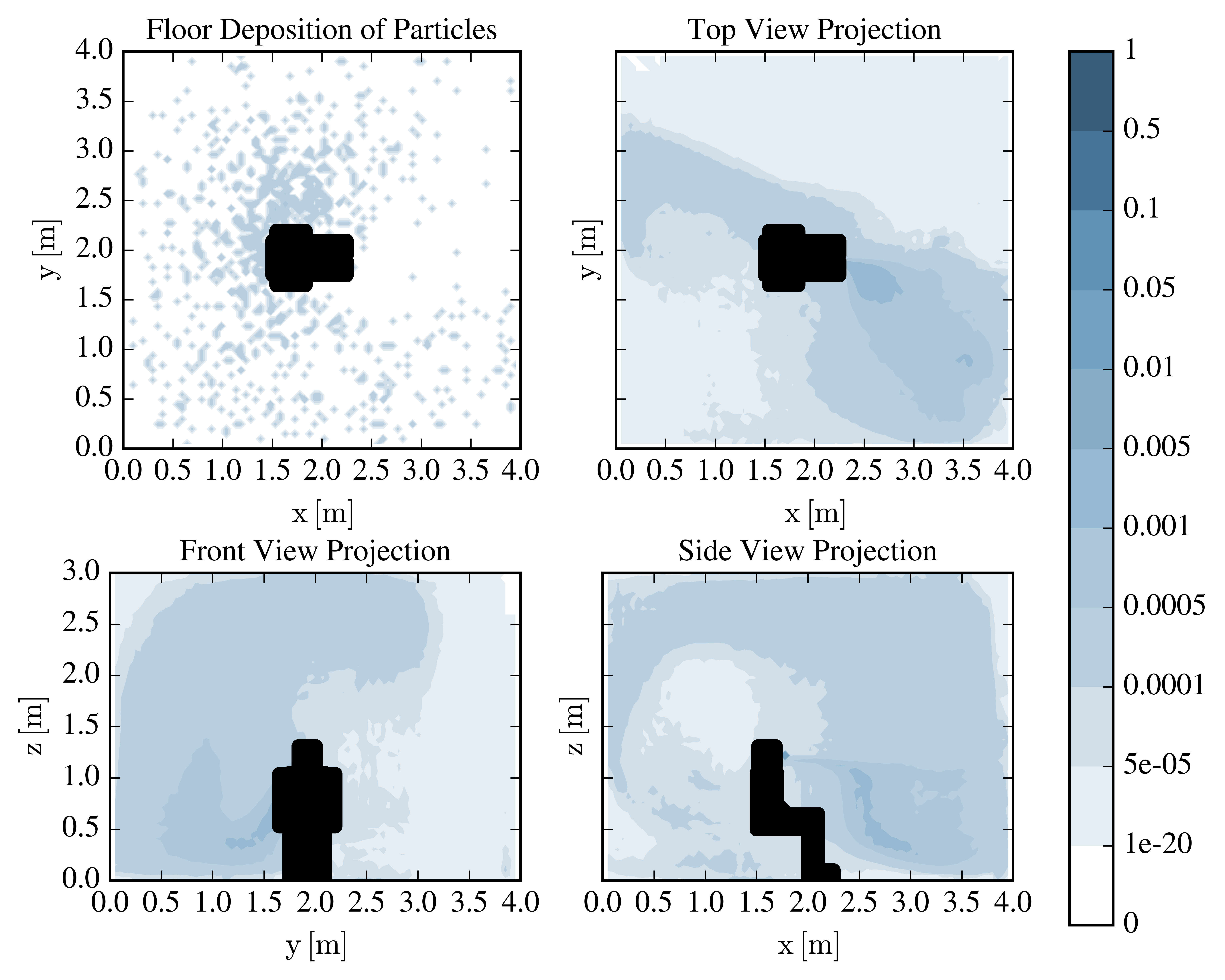}
	\caption{Unheated}
  \end{subfigure}
  
    \begin{subfigure}{1.0\linewidth}
  	\includegraphics[width=1.0\linewidth]{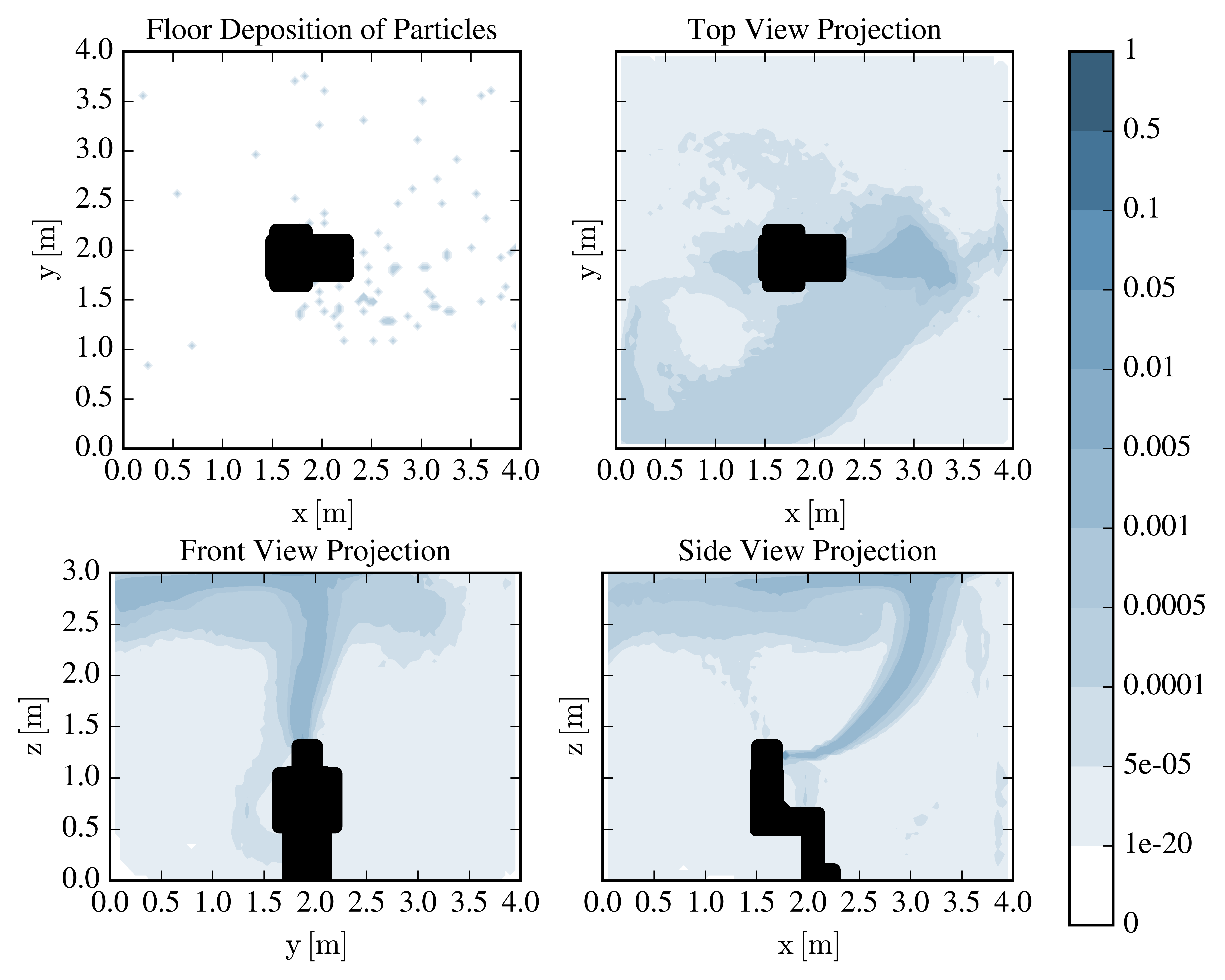}
	\caption{Heated}
  \end{subfigure}
  
  \caption{Probability distribution of droplet dispersion with 5 volume changes per hour, $35^{\circ}$ inflow angle and a (a) unheated and (b) heated body -- Cases~4 and 5 in Table~\ref{table:person_sim}.}
  \label{fig:S4S5}
\end{figure}

\begin{figure}[t]
  \centering
  
  \begin{subfigure}{1.0\linewidth}
  	\includegraphics[width=1.0\linewidth]{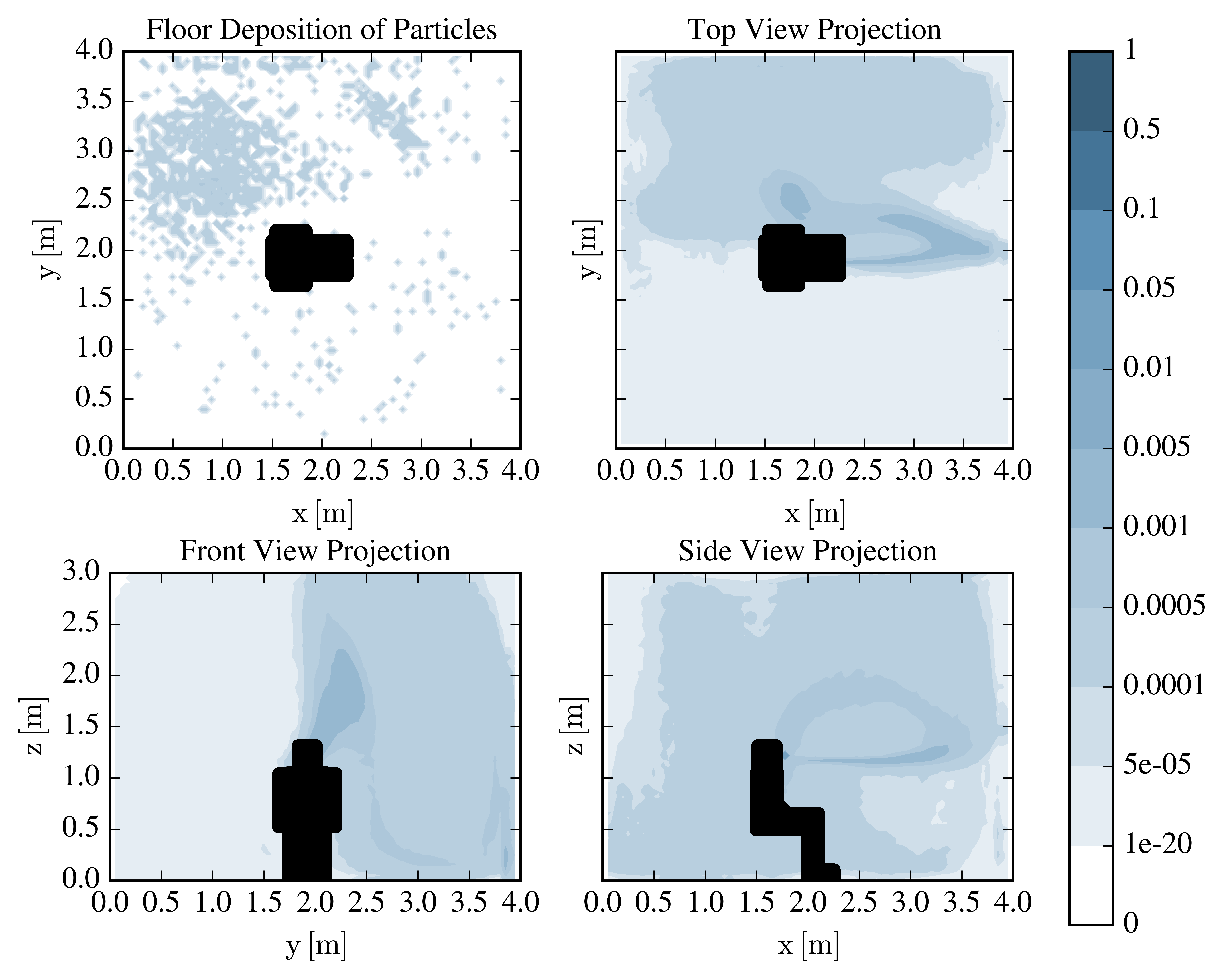}
	\caption{Unheated}
  \end{subfigure}
  
    \begin{subfigure}{1.0\linewidth}
  	\includegraphics[width=1.0\linewidth]{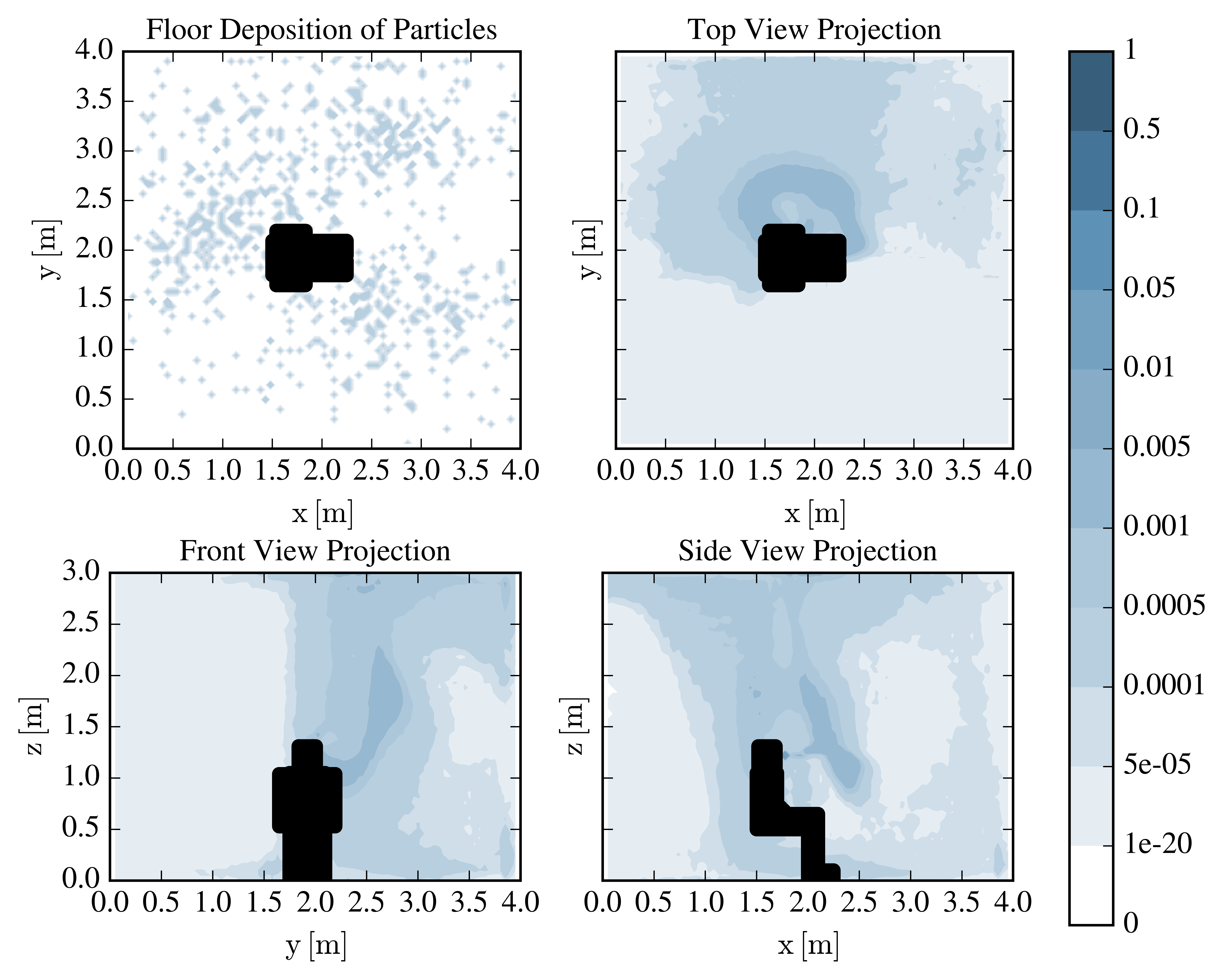}
	\caption{Heated}
  \end{subfigure}
  
  \caption{Probability distribution of droplet dispersion with 5 volume changes per hour, $20^{\circ}$ inflow angle and a (a) unheated and (b) heated body -- Cases~6 and 7 in Table~\ref{table:person_sim}.}
  \label{fig:S6S7}
\end{figure}

\begin{figure}[t]
  \centering
  
  \begin{subfigure}{1.0\linewidth}
  	\includegraphics[width=1.0\linewidth]{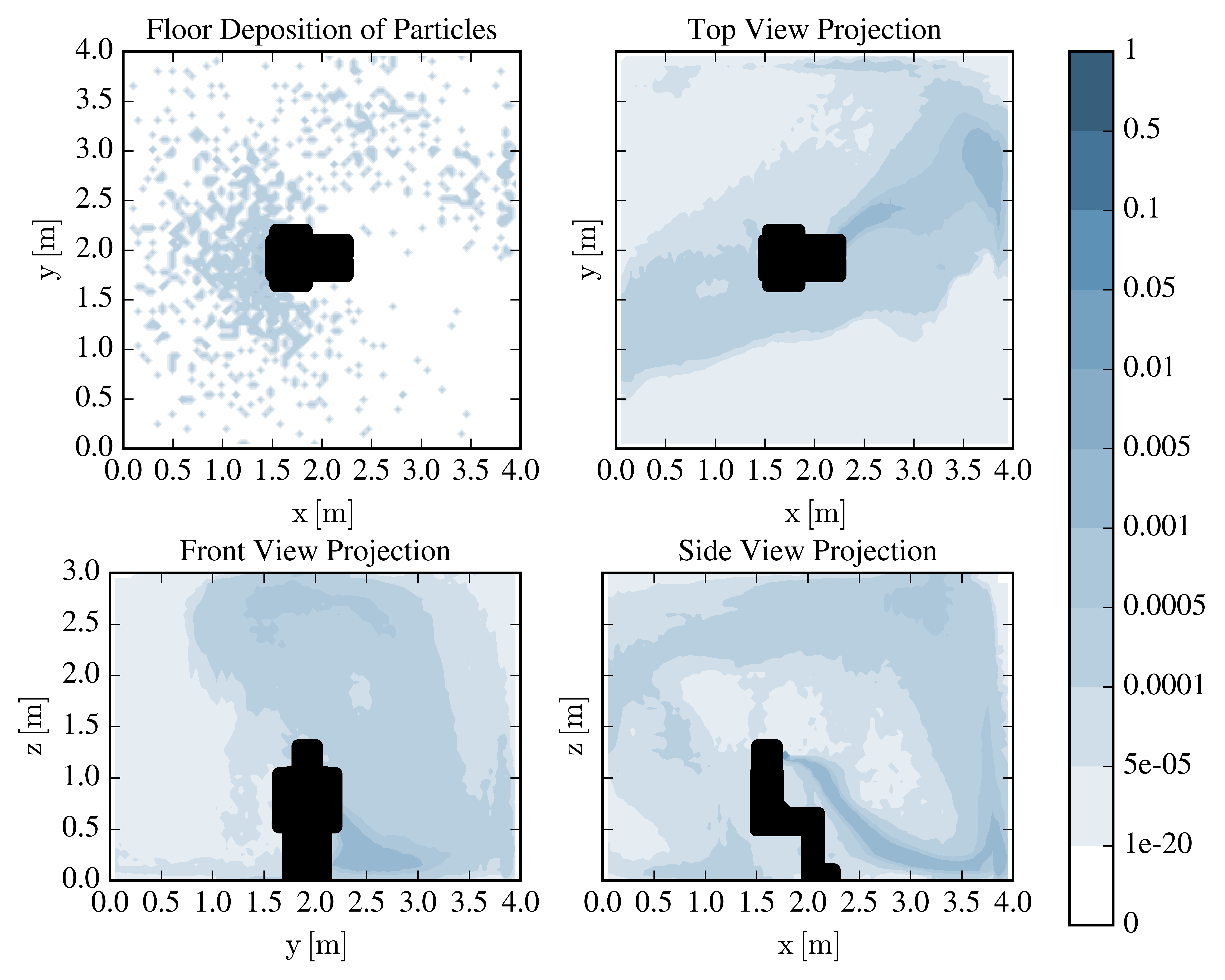}
	\caption{Unheated}
  \end{subfigure}
  
    \begin{subfigure}{1.0\linewidth}
  	\includegraphics[width=1.0\linewidth]{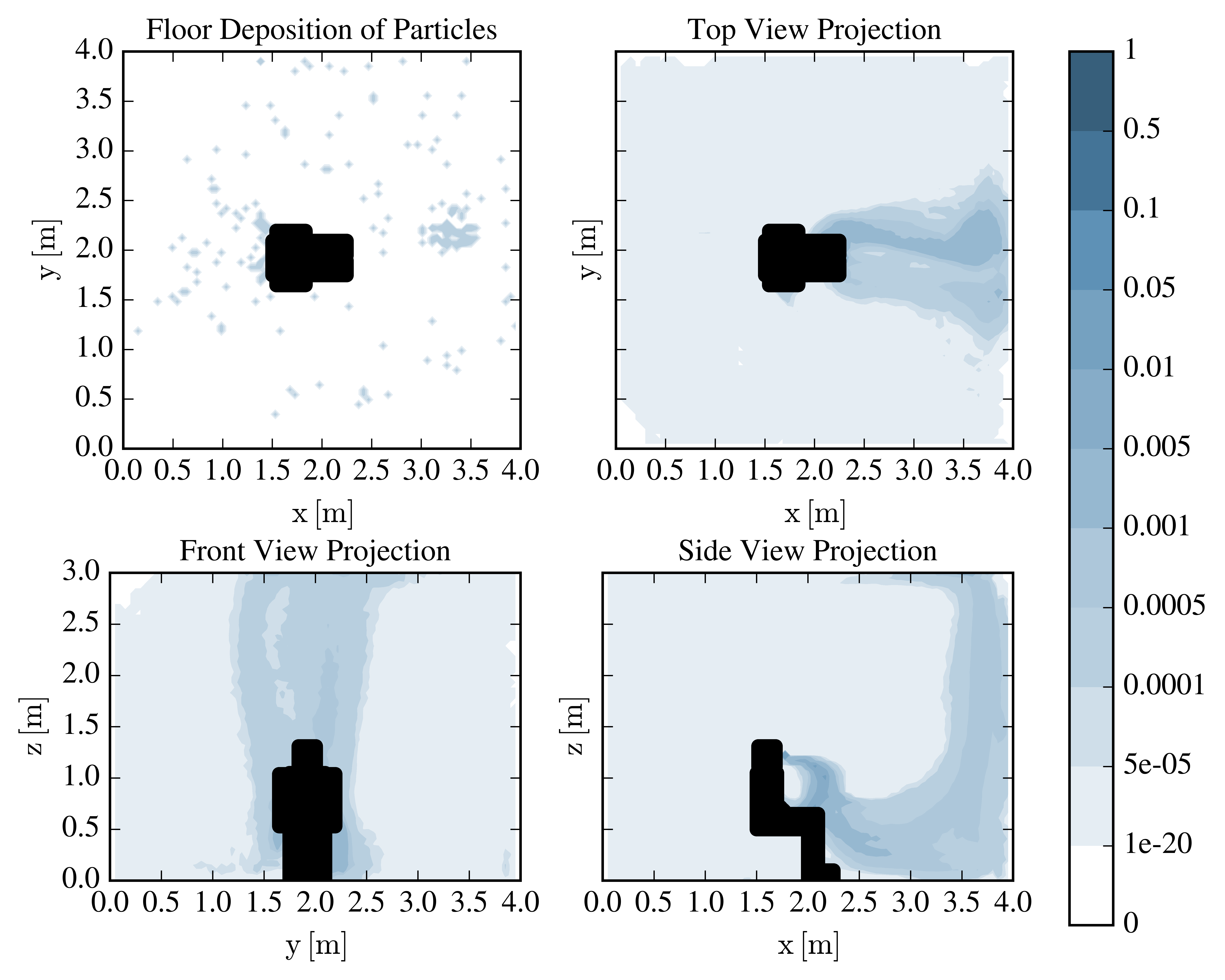}
	\caption{Heated}
  \end{subfigure}
  
  \caption{Probability distribution of droplet dispersion with 12 volume changes per hour, $35^{\circ}$ inflow angle and a (a) unheated and (b) heated body -- Cases~8 and 9 in Table~\ref{table:person_sim}.}
  \label{fig:S8S9}
\end{figure}

The results of the conducted simulations (the different cases are summarised in Section~\ref{sec:tPlume}, Table~\ref{table:person_sim}) are shown in Figures~\ref{fig:S1} to~\ref{fig:S8S9}. It is important to note that the floor deposition results are a non-dimensionalised plot of droplets that have come into contact with the floor. However, in none of the simulations is this fully statistically representative of reality, and should thus be used with caution and only as an indication of the surfaces that e.g., need to be cleaned more thoroughly.

The first simulation, displayed in Figure~\ref{fig:S1}, was conducted to visualise the effect of the thermal plume induced by the body with no ventilation flow. The results show that the droplets rise upwards and evidently follow the shape of the thermal plume. There are very few droplets from the simulation that come into contact with the ground (seen in the top left corner of \mbox{Figure~\ref{fig:S1}}). This indicates that the thermal plume is causing the droplets to stay suspended in the air for longer time periods if there is no ventilation.

Simulations 2 and 3, displayed in \mbox{Figure~\ref{fig:S2S3}}, were set up using a small ventilation rate of 1 volume change per hour. For the unheated body simulation, shown in Figure~\ref{fig:S2S3}~(a), most of the droplets were blown into the wall by the circular breath jet. This may be due to the limitations of the Gaussian circular jet model (Section~\ref{sec:Breath}) used to model the breath velocity as there is insufficient turbulence for the droplets to escape the jet. Therefore, this simulation suggests that the modification of the turbulent flow field due to breathing should be carefully addressed in future work, also considering the pulsed nature of the breath puffs. Note that the results of this particular simulation may not be statistically representative outside of the main circular breath jet, as the number of droplets outside the jet was too low to give statistically converged results.

In contrast, the simulation with the heated body shown in Figure~\ref{fig:S2S3}~(b) provided a much more statistically representative result, where the effect of the thermal plume on the dispersion of droplets is evident. Droplets rose upwards and dispersed, very similarly to the no ventilation simulation in Figure~\ref{fig:S1}. Hence, it can be concluded that for poorly ventilated environments, the human thermal plume dominates the dispersion of droplets, whereas the effect of the ventilation is less significant.

Simulations 4 and 5, displayed in Figure~\ref{fig:S4S5}, comprise the ventilation of 5 volume changes per hour with an inflow angle of $35^\circ{}$ from the negative vertical direction. This ventilation pattern seems to cause the droplets to follow a circular path, similar to that of a large eddy. Evidently, there is a large deposition on the floor in the unheated simulation. The velocity profile transports droplets to the floor behind the person, while in front of the person the ventilation exit causes droplets to rise upwards and later travel close along the ceiling. In general, however, there are many droplets located close to the ground, as can be seen in the Front View Projection.

In contrast, in the heated simulation, shown in Figure~\ref{fig:S4S5}~(b), most of the droplets move upwards and spend most of their lifetime close to the ceiling of the room. There is a much smaller deposition on the floor when compared to the unheated simulation. Even in these moderately ventilated conditions, the effect of a thermal plume on the dispersion of exhaled droplets is not negligible.

The only change in simulations 6 and 7, shown in Figure~\ref{fig:S6S7}, compared to the previous simulations 4 and 5, is the angle at which the ventilation stream flows into the room, i.e., $20^{\circ}$ instead of $35^{\circ}$. Once again, in the unheated body case, an eddy-like shape of the droplet dispersion can be observed. First, droplets are transported upwards towards the ceiling until they are recirculated back downwards, where many of the droplets come into contact with the floor, as seen in Figure~\ref{fig:S6S7}~(a). The ventilation is directed further towards the floor behind the person, which is likely the reason for a change in the velocity profile and the modified recirculation pattern. 

In the heated simulation, Figure~\ref{fig:S6S7}~(b), it can be observed that the ventilation is starting to impact (counteract) the effect of the plume. There is some significant downwards recirculation and many more droplets come into contact with the ground. The upward dispersion of the plume is not as clearly visible as in the previous ventilation patterns, indicating that the ventilation is playing a more significant role. Hence, the impact of the plume on the dispersion of droplets is also dependent on the angle at which the ventilation inflow is directed.

For the maximum ventilation rate of 12 volume changes per hour shown in Figure~\ref{fig:S8S9}, the thermal plume has a much smaller effect on the droplet dispersion than in the previous simulations. Likewise, there is a general eddy shape, clearly visible in the unheated simulation. Many of the droplets exit through the ventilation outlet, which may explain the strong upward motion towards the ceiling.

\subsubsection*{Conclusions}

After qualitative comparison, it can be concluded that the thermal plume of the human body does have a non-negligible effect on droplet dispersion. Prior to the investigation it was expected that at high ventilation rates, the effect of the thermal plume would diminish. This is confirmed by this study. However, it should be noted that only at very high ventilation rates, such as 12 volume changes per hour, the effect of the thermal plume diminishes. In normally ventilated environments, such as tutorial rooms or lecture theatres, the thermal effects of the human body should not be ignored, as they can significantly impact the dispersion of saliva droplets.

It is hence important for the post-processing of the CFD simulations undertaken in Section~\ref{sec:cfd} to include a model for buoyancy effects. Note that CFD simulations of the ventilation within a room are generally too slow to be performed in real-time speed. It would also not be feasible to create a CFD simulation database for each possible combination of occupants in the room (e.g., for lecture rooms this could be more than 120 factorial simulations). Therefore, an important step to be addressed in future work is to implement an analytical approximation to superimpose the thermal effects of bodies into the background ventilation velocity profile.

\clearpage

\section{Infection risk estimation}\label{sec:infectionRisk}

\vspace{3mm}

\noindent By \emph{A. Giusti, A. Kruse, N. Liniger and D. Fredrich}

\vspace{3mm}

\noindent The developed model allows for the estimation of saliva droplet deposition on surfaces and concentration in the air, as well as evaluation of the carbon dioxide concentration in indoor environments. To make the tool capable of estimating the infection risk, a model to predict (or monitor) the evolution of the viral activity is necessary. This is undoubtedly the most uncertain part of the model for a number of reasons: (i) the number of infected people in a room who are releasing saliva droplets containing the virus is not known (this number is likely to be correlated with the number of asymptomatic individuals in the population of students); (ii) the location of infected occupants is not known a-priori; (iii) the viral load (i.e., the amount of virus per unit volume of saliva) depends on a specific individual; (iv) the evolution of the virus on drying saliva droplets has not been studied yet, and how long the virus survives depending on the availability of saliva and ambient conditions is not known.

\subsection*{Modelling the viral activity}

Accurate and reliable models to predict the evolution of SARS-CoV-2 in saliva droplets, or on surfaces, have not been developed yet. In absence of an established model, following the work by de Oliveira et al.~\cite{Oliviera}, a simple evolutionary equation is used for each droplet:
\begin{equation}\label{eq:virus_decay}
    \frac{\mathrm{d}N_{k,v}}{\mathrm{d}t} = -\lambda N_{k,v},
\end{equation}
where $N_{k,v}$ is the total number of \emph{viable} viral copies in the $k$th droplet. The solution of this ordinary differential equation is an exponential decay. As an approximation, a decay rate of $\lambda=0.636$~$\mathrm{h^{-1}}$ is used for all droplet sizes. This value is based on the half-life of SARS-CoV-2 within droplets of $5$~$\mu \mathrm{m}$ diameter released in an environment with 65\% humidity~\cite{Doremalen2020}. To compute the time evolution of viral copies in each droplet, the initial value, $N_{k,v,0}$, of viral copies is required. This is computed from the initial viral load, $n_{v,0}$, as:
\begin{equation}
    N_{k,v,0} = n_{v,0}V_k,
\end{equation}
where $V_k$ is the volume of the saliva droplet. Values of the viral load (i.e., the number of viral copies per unit volume) range from $10^4$~copies per mL to $10^{14}$ copies per mL as the disease develops in sick individuals. Here, a value of $10^{8}$ copies per mL is assumed as a worst-case scenario for asymptomatic occupants (it is expected that individuals with symptoms are not attending in-person teaching). The total concentration of suspended copies, $N_{v,s}$, in the air domain can eventually be computed during the computation of droplet dispersion.

\subsubsection*{Estimation of infection risk}

To estimate the infection risk, the dose of viable copies needed per individual to lead to the disease must be defined. This value is usually defined in terms of infection risk, $P$, i.e., the probability that an individual gets the disease when a viral dose, $N_v$, is inhaled. Following the study of Watanabe et al.~\cite{Watanabe2010} for SARS-CoV-1, the following expression for $P$ is invoked:
\begin{equation}\label{eq:inf_risk}
    P=1-\mathrm{exp}\left( - \frac{N_v}{k_p}   \right),
\end{equation}
where $k_p=410$ is used. Once the concentration of suspended copies, $N_{v,s}$, in the air (computed as copies per unit volume) is known, and given the volume of air inhaled per breath, $V_{b}$, and the number of breaths per second, $\dot{n}_b$, the following quantities can be computed:
\begin{itemize}
    \item Number of viable viral copies inhaled per breath by each individual:
    \begin{equation}
        N_{v,b}^{\mathrm{ID}} = N_{v,s} V_{b},
    \end{equation}
    where ID indicates each specific occupant (the infection risk may be different depending on the location of the occupant, given that the concentration of suspended droplets changes in space.
    \item Number of cumulative viable viral copies inhaled by each individual, $\overline{N}_{v,b}^{\mathrm{ID}}$. To be conservative, it is assumed that once inhaled, the decay rate of viral copies tends to zero (i.e., the life of the virus tends to infinity):
    \begin{equation}
        \overline{N}_{v,b}^{\mathrm{ID}}(\overline{t}) = \int_0^{\overline{t}} N_{v,b}^{\mathrm{ID}} \dot{n}_b \mathrm{d}t.
    \end{equation}
    Since steady-state conditions are assumed, and also assuming that the presence of occupants does not significantly affect the local concentration in the air (this is also a conservative assumption), the \emph{integrand} is not a function of time. Therefore, the cumulative inhaled copies is a linear function of the residence time in a room. The related infection risk per individual can be computed from Equation~\ref{eq:inf_risk}.
\end{itemize}

\subsection*{Application to the tutorial room}

To provide an example of possible outputs from the developed model, useful for the management and design of indoor spaces, the infection risk in the tutorial room modelled in Section~\ref{sec:cfd} has been estimated. It is assumed that the room is occupied in full capacity (the use of sensors, e.g., on the chairs, would allow us to get the occupancy of the room autonomously and give it as input to the code). In addition, the conservative assumption that all the occupants are asymptomatic carriers is made. This assumption could be relaxed by assuming that only a given number of occupants are already infected (e.g., based on the probability of asymptomatic individuals in the student population). A Monte-Carlo approach may be used to find the final probability by performing a number of simulations with carriers randomly assigned to the occupant population (e.g., using a uniform probability distribution) and then averaging the results to obtain the final probability. This is part of future work.

\begin{figure}[t]
    \centering
    \includegraphics[trim={5cm 0 1cm 0},clip, width=\linewidth]{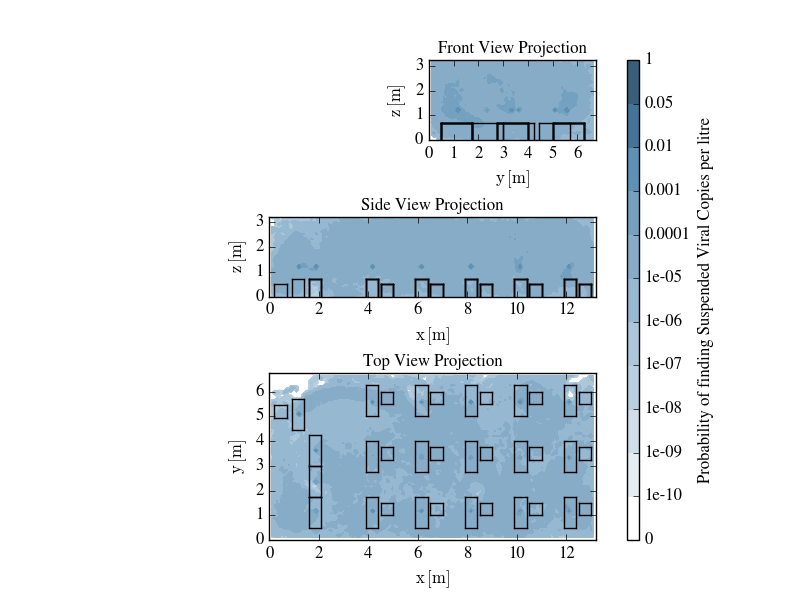}
    \caption{Probability of finding viable viral copies in the air (with respect to the total amount exhaled by the occupants). The quantities are expressed per unit volume of air (in $\mathrm{dm^3}$).}
    \label{fig:TutorialRoomDisp}
\end{figure}

The first quantity we analyse is the probability of finding suspended viral copies per unit volume (of air). This quantity is shown in Figure~\ref{fig:TutorialRoomDisp}. The probability is relatively uniform in the entire room, given the large amount of suspended droplets and the long lifetime of the virus in a single droplet according to Equation~\ref{eq:virus_decay}.

\begin{figure}[t]
    \centering
    \includegraphics[width=\linewidth]{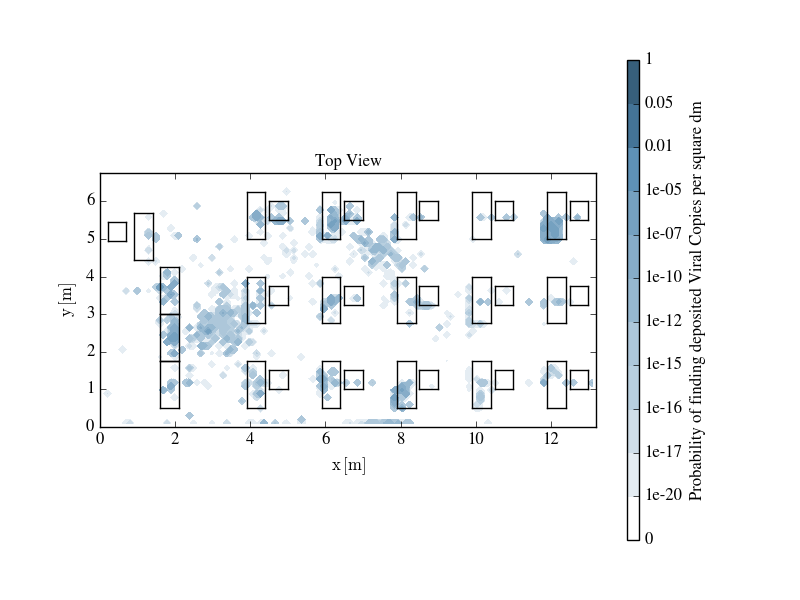}
    \caption{Probability of finding viable viral copies deposited on the bottom surfaces (with respect to the total amount exhaled by the occupants). The quantities are expressed per unit surface area (in $\mathrm{dm^2}$).}
    \label{fig:TutorialRoomDisp2}
\end{figure}

A similar plot can also be produced to show the probability of deposition of viral copies per unit surface area. This is shown in Figure~\ref{fig:TutorialRoomDisp2} for the tutorial room investigated here. Such information could be, for example, very useful to manage the cleaning of surfaces (lighter droplets may be found far from the places where students are sitting, depending on the ventilation field).

\begin{figure}[t]
    \centering
    \includegraphics[trim={5cm 0 1cm 0},clip, width=\linewidth]{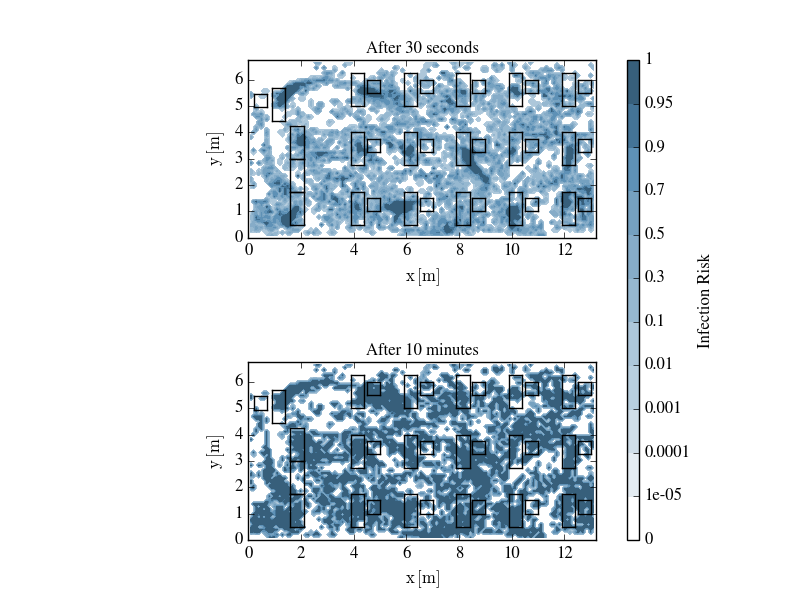}
    \caption{Two-dimensional map of the estimated infection risk for a well ventilated tutorial room at two different residence times (30 seconds and 10 minutes).}
    \label{fig:TutorialRoomDisp3}
\end{figure}

Finally, Figure~\ref{fig:TutorialRoomDisp3} shows the use of the developed model to compute a map of probability of being infected as a function of the exposure time. The map should be interpreted as the probability of infection for a new individual entering the room as a function of the location of their mouth. Curves as a function of residence time in the room can also be plotted for each of the occupants already in the room. The results suggest that for a well ventilated room, with all occupants breathing (small amounts of saliva emitted per breath) and asymptomatic carriers, the risk of infection is reasonably small only for very short exposure periods. The region of high risk is mainly around the infected occupants, however, viable viral copies survive for a long time also relatively far from them.

\clearpage

\section{Summary and future work}\label{sec:outlook}

\vspace{3mm}

\noindent By \emph{A. Giusti and D. Fredrich}

\subsection*{Summary}

A numerical framework for the `real-time' estimation of the infection risk from airborne diseases (e.g., SARS-CoV-2) in indoor environments, such as hospitals, restaurants, cinemas or teaching rooms, has been developed and assessed. The model comprises a pre-processor, which computes the time-averaged ventilation field (only once per room) using computational fluid dynamics. The obtained ventilation field is then used in a post-processing tool to compute the dispersion of saliva droplets suspended in the air, depending on the actual distribution of occupants inside the room. The spatially-resolved concentration and surface deposition rate of droplets, as well as the related infection risk, can thereby be estimated in (near) real time. Note that the model can thus also be used to assist the design of ventilation systems by associating an estimate of the potential infection risk to any given configuration. The model has been developed using open-source software and is based on a number of scripts written in Python that enable the user to compute the dispersion of droplets and to control the computational fluid dynamics software (OpenFOAM). The model is conceived to be user-friendly and highly automated, reducing the number of inputs that require experience with fluid mechanics to find a solution. Different methods to reproduce the effect of turbulence both on the mean ventilation field and the droplet dispersion have been evaluated. In addition to the implementation and assessment of the model, this project also investigated certain physical phenomena that could have an impact on the dispersion of human exhaled sprays and aerosols. In particular, the effect of thermal plumes caused by the human body has been analysed, showing that in general, the flow induced by buoyancy is not negligible for low to moderate ventilation rates. Methods to compute the dispersion of carbon dioxide have also been implemented and evaluated with the aim of extending the model to the computation of indoor air quality, regardless of the potential presence of any viable viral copies.

\subsection*{Future work}

Although substantial progress has been made towards the `real-time' prediction of infection risk in indoor environments, there are still a number of areas that require further work to have a final product to be delivered to users. In addition, this work suggested possible future investigations to improve the accuracy and robustness of the model. In the following, the key points for future activities are highlighted, grouped on the basis of their main purpose.

\subsubsection*{Fully-automated software}

The baseline model has been developed. However, to have a fully-automated software operating in e.g., lecture and tutorial rooms, further steps of development must be undertaken.
\begin{itemize}
    \item Improvement of the graphical user interface (GUI): The user should be able to easily sketch the geometry of the room and add items, such as chairs, desks, windows, etc., as well as specify ventilation rates, the location of the room's occupants and, optionally, certain simulation parameters.
    \item Improved coupling between the pre-processing and post-processing tools: The two input files should be merged so that the user only needs to edit a single file (if preferred over the use of a GUI) with all parameters in one place.
    \item Computation of the background ventilation pattern via CFD: To speed up the computation of the time-averaged ventilation field, simulations could be performed and stored in the cloud.
    \item Development of a library of ventilation fields: A number of possible ventilation rates should in principle be considered. The tool for the computation of the dispersion of droplets must be interfaced to this database. Interpolation for intermediate ventilation rates should be implemented. Note that the range of ventilation rates is defined in advance, depending on the ventilation system and management strategy used for each room.
    \item Computation of droplet dispersion in the cloud: To speed up the computation of droplet dispersion and reach `real-time' (or near `real-time') performance, multi-core computations in a dedicated cluster are foreseen. Adequate parallel computation must also be implemented.
    \item Interface to visualise the estimated infection risk on local screens: Remote infection risk computations (in the cloud or a dedicated cluster/server) should be visualised locally e.g., in the lectern of a lecture theatre or on dedicated screens in health and safety offices. To do so, it is necessary to develop an interface able to visualise and summarise the key results (e.g., infection risk of each occupant, status of the surfaces, etc.), ideally combined with the GUI used for specifying the simulation input parameters.
    \item Automated detection of occupants: The developed model requires the occupied seats as an input. With the objective of obtaining a fully automated method, the use of sensors (e.g, on each chair) interfaced with the software is recommended.
\end{itemize}

\subsubsection*{Modelling improvements}

To improve the accuracy of the estimation, further developments of the underlying models could be done. Note that the following list is meant to stress the most important points highlighted by this work. The modelling of SARS-CoV-2 is an ongoing activity and many organisations and researchers all over the world are investing considerable resources in it. Each study offers new insight into this intricate subject.
\begin{itemize}
    \item Effect of local turbulence generated by breathing: The local effect of breathing has been considered by modifying the ventilation velocity field. However, the turbulence induced by the breath has not been considered. This may have an impact on the dispersion of droplets.
    \item Pulsed nature of breathing and coughing: The breath is not a continuous jet. The average velocity field and induced turbulence may drastically differ from the values of a steady jet. Therefore, it is recommended to further investigate the local velocity field and turbulence induced by human breath and improve their modelling.
    \item Anisotropy of carbon dioxide dispersion: Dispersion coefficients used to model pulsed puffs of carbon dioxide should be modelled as a function of the direction.
    \item Effect of the body's thermal plume on the ventilation field: This study has found that in the case of low and moderate ventilation rates, the body's thermal plume significantly affects the ventilation velocity field, with a subsequent impact on the droplet dispersion. Since running CFD simulations for every possible combination of room occupants is not realistic, `engineered' models to locally modify the velocity field computed in absence of occupants is foreseen.
    \item Effect of humidity: The presence of water vapour in the air has been found to affect the evaporation rate of saliva as well as the decay rate of the virus. The presence of humid air should be modelled to improve the accuracy of the estimations. Further research must be done to investigate the decay rate of the virus in droplets at different ambient conditions.
    \item Composition of the saliva: In this work, the saliva evaporation has been modelled with a modified $\mathscrsfs{D}^2$-law. Further improvements must look at the multi-component nature of the saliva. Experiments may be useful here, as well as a statistical approach to take into account the variability of composition between different individuals.
    \item Development of guidelines for the mesh size: Further simulations of a number of different rooms must be performed to develop a robust best-practice for the computation of the background ventilation field. In particular, the focus should be on the development of guidelines for the mesh resolution. Local refinements around smaller objects should be evaluated.
    \item Experimental validation: An experimental campaign, with measurements of the concentration of sprays, is recommended to fully validate the model. The main obstacles for the experimental campaign are mainly related to intrinsic difficulties of applying laser techniques to measure droplet dispersion in large environments. Methods based on smoke (or carbon dioxide) concentration should be evaluated.
\end{itemize}

\end{document}